\documentclass[fleqn,usenatbib]{mnras}

\usepackage{mathptmx}
\usepackage{txfonts}
\usepackage{pdflscape}
\usepackage[T1]{fontenc}
\usepackage{ae,aecompl}

\usepackage{graphicx}	
\usepackage[normalem]{ulem} 

\usepackage{float}

\usepackage{subcaption}

\newcommand{\beq}{\begin{equation}}
\newcommand{\eeq}{\end{equation}}
\newcommand{\beqa}{\begin{eqnarray}}
\newcommand{\eeqa}{\end{eqnarray}}

\newcommand{\srg}{{\it SRG}}
\newcommand{\erosita}{eROSITA}
\newcommand{\rosat}{{\it ROSAT}}
\newcommand{\xmm}{{\it XMM-Newton}}
\newcommand{\gaia}{{\it Gaia}}
\newcommand{\wise}{{\it WISE}}

\newcommand{\mbh}{M_{\rm BH}} 
\newcommand{\msun}{M_\odot} 
\newcommand{\nh}{N_{\rm H}} 
\newcommand{\nhgal}{N_{\rm H,Gal}} 
 
\newcommand{\lx}{L_{\rm X}} 
\newcommand{\lxmax}{L_{\rm X,\,max}} 
\newcommand{\lopt}{L_g} 
\newcommand{\loiii}{L_{[{\rm OIII}]}}
\newcommand{\tdet}{t_{\rm det}} 
 
\newcommand{\fmax}{F_{\rm max}} 
\newcommand{\becl}{b_{\rm ecl}} 
\newcommand{\tin}{T_{\rm in}} 
\newcommand{\rin}{R_{\rm in}} 
\newcommand{\ledd}{\lambda_{\rm Edd}} 
\newcommand{\spin}{a_\ast} 
\newcommand{\vmax}{V_{\rm max}} 
\newcommand{\dmax}{D_{\rm max}} 
\newcommand{\zmax}{z_{\rm max}} 
\newcommand{\crmin}{CR_{\rm min}}

\usepackage{color}
\usepackage{tcolorbox}

\title[First SRG/eROSITA TDEs]{First tidal disruption events discovered by \srg/\erosita: X-ray/optical properties and X-ray luminosity function at $z<0.6$}
\author[Sazonov et al.]{Sazonov S.$^{1,2}$\thanks{Contact e-mail: \href{mailto:sazonov@iki.rssi.ru}{sazonov@iki.rssi.ru}},
Gilfanov M.$^{1,3}$,
Medvedev P.$^1$,
Yao Y.$^4$,
Khorunzhev G.$^1$,
Semena A.$^1$,
Sunyaev R.$^{1,3}$,
\newauthor
Burenin R.$^1$,
Lyapin A.$^1$,
Meshcheryakov A.$^1$,
Uskov G.$^1$,
Zaznobin I.$^1$,
Postnov K.A.$^5$,
Dodin A.V.$^5$,
\newauthor
Belinski A.A.$^5$,
Cherepashchuk A.M.$^5$,
Eselevich M.$^6$,
Dodonov S.N.$^{7,8}$,
Grokhovskaya A.A.$^{7,8}$,
\newauthor
Kotov S.S.$^{7,8}$,
Bikmaev I.F.$^{9,10}$,
Zhuchkov R.Ya.$^{9,10}$, 
Gumerov R.I.$^{9,10}$,
van Velzen S.$^{11}$
\newauthor
and Kulkarni S.$^4$\\
$^1$Space Research Institute, Russian Academy of Sciences, Profsoyuznaya ul. 84/32, Moscow, 117997, Russia \\
$^2$National Research University Higher School of Economics, Myasnitskaya ul. 20, Moscow, 101000, Russia\\
$^3$Max-Planck-Institut f\"{u}r Astrophysik, Karl-Schwarzschild-Str. 1, D-85741 Garching, Germany \\
$^4$Cahill Center for Astrophysics, California Institute of Technology, MC 249-17, 1200 E California Boulevard, Pasadena, CA 91125, USA\\
$^5$Sternberg Astronomical Institute, Moscow M.V. Lomonosov State University, Universitetskij pr. 13, Moscow, 119992, Russia\\
$^6$Institute of Solar-Terrestrial Physics, Russian Academy of Sciences, Siberian Branch, Lermontov st. 126a, Irkutsk, 664033, Russia\\
$^7$Special Astrophysical Observatory, N. Arkhyz, Karachaevo-Cherkesia, 369167, Russia\\
$^8$Institute of Applied Astronomy of the Russian Academy of Sciences, Kutuzov Quay 10, St. Petersburg, 191187, Russia\\
$^9$Kazan Federal University, Kremlevskaya str.18, 420008 Kazan, Russia\\
$^{10}$Academy of Sciences of Tatarstan, Baumana Str., 20, 420111 Kazan, Russia\\
$^{11}$Leiden Observatory, Leiden University, Postbus 9513, 2300 RA, Leiden, The Netherlands
}

\begin{document}

\maketitle

\begin{abstract}
We present the first sample of tidal disruption events (TDEs) discovered during the \srg\ all-sky survey. These 13 events were selected among X-ray transients detected in the $0<l<180^\circ$ hemisphere by \erosita\ during its second sky survey (10 June -- 14 December 2020) and confirmed by optical follow-up observations. The most distant event occurred at $z=0.581$. One TDE continued to brighten at least 6 months. The X-ray spectra are consistent with nearly critical accretion onto black holes of a few $\times 10^3$ to $10^8\,\msun$, although supercritical accretion is possibly taking place. In two TDEs, a spectral hardening is observed 6 months after the discovery. Four TDEs showed an optical brightening apart from the X-ray outburst. The other 9 TDEs demonstrate no optical activity. All 13 TDEs are optically faint, with $\lopt/\lx<0.3$ ($\lopt$ and $\lx$ being the $g$-band and 0.2--6~keV luminosity, respectively). We have constructed a TDE X-ray luminosity function, which can be fit by a power law with a slope of $-0.6\pm 0.2$, similar to the trend observed for optically selected TDEs. The total rate is estimated at $(1.1\pm 0.5)\times 10^{-5}$ TDEs per galaxy per year, an order of magnitude lower than inferred from optical studies. This suggests that X-ray bright events constitute a minority of TDEs, consistent with models predicting that X-rays can only be observed from directions close to the axis of a thick accretion disk formed from the stellar debris. Our TDE detection threshold can be lowered by a factor of $\sim 2$, which should allow a detection of $\sim 700$ TDEs by the end of the \srg\ survey. 
\end{abstract}


\section{Introduction}
\label{s:intro}

Stellar disruptions by the gravitation of supermassive black holes (SMBHs) -- tidal disruption events (TDEs) -- provide valuable and largely unique information on relatively small ($\mbh\lesssim 10^8\,\msun$) SMBHs in, usually, dormant galactic nuclei and allow us to explore various regimes of accretion onto black holes. Predicted by theorists \citep{hills1975,lidskii1979,gurzadian1981,rees1988}, TDEs were first discovered as soft X-ray transients \citep{Komossa_1999} by the \rosat\ satellite during its all-sky survey in 1990--1991. Soft X-ray emission is one of two expected distinctive features (together with a $t^{-5/3}$ flux decline) of TDEs, since the thermal emission of an accretion disk that forms around the SMBH from the debris of the disrupted star is expected to have a characteristic temperature of $\sim 10^{6}$\,K. Accordingly, X-ray searches remained the main channel of discovering TDEs until recently (e.g. \citealt{Donley_2002,Esquej_2008,maksym2010,Khabibullin_2014a}), and the current list of TDEs discovered in X-rays comprises some 20 events (see \citealt{saxton2021} for a recent review). 

In the last $\sim 15$ years, a new influx of TDE discoveries has emerged from optical/UV surveys, which have provided an additional sample of TDEs that is comparable in size to the X-ray based one (e.g., \citealt{Velzen_2011,Gezari_2012,Blagorodnova_2019,Holoien_2019,Velzen_2020,Velzen_2021}). Interestingly, these optical/UV selected objects seem to have, on average, quite different properties compared to the X-ray selected ones. Specifically, the spectral energy distribution (SED) of the former can be described to a first approximation as thermal emission with a temperature $\sim {\rm a~few~} 10^{4}$\,K, which is at least an order of magnitude lower than expected for near-Eddington standard accretion disks \citep{shakura1973} around moderately massive black holes. 

This apparent dichotomy in TDE SEDs resembles the distinction between type 1 and type 2 active galactic nuclei (AGN), which has led to suggestions that we might be dealing with a similar orientation-driven effect. In particular, it was proposed \citep{dai2018,Curd_2019} that X-ray rich TDEs are observed from directions close to the axis of a thick accretion disk with a powerful wind, whereas X-ray weak ones are viewed from larger inclination angles. In the latter case, the central X-ray source is obscured by the disk so that we can only see the reprocessed optical/UV emission. However, in addition to the viewing direction, other factors, in particular the black hole mass, are also likely to strongly affect TDE SEDs and light curves \citep{Mummery_2021}. In other proposed models, the conversion of the accretion disk emission to optical wavelengths occurs in the unbound part of the stellar debris \citep{Metzger_2016,Lu_2020}, or the optical emission is directly powered by energy liberated during the formation of the accretion disk rather than by energy released during subsequent accretion onto the black hole \citep{Piran_2015}.

It is clear that more observational data in various energy bands are needed for a better understanding of the physics of the TDE phenomenon. Therefore, the launch of the \erosita\ telescope \citep{Predehl2021} on board the \srg\ observatory \citep{Sunyaev_2021} was eagerly awaited, because it was expected to find hundreds to thousands of TDEs during its revolutionary all-sky X-ray survey \citep{khabibullin2014b}. On 13 July 2019, \srg\ was successfully launched from the Baikonur Cosmodrome and on 12 December 2019 it started its all-sky X-ray survey from a halo orbit around the Sun--Earth L2 point. The survey is to consist of 8 consecutive full scans of the sky, each lasting 6 months. Already during the first weeks of the survey a few TDE candidates were found in the \erosita\ data through comparison with archival observations by previous X-ray missions \citep{Khabibullin_2020,Khabibullin_2020b}. On 10 June 2020, the second \erosita\ all-sky survey began, which allowed us to begin a regular search for TDEs over the sky. Specifically, TDE candidates are sought among the multitude of transient X-ray sources detected in a given \erosita\ all-sky survey and undetected in the preceding survey. The second all-sky survey was completed on 14 Dec. 2020, and the third survey has also been finished by now. 

In this paper, we present an initial sample of 13 relatively bright TDEs discovered by \srg/\erosita\ during its second all-sky survey at $0<l<180^\circ$\footnote{Analysis of \erosita\ data in this half of the sky is performed by the Russian \erosita\ consortium.}. The TDE nature of these transients was suggested by their X-ray properties and then confirmed by our follow-up optical observations. Below we discuss the X-ray and optical properties of this TDE sample and use it to draw inferences on the statistical properties of TDEs in the $z<0.6$ Universe. 

In what follows we adopt a flat $\Lambda$ cold dark matter cosmological model with $h=0.70$ and $\Omega_\Lambda=0.7$. 

\section{Selection of TDE candidates}
\label{s:selection}

\begin{table*}
  \caption{The first \srg\ TDE sample.} 
  \label{tab:sample}
  \begin{tabular}{rlllrlrlr}
  \hline
  \multicolumn{1}{c}{No.} &
  \multicolumn{1}{c}{X-ray source} &
  \multicolumn{1}{c}{$R_{98}$$^1$} &
  \multicolumn{2}{c}{eRASS1:} &
  \multicolumn{2}{c}{eRASS2:} &
  \multicolumn{2}{c}{eRASS3:} \\
  & & & \multicolumn{1}{c}{Dates} & \multicolumn{1}{c}{X-ray flux$^2$} & \multicolumn{1}{c}{Dates} & \multicolumn{1}{c}{X-ray flux} & \multicolumn{1}{c}{Dates} & \multicolumn{1}{c}{X-ray flux}\\
  \hline
1 & SRGE\,J135514.8+311605 & $ 4.9 $ & 2019 Dec 22--24 & $< 2.7 \times 10^{-14} $ & 2020 Jun 23--24 & $ 3.8 \times 10^{-13} $ & 2020 Dec 24--25 & $< 2.0 \times 10^{-14} $ \\[0.1cm]
2 & SRGE\,J013204.6+122236 & $ 4.8 $ & 2020 Jan 08--09 & $< 2.3 \times 10^{-14} $ & 2020 Jul 08--09 & $ 2.8 \times 10^{-13} $ & 2021 Jan 07--08 & $< 2.5 \times 10^{-14} $ \\[0.1cm]
3 & SRGE\,J153503.4+455056 & $ 3.8 $ & 2020 Jan 11--14 & $< 1.1 \times 10^{-14} $ & 2020 Jul 13--16 & $ 3.2 \times 10^{-13} $ & 2021 Jan 09--11 & $ 1.6 \times 10^{-13} $ \\[0.1cm]
4 & SRGE\,J163831.7+534020 & $ 4.5 $ & 2020 Jan 30--04 & $< 6.7 \times 10^{-15} $ & 2020 Aug 01--08 & $ 1.2 \times 10^{-13} $ & 2021 Jan 22--26 & $< 8.4 \times 10^{-15} $ \\[0.1cm]
5 & SRGE\,J163030.2+470125 & $ 4.0 $ & 2020 Feb 08--16 & $< 1.4 \times 10^{-14} $ & 2020 Aug 06--14 & $ 3.7 \times 10^{-13} $ & 2021 Jan 29--31 & $< 1.7 \times 10^{-14} $ \\[0.1cm]
6 & SRGE\,J021939.9+361819 & $ 5.5 $ & 2020 Feb 05--07 & $< 1.8 \times 10^{-14} $ & 2020 Aug 07--09 & $ 2.4 \times 10^{-13} $ & 2021 Jan 27--28 & $ 1.5 \times 10^{-13} $ \\[0.1cm]
7 & SRGE\,J161001.2+330121 & $ 5.1 $ & 2020 Feb 12--18 & $< 1.2 \times 10^{-14} $ & 2020 Aug 14--16 & $ 1.8 \times 10^{-13} $ & 2021 Feb 01--02 & $< 2.3 \times 10^{-14} $ \\[0.1cm]
8 & SRGE\,J171423.6+085236 & $ 3.5 $ & 2020 Mar 18--19 & $< 3.0 \times 10^{-14} $ & 2020 Sep 18--19 & $ 1.2 \times 10^{-12} $ & 2021 Mar 14--16 & $ 7.7 \times 10^{-14} $ \\[0.1cm]
9 & SRGE\,J071310.6+725627 & $ 4.0 $ & 2020 Apr 07--08 & $< 2.8 \times 10^{-14} $ & 2020 Oct 11--12 & $ 1.4 \times 10^{-12} $ & 2021 Apr 06--07 & $ 2.0 \times 10^{-13} $ \\[0.1cm]
10 & SRGE\,J095928.6+643023 & $ 4.8 $ & 2020 Apr 25--26 & $< 2.0 \times 10^{-14} $ & 2020 Oct 28--29 & $ 3.9 \times 10^{-13} $ & 2021 Apr 28--30 & $ 3.0 \times 10^{-13} $ \\[0.1cm]
11 & SRGE\,J091747.6+524821 & $ 5.9 $ & 2020 Apr 26--27 & $< 4.5 \times 10^{-14} $ & 2020 Oct 28--29 & $ 4.6 \times 10^{-13} $ & 2021 Apr 29--30 & $< 4.5 \times 10^{-14} $ \\[0.1cm]
12 & SRGE\,J133053.3+734824 & $ 5.3 $ & 2020 May 01--03 & $< 1.9 \times 10^{-14} $ & 2020 Nov 02--04 & $ 2.8 \times 10^{-13} $ & 2021 May 05--07 & $< 1.9 \times 10^{-14} $ \\[0.1cm]
13 & SRGE\,J144738.4+671821 & $ 5.4 $ & 2020 May 20--22 & $< 2.4 \times 10^{-14} $ & 2020 Nov 19--21 & $ 2.6 \times 10^{-13} $ & 2021 May 20--22 & $ 6.3 \times 10^{-13} $ \\[0.1cm]

  \hline
  \end{tabular}

Notes: (1) Radius of the 98\% localization region in eRASS2 in units of arcsec; (2) observed fluxes or upper limits (3$\sigma$) in the 0.3--2.2\,keV energy band in units of erg\,s$^{-1}$\,cm$^{-2}$. 

\end{table*}

The initial sample of X-ray transients for this pilot study consisted of sources that were undetected in the first \erosita\ all-sky survey (hereafter eRASS1) but were detected during the second scan (hereafter eRASS2) at a flux level exceeding at least tenfold the upper limit at a likelihood of 6 ($\approx 3\sigma$) on their flux in the 0.3--2.2\,keV energy band during eRASS1. The search for transients spanned the entire period of eRASS2 (10 June -- 14 Dec. 2020). By the time of this writing, all transients found in eRASS2 have been scanned by \erosita\ for the third time during eRASS3. 

Among these X-ray transients, we filtered out objects of likely Galactic origin based on positional coincidence of the X-ray source with a star having a statistically significant ($>5\sigma$) parallax and/or proper motion in the \gaia\ astrometric catalog \citep{Gaia2016, GaiaDR3}. For the majority of the remaining, potentially extragalactic, objects, a likely counterpart was readily found in archival optical and/or infrared images within the \erosita\ localization region (of $\sim 5$\,arcsec radius). Some of these counterparts are known AGN or have signatures suggestive of an AGN origin, namely, archival detections by previous X-ray missions, a $W1-W2>0.5$ color \citep{Assef_2013} in the \wise\ infrared all-sky survey \citep{Wright_2010}, or irregular optical variability in the Zwicky Transient Facility (ZTF; \citealt{Bellm2019b, Graham2019, Masci2019}) data many months before the \erosita\ X-ray discovery. In addition, we analyzed the \erosita\ data and found that a number of transients had very soft X-ray spectra, suggesting a TDE origin. 

For a final identification, we have been following up the presumably extragalactic \erosita\ transients detected in eRASS2, apart from the indisputable AGN, with optical spectroscopy (\S\ref{s:opt}). This campaign has now come to an end and allowed us to discriminate TDEs from AGN. The new AGN discovered during this campaign will be discussed elsewhere.

As a result, we have obtained a sample of firmly established TDEs detected at $0<l<180^\circ$ during eRASS2. The limiting X-ray flux of our sample varies across the sky depending on the sensitivity achieved at a given location in the first \srg\ sky survey. In the present paper, we discuss the 13 events from this catalog that have revealed themselves as bright transients in X-rays only (although a few of them also showed moderate activity in the optical, as will be discussed in \S\ref{s:opt}). In a companion paper (Gilfanov et al., in preparation), we discuss the subset of \erosita\ TDEs with prominent optical transient counterparts.

\section{The sample and its X-ray properties}
\label{s:xray}

Table~\ref{tab:sample} presents our TDE sample. Specifically, the following information is provided: (1) the source name in the \erosita\ catalog (``SRGE'' followed by the equatorial coordinates of the source measured in eRASS2), (2) the radius of the localization region (at the 98\% confidence level) in eRASS2, (3) the dates when the source was scanned during eRASS1, (4) the upper limit (3$\sigma$) on the source flux (0.3--2.2\,keV) during this period, (5) the dates when the source was scanned during eRASS2, (6) the average X-ray flux during this period, (7) the dates when the source was scanned during eRASS3, and (8) the average X-ray flux or 3$\sigma$ upper limit for this period. The fluxes and upper limits were determined from the measured count rates using the results of our X-ray spectral analysis, discussed in \S\ref{s:xrayspec} below, namely the best-fitting parameters of eRASS2 and eRASS3 spectra by a multi-blackbody accretion disk emission model if available or the corresponding eRASS2 spectral parameters in the remaining cases. 

Weak, marginally significant ($\approx 3.4\sigma$) X-ray emission at the position of SRGE\,J135514.8+311605 was registered by \erosita\ during eRASS1. This emission may be associated with a quasar (which was revealed by our follow-up optical spectroscopy), J135514.83+311612.7, that is located 8\arcsec\ away, i.e. close to the \erosita\ localization region of SRGE\,J135514.8+311605. In computing the upper limit on the eRASS1 flux of this TDE, we took into account all counts registered near this source.

\subsection{Data reduction}
\label{s:xraydata}

\erosita\ raw data were processed by the calibration pipeline developed at IKI based on the \erosita\ Science Analysis Software System (eSASS\footnote{https://erosita.mpe.mpg.de/}, \citealt{Brunner2021}) and using early in-flight calibration data. Source spectra and light curves were extracted using a circular aperture of 60\arcsec\ (corresponding to $\approx$90\% encircled energy) centered on the best-fitting X-ray source position. An annulus with the inner and outer radii of 150\arcsec\ and 450\arcsec\ around the source position was used for background extraction. Any faint sources detected in the background extraction area were masked out using a 15\arcsec\ circular mask. 

We took advantage of the soft X-ray response of \erosita\ to perform an X-ray spectral analysis of the TDEs in the 0.2--6\,keV energy band. To this end, we used the 5 (out of 7) telescope modules that are equipped with the on-chip filter. Source detection and construction of X-ray light curves was done in the 0.3--2.2\,keV energy band, using the data from all operational telescope modules.  

The spectral analysis was done using \textsc{xspec}, version 12.11.1 \citep{arnaud1996}. The quality of spectral fits was accessed using W-statistic for the source and background spectra with Poisson statistics. To avoid the well-known bias in a profile likelihood, we bin the source spectrum so that every bin in the corresponding background spectrum contains at least 5 counts. The binning was done using the ftgrouppha tool of the \textsc{HEASoft} package (v. 6.28). Errors for the best-fitting parameters are quoted at the 90\% confidence level.

\begin{figure*}
\centering
\includegraphics[width=0.95\textwidth]{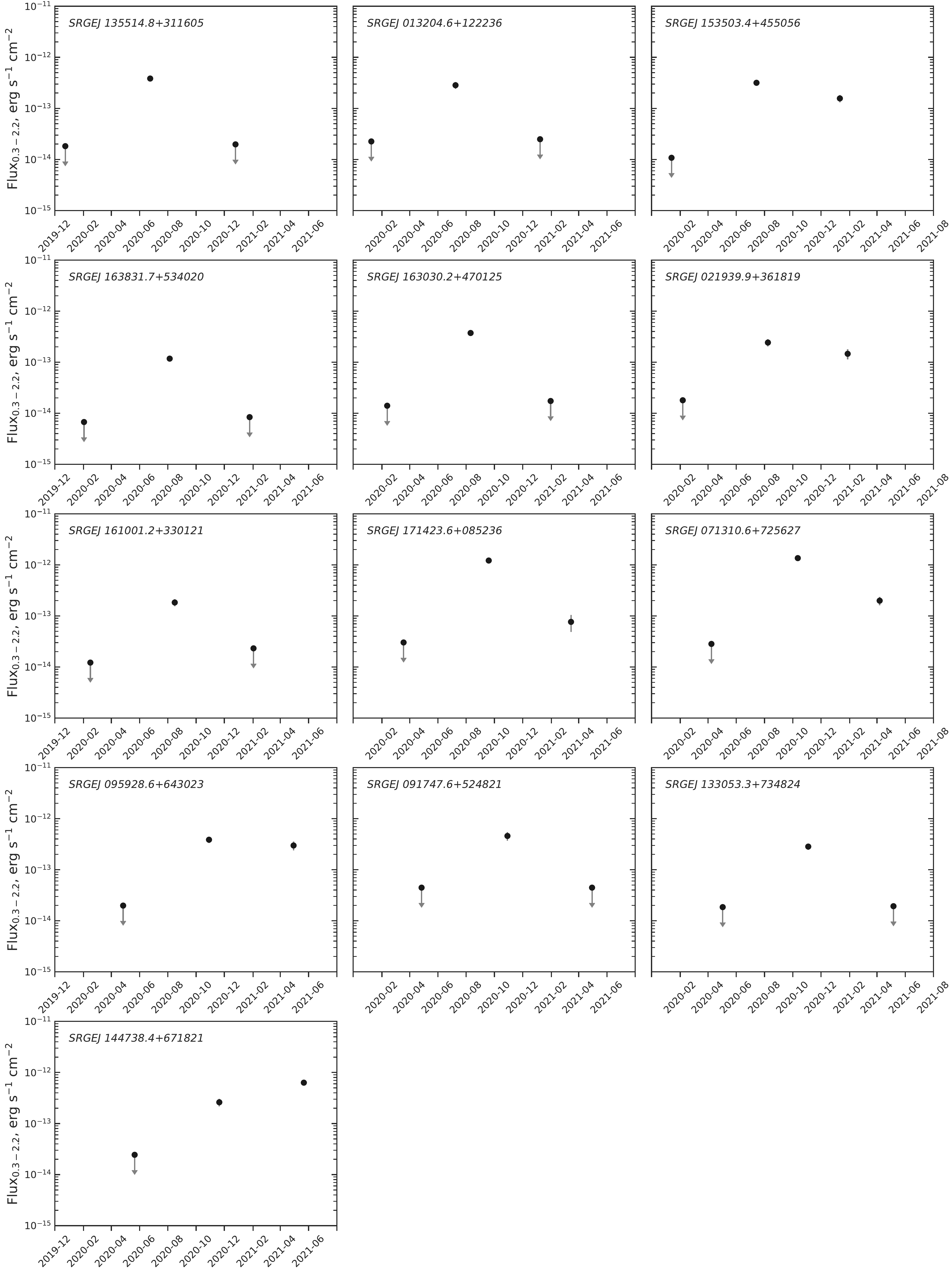}
    \caption{X-ray light curves of the TDEs obtained by \erosita\ in the 0.3--2.2\,keV energy range. 
    }
    \label{fig:xraylc}
\end{figure*}

\subsection{X-ray light curves}
\label{s:xraylc}


\begin{table}
  \caption{Time of TDE onset, assuming a $t^{-5/3}$ law.} 
  \label{tab:peak}
  \begin{tabular}{lr}
  \hline
  \multicolumn{1}{c}{Object} & 
  \multicolumn{1}{c}{$\tdet-t_0$, month} \\
  \hline 
SRGE\,J135514.8+311605 & $<1.7$ \\
SRGE\,J013204.6+122236 & $<1.9$ \\
SRGE\,J153503.4+455056 & $11.4$ \\
SRGE\,J163831.7+534020 & $<1.1$ \\
SRGE\,J163030.2+470125 & $<1.3$ \\
SRGE\,J021939.9+361819 & $16.9$ \\
SRGE\,J161001.2+330121 & $<1.7$ \\
SRGE\,J171423.6+085236 & $5.6$ \\
SRGE\,J071310.6+725627 & $2.8$ \\
SRGE\,J095928.6+643023 & $35.3$ \\
SRGE\,J091747.6+524821 & $<2.2$ \\
SRGE\,J133053.3+734824 & $<1.5$ \\
SRGE\,J144738.4+671821 & undefined \\
  \hline
  \end{tabular}

\end{table}

Figure~\ref{fig:xraylc} shows the X-ray (0.3--2.2\,keV) light curves of the TDEs based on \erosita\ data. Specifically, these light curves  consist of three flux measurements or upper limits (see Table~\ref{tab:sample}) taken at 6-month intervals during eRASS1, eRASS2, and eRASS3. The fluxes and upper limits for each source were determined from the measured count rates using the spectral parameters derived from our X-ray spectral analysis (see \S\ref{s:xrayspec} below). In addition, we present in Appendix~\ref{s:shortlc} the short-term X-ray light curves of the TDEs obtained during their visits by \erosita\ in eRASS2 and eRASS3, which last between $\sim 1$ and $\sim 7$ days. None of the transients have demonstrated substantial variability on these short time scales.

Twelve out of the 13 TDEs have faded during the 6-month interval between their passages by \erosita\ in eRASS2 and eRASS3, with 7 of them not detected anymore in the third scan. The amplitude of the flux drop varies between a factor of $\sim 1.3$ and $\gtrsim 20$. However, SRGE\,J144738.4+671821 has instead become brighter by a factor of $\sim 2$ in eRASS3 compared to eRASS2. 

Assuming that \erosita\ caught all these events at a canonical $t^{-5/3}$ TDE decay phase \citep{Evans_1989}, we can use the eRASS2 and eRASS3 flux measurements/upper limits to evaluate the time of onset of a given TDE relative to the date of its discovery during eRASS2, $\tdet-t_0$. Specifically, we assume that the TDE X-ray light curve consists of an initial rise/peak (possibly supercritical) phase with duration $\tau$ and flux $\fmax$ and a subsequent phase when the flux (0.3--2.2\,keV) decreases as
\begin{equation}
F(t)=\fmax\left(\frac{t_0-t}{\tau}\right)^{-5/3}.
\label{eq:decay}
\end{equation}
The resulting estimates are presented in Table~\ref{tab:peak}. The maximum flux ($\fmax$) depends on the unknown $\tau$, and we thus do not try to estimate it.

For the majority of the TDEs, $\tdet-t_0<6$\,months, and, moreover, $\tdet-t_0<2$\,months. This implies that these events started somewhere between their first and second visits by \srg, and in fact shortly before the eRASS2 observations. This is consistent with the non-detection of these transients in eRASS1. However, $\tdet-t_0>6$\,months for three events. This implies that their X-ray temporal behavior cannot be described by a $t^{-5/3}$ law and that \erosita\ probably caught these TDEs during eRASS2 at a rise/peak phase. An even more extreme case is the already mentioned SRGE\,J144738.4+671821, which has brightened between eRASS2 and eRASS3.

We stress that the above X-ray variability analysis is currently based on just two flux measurements/upper limits per TDE. As five more scans of the entire sky are planned during the \srg\ mission, it should be possible to add a few more data points to the X-ray light curves of at least some of these TDEs and draw firmer conclusions about their long-term behavior. Dedicated follow-up observations by other X-ray observatories can also be helpful in studying this unique TDE sample. 

The above discussion was based on the assumption that the observed luminosity in the 0.3--2.2\,keV band (which corresponds to a factor of $(1+z)$ harder X-ray band in the TDE rest frame) is proportional to the accretion rate, $\dot{M}$. In reality, when the accretion rate drops below the critical Eddington rate, the maximum temperature of the accretion disk \citep{shakura1973} is expected to decline as $\dot{M}^{-1/4}$. As a result, \erosita\ should at some point start probing the Wien tail of the disk thermal emission, with the X-ray flux declining exponentially \citep{Lodato2011}. This might affect some of the the \erosita\ light curves presented here.

We also note that optically selected TDEs often show dramatic soft X-ray variability in contrast to the smooth $t^{-5/3}$ power-law decline of the UV/optical light curve (e.g., \citealt{Velzen_2021}). In particular, TDEs ASASSN-15oi \citep{Gezari2017} and AT~2019azh/ASASSN-19dj \citep{Liu2019,Velzen_2021,Hinkle2021} showed a prominent (by a factor of $\sim 20$ and $\sim 100$, respectively) brightening in X-rays over a period of $\sim 250$ days since the peak of the initial optical-UV flare, which might be attributed to a delayed formation of an accretion disk. Something similar might have been occuring in SRGE\,J144738.4+671821, for which we see an X-ray brightening on a $\sim 180$~day timescale, although there is no evidence of prominent optical activity for this TDE (see \S\ref{s:opt} below).

\begin{figure*}
\centering
\includegraphics[width=0.95\textwidth]{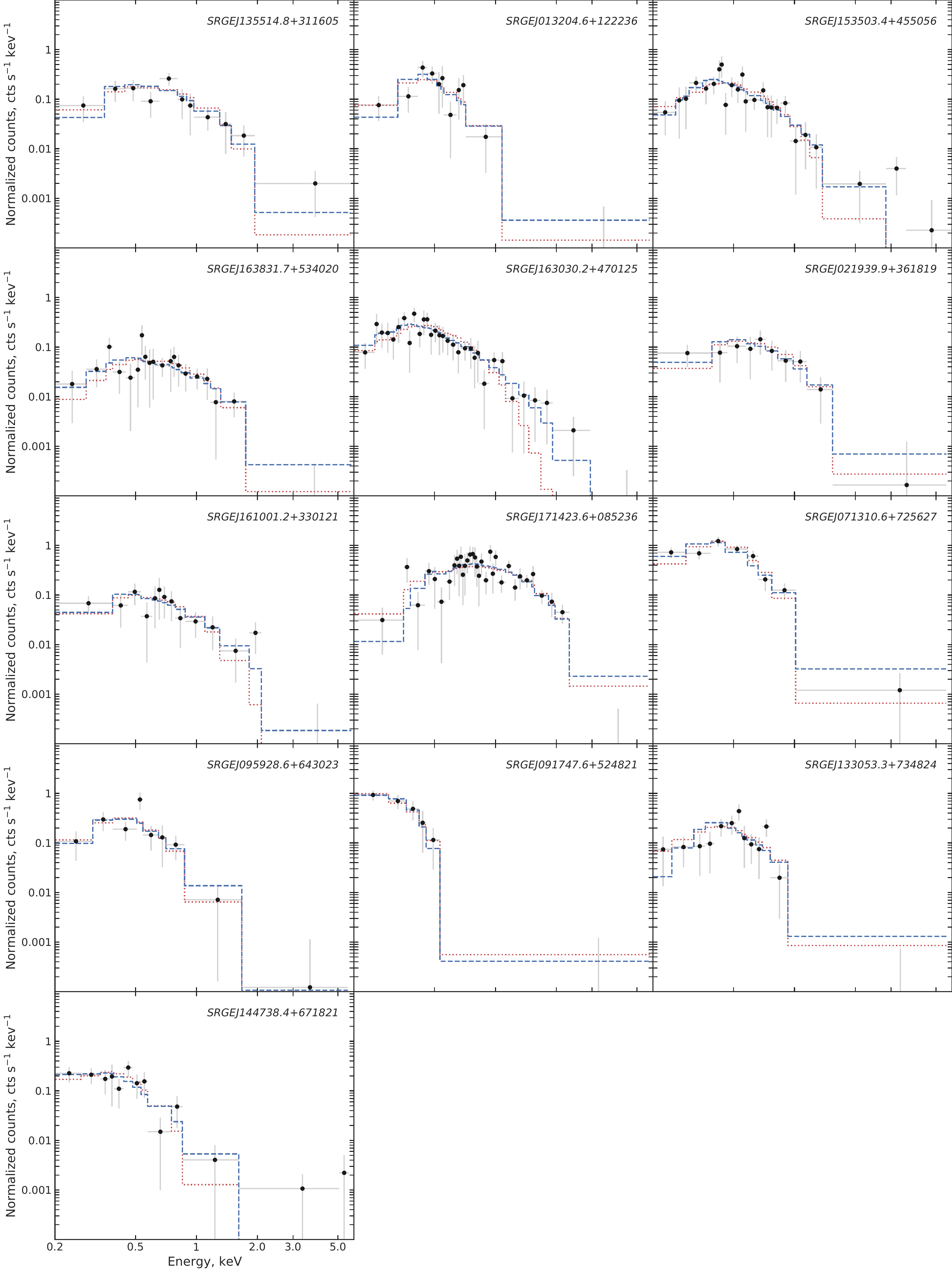}
\caption{X-ray spectra of the TDEs obtained by \erosita\ during the second \srg\ all-sky survey, fitted by the absorbed power-law model (blue dashed line) and the color temperature corrected disk model (red dotted line) in the 0.2--6\,keV energy range; see Tables~\ref{tab:bestfit} and \ref{tab:optxagnf} for the corresponding best-fitting parameters. 
The spectral channels are combined in bins with at least $1\sigma$ significance (for plotting purposes only).
}
\label{fig:xrayspec}
\end{figure*}

\begin{table}
\setlength{\tabcolsep}{1pt}
  \caption{Parameters of simple models applied to eRASS2 spectra.} 
  \label{tab:bestfit}
  \begin{tabular}{lcccc}
  \hline
  \multicolumn{5}{c}{Absorbed power-law model: \textsc{tbabs*zphabs*zpowerlw}} \\[0.1cm]
  \multicolumn{1}{c}{Object (SRGE)} &
  \multicolumn{1}{c}{$\nh$} &
  \multicolumn{1}{c}{$K_{1\,{\rm keV}}^{a}$} &
  \multicolumn{1}{c}{$\Gamma$} & 
  \multicolumn{1}{c}{cstat/} \\
  & \multicolumn{1}{c}{$10^{21}$\,cm$^{-2}$} & & & \multicolumn{1}{c}{d.o.f} \\
  \hline

J135514.8+311605 & $ 1.9_{-1.5}^{+2.7} $ & $ 3.1_{-1.2}^{+2.8} \times 10^{-4} $ & $ 3.8_{-1.4}^{+1.6} $ & 28.5/25 \\[0.1cm]
J013204.6+122236 & $ 2.3_{-1.7}^{+3.9} $ & $ 7.2_{-4.0}^{+7.0} \times 10^{-5} $ & $ 7.1_{-2.2}^{+3.8} $ & 35.7/43 \\[0.1cm]
J153503.4+455056 & $ 1.2_{-0.8}^{+1.2} $ & $ 1.5_{-0.4}^{+0.6} \times 10^{-4} $ & $ 4.6_{-0.9}^{+1.1} $ & 68.4/64 \\[0.1cm]
J163831.7+534020 & $ 0.7_{-0.7}^{+2.9} $ & $ 1.6_{-0.5}^{+1.6} \times 10^{-4} $ & $ 3.0_{-0.6}^{+1.2} $ & 91.1/99 \\[0.1cm]
J163030.2+470125 & $ 0.6_{-0.6}^{+0.7} $ & $ 1.9_{-0.4}^{+0.5} \times 10^{-4} $ & $ 4.3_{-0.7}^{+0.8} $ & 59.8/77 \\[0.1cm]
J021939.9+361819 & $ 1.8_{-1.8}^{+5.0} $ & $ 3.4_{-1.7}^{+7.2} \times 10^{-4} $ & $ 4.1_{-1.4}^{+2.5} $ & 19.5/29 \\[0.1cm]
J161001.2+330121 & $ 0.7_{-0.7}^{+1.5} $ & $ 9.5_{-3.3}^{+4.8} \times 10^{-5} $ & $ 3.4_{-1.1}^{+1.5} $ & 44.9/43 \\[0.1cm]
J171423.6+085236 & $ 2.9_{-1.7}^{+2.3} $ & $ 1.1_{-0.4}^{+0.8} \times 10^{-3} $ & $ 3.1_{-0.8}^{+1.0} $ & 62.8/55 \\[0.1cm]
J071310.6+725627 & $ 0.4_{-0.4}^{+0.5} $ & $ 1.9_{-0.6}^{+0.7} \times 10^{-4} $ & $ 5.4_{-0.8}^{+0.9} $ & 27.5/19 \\[0.1cm]
J095928.6+643023 & $ 2.4_{-2.4}^{+21.2} $ & $ 4.6_{-2.3}^{+148.4} \times 10^{-4} $ & $ 5.7_{-2.1}^{+8.9} $ & 18.6/17 \\[0.1cm]
J091747.6+524821 & $ 1.6_{-1.2}^{+1.9} $ & $ 9.5_{-9.3}^{+91.0} \times 10^{-7} $ & $ 12.4_{-4.3}^{+5.9} $ & 9.9/20 \\[0.1cm]
J133053.3+734824 & $ 1.7_{-1.1}^{+4.8} $ & $ 8.7_{-3.5}^{+7.1} \times 10^{-5} $ & $ 5.7_{-1.4}^{+3.8} $ & 36.3/38 \\[0.1cm]
J144738.4+671821 & $ 0.3_{-0.3}^{+0.6} $ & $ 3.0_{-1.6}^{+2.3} \times 10^{-5} $ & $ 5.2_{-1.1}^{+1.6} $ & 34.9/41 \\[0.1cm]

\multicolumn{5}{c}{Multi-blackbody accretion disk model: \textsc{tbabs*zashift*diskbb}} \\[0.1cm]
  \multicolumn{1}{c}{Object (SRGE)} &
  \multicolumn{1}{c}{$\rin\sqrt{\cos\theta}$} &
  \multicolumn{1}{c}{$k\tin$} & 
  \multicolumn{1}{c}{cstat/} \\ 
  & \multicolumn{1}{c}{km} & \multicolumn{1}{c}{eV} & \multicolumn{1}{c}{d.o.f}\\
  \hline
  
J135514.8+311605 & ~~$ 1.9_{-0.8}^{+1.3} \times 10^{5} $ & $ 324_{-66}^{+100} $ & 30.3/26 \\[0.1cm]
J013204.6+122236 & ~~$ 1.3_{-0.7}^{+1.4} \times 10^{6} $ & $ 121_{-24}^{+31} $ & 36.9/44 \\[0.1cm]
J153503.4+455056 & ~~$ 7.6_{-2.5}^{+3.5} \times 10^{5} $ & $ 188_{-24}^{+30} $ & 69.2/65 \\[0.1cm]
J163831.7+534020 & ~~$ 3.1_{-1.4}^{+2.4} \times 10^{5} $ & $ 413_{-90}^{+128} $ & 93.5/100 \\[0.1cm]
J163030.2+470125 & ~~$ 1.4_{-0.4}^{+0.6} \times 10^{6} $ & $ 178_{-23}^{+28} $ & 73.6/78 \\[0.1cm]
J021939.9+361819 & ~~$ 5.7_{-3.0}^{+6.1} \times 10^{5} $ & $ 288_{-69}^{+105} $ & 20.4/30 \\[0.1cm]
J161001.2+330121 & ~~$ 1.4_{-0.8}^{+1.5} \times 10^{5} $ & $ 247_{-64}^{+117} $ & 49.4/44 \\[0.1cm]
J171423.6+085236 & ~~$ 1.8_{-0.6}^{+0.8} \times 10^{4} $ & $ 527_{-87}^{+126} $ & 60.7/56 \\[0.1cm]
J071310.6+725627 & ~~$ 2.8_{-0.9}^{+1.2} \times 10^{6} $ & $ 109_{-11}^{+13} $ & 26.3/20 \\[0.1cm]
J095928.6+643023 & ~~$ 3.9_{-2.0}^{+3.9} \times 10^{6} $ & $ 164_{-32}^{+49} $ & 19.4/18 \\[0.1cm]
J091747.6+524821 & ~~$ 4.4_{-2.2}^{+4.6} \times 10^{7} $ & $ 54_{-8}^{+10} $ & 10.1/21 \\[0.1cm]
J133053.3+734824 & ~~$ 7.4_{-3.0}^{+4.9} \times 10^{5} $ & $ 150_{-24}^{+31} $ & 30.3/39 \\[0.1cm]
J144738.4+671821 & ~~$ 1.7_{-0.8}^{+1.4} \times 10^{6} $ & $ 100_{-17}^{+22} $ & 36.3/42 \\[0.1cm]

\hline
 \end{tabular}
 
Notes: $^{a}$ --- power-law normalization at 1\,keV in the source's rest frame, in units of photons\,keV$^{-1}$\,cm$^{-2}$\,s$^{-1}$.
\end{table}
 
\subsection{X-ray spectra}
\label{s:xrayspec}

We analyzed the X-ray spectra of the TDEs in their ``bright'' phase, using the \erosita\ data obtained during eRASS2 (Fig.~\ref{fig:xrayspec}). Four transients after their discovery in eRASS2 remained sufficiently bright (at least 30 detector counts) six months later (during eRASS3) to allow us to analyze their spectra obtained at this late phase.  

\begin{table}
\setlength{\tabcolsep}{3pt}
 \caption{Parameters of the color temperature corrected disk model applied to eRASS2 spectra.} 
 \label{tab:optxagnf}
 \begin{tabular}{lcccc}
  \hline
  \multicolumn{5}{c}{\textsc{tbabs*optxagnf}} \\[0.1cm]
  \multicolumn{1}{l}{Object (SRGE)} &
  \multicolumn{1}{c}{$\spin$} &
  \multicolumn{1}{c}{$\log\ledd$} &
  \multicolumn{1}{c}{$M,\,\msun$ } &
  \multicolumn{1}{c}{cstat/d.o.f} \\ 
  \hline
J135514.8+311605 & $ 0 $ & $ 1.13_{-0.20}^{+0.25} $ & $ 3.3_{-1.6}^{+2.7} \times 10^{4} $ & 30.4/26 \\[0.1cm]
J013204.6+122236 & $ 0 $ & $ 0.20_{-0.12}^{+0.14} $ & $ 3.2_{-1.7}^{+3.7} \times 10^{5} $ & 36.8/44 \\[0.1cm]
J153503.4+455056 & $ 0 $ & $ 0.73_{-0.10}^{+0.11} $ & $ 1.5_{-0.5}^{+0.8} \times 10^{5} $ & 69.3/65 \\[0.1cm]
J163831.7+534020 & $ 0 $ & $ 1.68_{-0.21}^{+0.24} $ & $ 3.9_{-1.9}^{+3.5} \times 10^{4} $ & 93.5/100 \\[0.1cm]
J163030.2+470125 & $ 0 $ & $ 0.87_{-0.10}^{+0.11} $ & $ 2.6_{-0.9}^{+1.3} \times 10^{5} $ & 74.0/78 \\[0.1cm]
J021939.9+361819 & $ 0 $ & $ 1.35_{-0.21}^{+0.26} $ & $ 9.1_{-5.1}^{+10.8} \times 10^{4} $ & 20.5/30 \\[0.1cm]
J161001.2+330121 & $ 0 $ & $ 0.54_{-0.25}^{+0.35} $ & $ 2.9_{-1.8}^{+3.6} \times 10^{4} $ & 49.4/44 \\[0.1cm]
J171423.6+085236 & $ 0 $ & $ 1.07_{-0.18}^{+0.22} $ & $ 3.3_{-1.2}^{+1.6} \times 10^{3} $ & 60.7/56 \\[0.1cm]
J071310.6+725627 & $ 0 $ & $ 0.34_{-0.06}^{+0.07} $ & $ 7.0_{-2.3}^{+3.3} \times 10^{5} $ & 26.5/20 \\[0.1cm]
J095928.6+643023 & $ 0 $ & $ 1.12_{-0.13}^{+0.17} $ & $ 6.8_{-3.8}^{+7.3} \times 10^{5} $ & 19.4/18 \\[0.1cm]
J091747.6+524821 & $ 0 $ & $ 0.21_{-0.06}^{+0.06} $ & $ 1.2_{-0.6}^{+1.3} \times 10^{7} $ & 10.1/21 \\[0.1cm]
J133053.3+734824 & $ 0 $ & $ 0.33_{-0.12}^{+0.13} $ & $ 1.7_{-0.7}^{+1.2} \times 10^{5} $ & 30.2/39 \\[0.1cm]
J144738.4+671821 & $ 0 $ & $ -0.04_{-0.11}^{+0.12} $ & $ 4.3_{-2.0}^{+3.9} \times 10^{5} $ & 36.4/42 \\[0.1cm]  
\hline
J135514.8+311605 & $ 0.998 $ & $ 0.22_{-0.20}^{+0.25} $ & $ 2.5_{-1.2}^{+2.0} \times 10^{5} $ & 30.4/26 \\[0.1cm]
J013204.6+122236 & $ 0.998 $ & $ -0.72_{-0.12}^{+0.14} $ & $ 2.4_{-1.3}^{+2.8} \times 10^{6} $ & 36.8/44 \\[0.1cm]
J153503.4+455056 & $ 0.998 $ & $ -0.19_{-0.10}^{+0.11} $ & $ 1.2_{-0.4}^{+0.6} \times 10^{6} $ & 69.3/65 \\[0.1cm]
J163831.7+534020 & $ 0.998 $ & $ 0.76_{-0.21}^{+0.25} $ & $ 2.9_{-1.4}^{+2.7} \times 10^{5} $ & 93.5/100 \\[0.1cm]
J163030.2+470125 & $ 0.998 $ & $ -0.05_{-0.10}^{+0.11} $ & $ 2.0_{-0.7}^{+1.0} \times 10^{6} $ & 74.0/78 \\[0.1cm]
J021939.9+361819 & $ 0.998 $ & $ 0.43_{-0.21}^{+0.26} $ & $ 6.8_{-3.9}^{+8.1} \times 10^{5} $ & 20.4/30 \\[0.1cm]
J161001.2+330121 & $ 0.998 $ & $ -0.38_{-0.25}^{+0.36} $ & $ 2.2_{-1.3}^{+2.7} \times 10^{5} $ & 49.4/44 \\[0.1cm]
J171423.6+085236 & $ 0.998 $ & $ 0.16_{-0.18}^{+0.23} $ & $ 2.4_{-0.9}^{+1.2} \times 10^{4} $ & 60.8/56 \\[0.1cm]
J071310.6+725627 & $ 0.998 $ & $ -0.58_{-0.06}^{+0.07} $ & $ 5.3_{-1.7}^{+2.5} \times 10^{6} $ & 26.5/20 \\[0.1cm]
J095928.6+643023 & $ 0.998 $ & $ 0.20_{-0.13}^{+0.17} $ & $ 5.1_{-2.8}^{+5.5} \times 10^{6} $ & 19.4/18 \\[0.1cm]
J091747.6+524821 & $ 0.998 $ & $ -0.71_{-0.06}^{+0.06} $ & $ 8.9_{-4.6}^{+9.9} \times 10^{7} $ & 10.1/21 \\[0.1cm]
J133053.3+734824 & $ 0.998 $ & $ -0.59_{-0.12}^{+0.13} $ & $ 1.3_{-0.6}^{+0.9} \times 10^{6} $ & 30.2/39 \\[0.1cm]
J144738.4+671821 & $ 0.998 $ & $ -0.96_{-0.11}^{+0.12} $ & $ 3.2_{-1.5}^{+2.9} \times 10^{6} $ & 36.4/42 \\[0.1cm]

\hline
\end{tabular}

\end{table}

\subsubsection{Simple models}

We first tried to describe the spectra by two alternative simple models, modified by Galactic and intrinsic absorption: (i) power law (\textsc{tbabs*zphabs*zpowerlw}) and (ii) multi-blackbody accretion disk emission (\textsc{tbabs*zphabs*zashift*diskbb}). The Galactic absorption was adopted from the HI4PI survey \citep{hi4pi2016}, while the redshifts of the TDEs have been measured during our optical spectroscopy program (see \S\ref{s:opt} below).

Most of the studied spectra can be described similarly well by the power-law and accretion disk models, with some intrinsic absorption ($\nh\sim 10^{21}$\,cm$^{-2}$) required for the former. Intrinsic absorption does not improve the quality of approximation of the spectra by \textsc{diskbb}, i.e. $\nh$ is consistent with zero for this model. We thus omitted intrinsic absorption from subsequent consideration. The best-fitting parameters of the two models applied to the eRASS2 spectra are given in Table~\ref{tab:bestfit}. 

The inferred temperatures at the inner boundary of the accretion disk vary between $k\tin\approx 0.05$\,keV for SRGE\,J091747.6+524821 and $\tin\approx 0.5$\,keV for SRGE\,J171423.6+085236. The significantly better fit quality provided by the power-law model compared to \textsc{diskbb} for SRGE\,J163030.2+470125 suggests the presence of a harder component in addition to the soft thermal emission in the spectrum of this object, which is possibly associated with comptonized emission from a hot corona of the accretion disk. 

\subsubsection{Black hole masses and Eddington ratios}

\begin{figure*}
\centering
\includegraphics[width=0.85\textwidth]{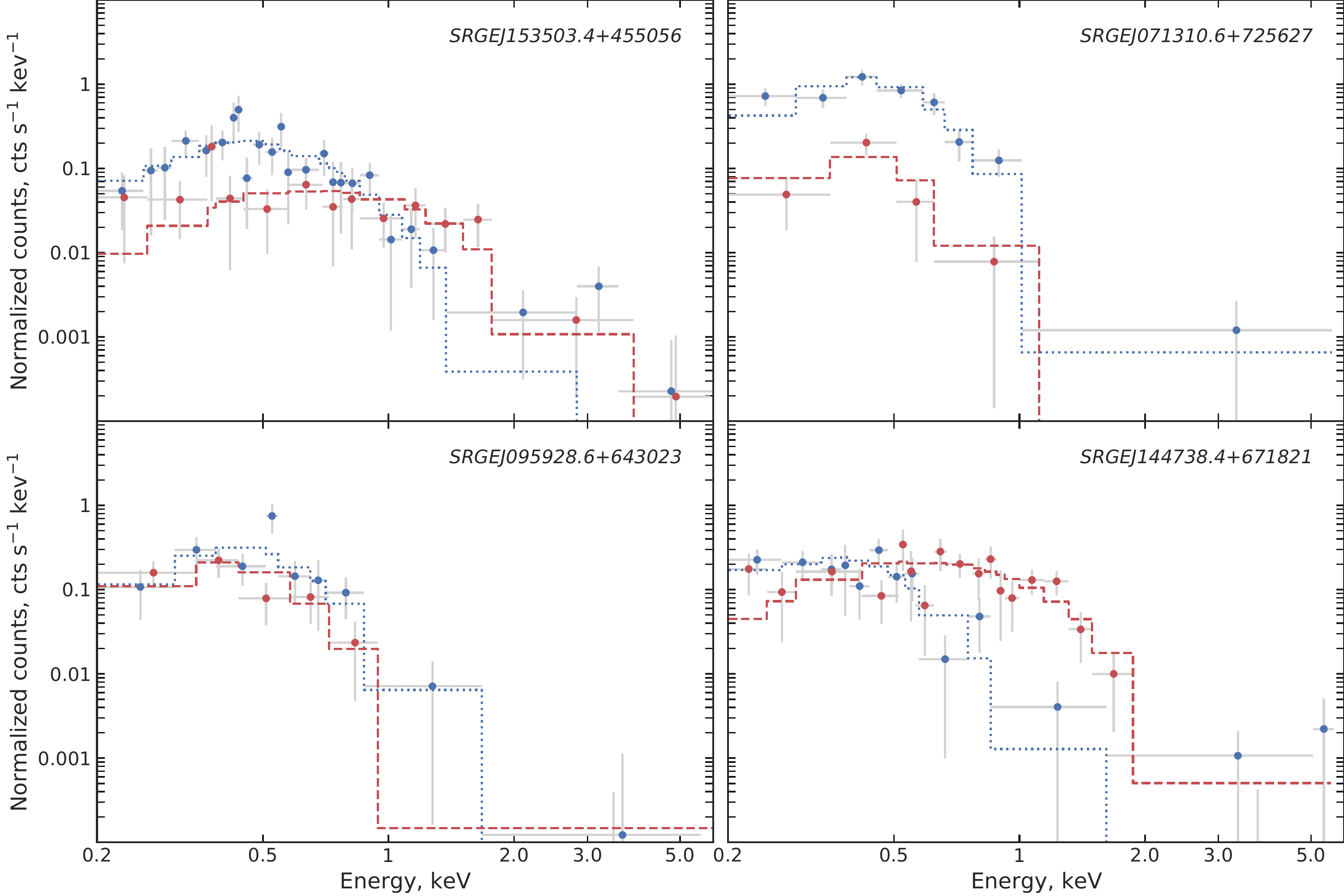}
\caption{Comparison of the X-ray spectra of four TDEs obtained by \erosita\ in the second (blue) and third (red) \srg\ all-sky surveys. The dotted and dashed lines show the corresponding best fits by the color temperature corrected disk model.}
\label{fig:xrayspec_e3}
\end{figure*}

\begin{table}
\setlength{\tabcolsep}{0.5pt}
  \caption{Best-fitting model parameters for eRASS3 spectra. 
  } 
  \label{tab:bestfit_e3}
  \begin{tabular}{lcccc}
  \hline
  \multicolumn{5}{c}{Absorbed power-law model: \textsc{tbabs*zphabs*zpowerlw}} \\[0.1cm]
  \multicolumn{1}{c}{Object (SRGE)} &
  \multicolumn{1}{c}{$\nh$} &
  \multicolumn{1}{c}{$K_{1\,{\rm keV}}^{a}$} &
  \multicolumn{1}{c}{$\Gamma$} & 
  \multicolumn{1}{c}{cstat} \\
  & \multicolumn{1}{c}{$10^{21}$\,cm$^{-2}$} & & & \multicolumn{1}{c}{d.o.f} \\
  \hline
J153503.4+455056 & $ <0.5 $ & $ 8.7_{-2.4}^{+2.9} \times 10^{-5} $ & $ 2.0_{-0.6}^{+0.6} $ & 47.9/42 \\[0.1cm]
J071310.6+725627 & $ 3.0_{-2.2}^{+6.2} $ & $ 9.3_{-9.3}^{+34.1} \times 10^{-6} $ & $ 10.1_{-4.1}^{+6.0} $ & 8.7/19 \\[0.1cm]
J095928.6+643023 & $ 1.3_{-1.3}^{+3.9} $ & $ 1.8_{-0.9}^{+2.2} \times 10^{-4} $ & $ 6.0_{-1.7}^{+3.3} $ & 8.5/16 \\[0.1cm]
J144738.4+671821 & $ <0.6 $ & $ 2.6_{-0.5}^{+0.7} \times 10^{-4} $ & $ 2.4_{-0.3}^{+0.6} $ & 34.9/27 \\[0.1cm]
\hline
 \multicolumn{5}{c}{Multi-blackbody accretion disk model: \textsc{tbabs*zashift*diskbb}} \\[0.1cm]
  \multicolumn{1}{c}{Object (SRGE)} &
  \multicolumn{1}{c}{$\rin\sqrt{\cos\theta}$} &
  \multicolumn{1}{c}{$k\tin$} & 
  \multicolumn{1}{c}{cstat/} \\ 
  & \multicolumn{1}{c}{km} & \multicolumn{1}{c}{eV} & \multicolumn{1}{c}{d.o.f} \\ 
  \hline
J153503.4+455056 & $ ~~5.7_{-4.4}^{+7.9} \times 10^{4} $ & $ 499_{-155}^{+640} $ & 52.1/43 \\[0.1cm]
J071310.6+725627 & $ ~~1.5_{-0.9}^{+3.1} \times 10^{6} $ & $ 98_{-27}^{+37} $ & 12.6/20 \\[0.1cm]
J095928.6+643023 & $ ~~5.7_{-3.2}^{+8.2} \times 10^{6} $ & $ 138_{-33}^{+42} $ & 10.9/17 \\[0.1cm]
J144738.4+671821 & $ ~~1.2_{-0.4}^{+0.6} \times 10^{5} $ & $ 335_{-54}^{+75} $ & 30.5/28 \\[0.1cm]
\hline
  \multicolumn{5}{c}{Color temperature corrected disk model: \textsc{tbabs*optxagnf}} \\[0.1cm]
  \multicolumn{1}{c}{Object (SRGE)} &
  \multicolumn{1}{c}{$\spin$} &
  \multicolumn{1}{c}{$\log\ledd$} &
  \multicolumn{1}{c}{$M,\,\msun$} &
  \multicolumn{1}{c}{cstat} \\ 
  & & & & \multicolumn{1}{c}{d.o.f} \\ 
  \hline
J153503.4+455056 & $ 0 $ & $ 1.40_{-0.35}^{+0.85} $ & $ 9.1_{-7.4}^{+12.9} \times 10^{3} $ & 52.1/43 \\[0.1cm]
J071310.6+725627 & $ 0 $ & $ -0.13_{-0.15}^{+0.18} $ & $ 3.8_{-2.5}^{+9.0} \times 10^{5} $ & 12.6/20 \\[0.1cm]
J095928.6+643023 & $ 0 $ & $ 0.97_{-0.14}^{+0.16} $ & $ 1.0_{-0.6}^{+1.6} \times 10^{6} $ & 10.9/17 \\[0.1cm]
J144738.4+671821 & $ 0 $ & $ 1.04_{-0.16}^{+0.19} $ & $ 2.3_{-0.9}^{+1.3} \times 10^{4} $ & 30.5/28 \\[0.1cm]
\hline
J153503.4+455056 & $ 0.998 $ & $ 0.49_{-0.36}^{+0.90} $ & $ 6.7_{-5.5}^{+9.7} \times 10^{4} $ & 52.0/43 \\[0.1cm]
J071310.6+725627 & $ 0.998 $ & $ -1.05_{-0.15}^{+0.18} $ & $ 2.9_{-1.9}^{+6.7} \times 10^{6} $ & 12.6/20 \\[0.1cm]
J095928.6+643023 & $ 0.998 $ & $ 0.05_{-0.14}^{+0.16} $ & $ 7.7_{-4.5}^{+12.3} \times 10^{6} $ & 10.9/17 \\[0.1cm]
J144738.4+671821 & $ 0.998 $ & $ 0.12_{-0.16}^{+0.19} $ & $ 1.7_{-0.7}^{+1.0} \times 10^{5} $ & 30.4/28 \\[0.1cm]
\hline
\end{tabular}

$^{a}$ --- power-law normalization at 1\,keV in the source's rest frame, in units of photons\,keV$^{-1}$\,cm$^{-2}$\,s$^{-1}$.

\end{table}

Under the assumption that the X-ray emission in TDEs is produced in a standard accretion disk \citep{shakura1973}, the \textsc{diskbb} model \citep{Makishima_1986} is more physically motivated than \textsc{powerlaw}. However, the real situation is likely more complicated. First, the shape of the spectrum emergent from a standard accretion disk can significantly deviate from a sum of blackbodies (e.g. \citealt{Koratkar_1999,Davis_2005}), assumed in \textsc{diskbb}. Second, at least some of the TDEs discussed here may have been caught by \erosita\ in their early super-Eddington phase, when the accretion disk is expected to be geometrically thick, with the radial distribution of X-ray surface brightness being different than in the case of a thin disk \citep{shakura1973,Abramowicz_1988,Watarai_2000}. Finally, general relativity can significantly affect the properties of emission from the inner accretion disk (e.g. \citealt{Wen_2021}). On the other hand, at a later stage in the evolution of a TDE, the X-ray spectrum might experience a transition from a purely thermal state to a combination of thermal and comptonized emission, as suggested by our knowledge of X-ray binary systems and AGN and has been observed in a number of TDEs (e.g. \citealt{Komossa_2004,Jonker_2020,Wevers_2020,Wevers_2021}).

We thus next tried to describe the \erosita\ spectra by the \textsc{optxagnf} model \citep{Done_2012}, designed for AGN SEDs. In its simplest version, it represents the spectrum of emission from a standard (i.e. geometrically thin and optically thick) accretion disk around a SMBH, taking into account the expected deviations from a simple multitemperature blackbody shape due to the incomplete thermalization of the radiation in the disk. The main parameters of the model are black hole mass ($\mbh$), spin ($\spin$), and accretion rate in terms of the Eddington critical luminosity ($\ledd$). Essentially, this model is a modification of \textsc{diskbb}, with the blackbody spectrum of each annulus of the disk modified by a temperature-dependent hardening factor $\sim 2$. In addition, it takes into account the radiative efficiency of the accretion disk as a function of spin \citep{Novikov1973}. As a result, \textsc{optxagnf} retains the key property of the \textsc{diskbb} model that, for a given $\spin$, the peak energy of the spectrum is proportional to $\mbh^{-1/4}\ledd^{1/4}$ \citep{shakura1973}, while the observed bolometric flux is of course proportional to $\mbh\ledd$ (divided by the distance squared). Therefore, by fitting a measured spectrum by \textsc{optxagnf} it is possible to infer both $\mbh$ and $\ledd$ for an assumed $\spin$.

We note, however, that \textsc{optxagnf} is not applicable to the case of a slim accretion disk. Hence, whenever our spectral analysis implies a super-Eddington accretion rate for a given TDE, the inferred parameter values should be taken with great caution. 

The best fits of the eRASS2 spectra by the \textsc{optxagnf} model are shown in Fig.~\ref{fig:xrayspec} in comparison with those by the power-law model. The quality of approximation by the \textsc{optxagnf} model proves to be nearly insensitive to the black hole spin. We thus fixed this parameter at two extreme values: $\spin=0$ (Schwarzschild black hole) and $\spin=0.998$ (maximally rotating Kerr black hole)\footnote{Assuming that the accretion disk rotates in the same direction as the black hole.}. The resulting best-fitting parameters are given in Table~\ref{tab:optxagnf}. The quality of approximation by the \textsc{optxagnf} model is nearly identical to that by the \textsc{diskbb} model (Table~\ref{tab:bestfit}), which is not surprizing since both models describe multi-blackbody accretion disk emission. 

As seen from Table~\ref{tab:optxagnf}, the black hole mass for each of our objects may range by a factor of $\sim 8$ depending on the unknown spin of the black hole, with $\mbh$ increasing with $\spin$. The corresponding accretion rates are close to the Eddington limit if the black holes are rapidly spinning ($\spin\approx 0.998$) and are supercritical if the rotation is not extreme. As already noted, the inferred parameter values, including the black hole masses, become unreliable in the latter case. 

Although our X-ray spectral analysis is admittedly simplistic, we can nonetheless conclude that the available X-ray spectral data are consistent with the studied events being stellar disruptions by black holes with masses between $\sim 3\times 10^3$ and $\sim 10^8\,\msun$. 

\subsubsection{Late-phase X-ray spectra}

As was mentioned before, we have also analyzed the spectra of four TDEs obtained in eRASS3, six months after the discovery of these transients by \erosita. These spectra are compared in Fig.~\ref{fig:xrayspec_e3} with the corresponding spectra taken during eRASS2. We similarly applied the absorbed power-law, multi-blackbody accretion disk, and the color temperature corrected disk models to these late-phase TDE spectra; see the resulting best-fitting parameters in Table~\ref{tab:bestfit_e3}. 

Individual TDEs show significantly different evolution. In the case of SRGE\,J071310.6+725627 and SRGE\,J095928.6+643023, the black hole masses estimated from the late-phase spectra are consistent with those inferred from the early-phase ones and the observed spectral evolution can be accounted for by moderate decreases in the accretion rates. 

The situation is quite different for SRGE\,J153503.4+455056. Here, fitting the eRASS3 spectrum by \textsc{optxagnf} leads to a much lower black hole mass compared to the eRASS2 spectrum, which indicates that this model is inadequate for description of the observed spectral evolution. We might be witnessing a change from purely thermal accretion disk emission to a harder spectral state, where a major contribution is provided by comptonized emission associated with a freshly formed hot corona of the accretion disk. This is also suggested by the somewhat better fit quality provided by the power-law model compared to \textsc{optxagnf} for the eRASS3 spectrum (see Table~\ref{tab:bestfit_e3}). 

The least clear situation is with SRGE\,J144738.4+671821. Here too, the eRASS3 spectrum is much harder than the eRASS2 one and the early- and late-phase $\mbh$ estimates are inconsistent with each other, despite the good fit quality provided by the \textsc{optxagnf} model for both spectra. We recall that this TDE is unique in that \erosita\ seems to have caught it during an usually long rising phase, so that it has become brighter in eRASS3 compared to eRASS2. Hopefully, if SRGE\,J144738.4+671821 remains bright in several subsequent \srg/\erosita\ scans, we will be able to better understand its spectral evolution and nature. 

\begin{figure*}
\begin{subfigure}[t]{0.19\textwidth}
\centering
\includegraphics[width=\linewidth]{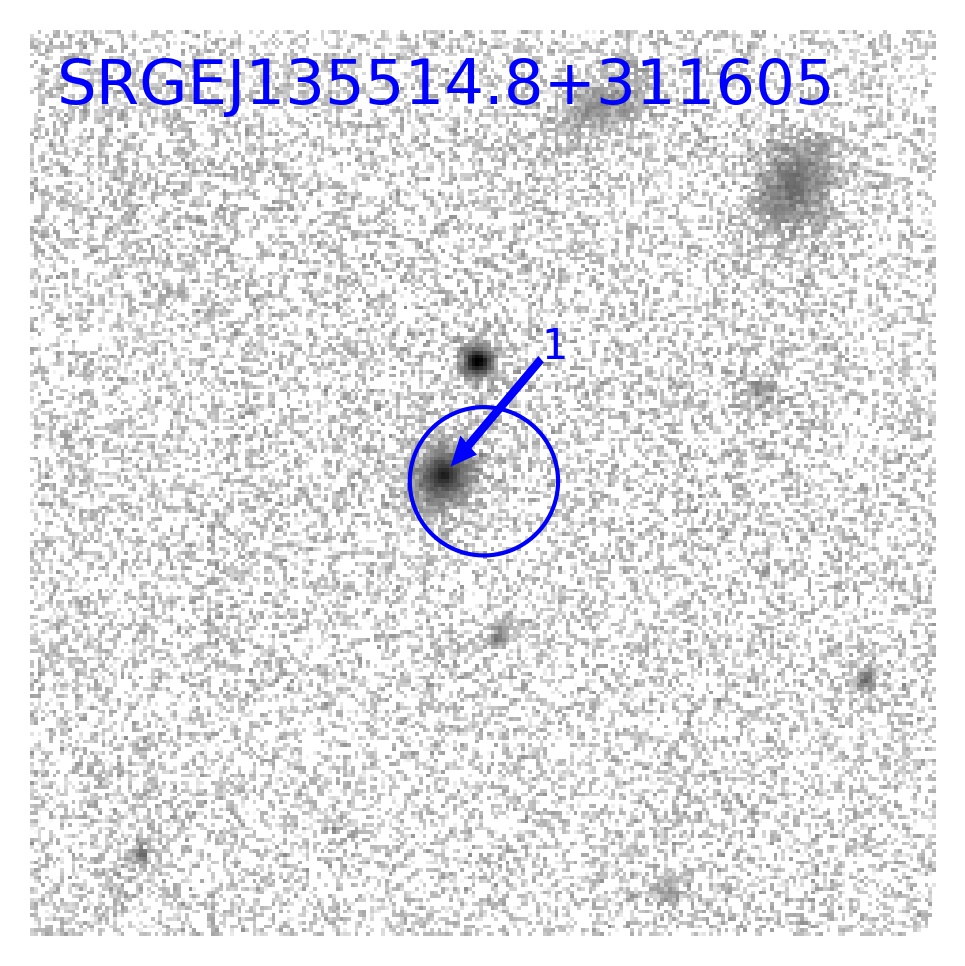}
\end{subfigure}
\begin{subfigure}[t]{0.19\textwidth}
\centering
\includegraphics[width=\linewidth]{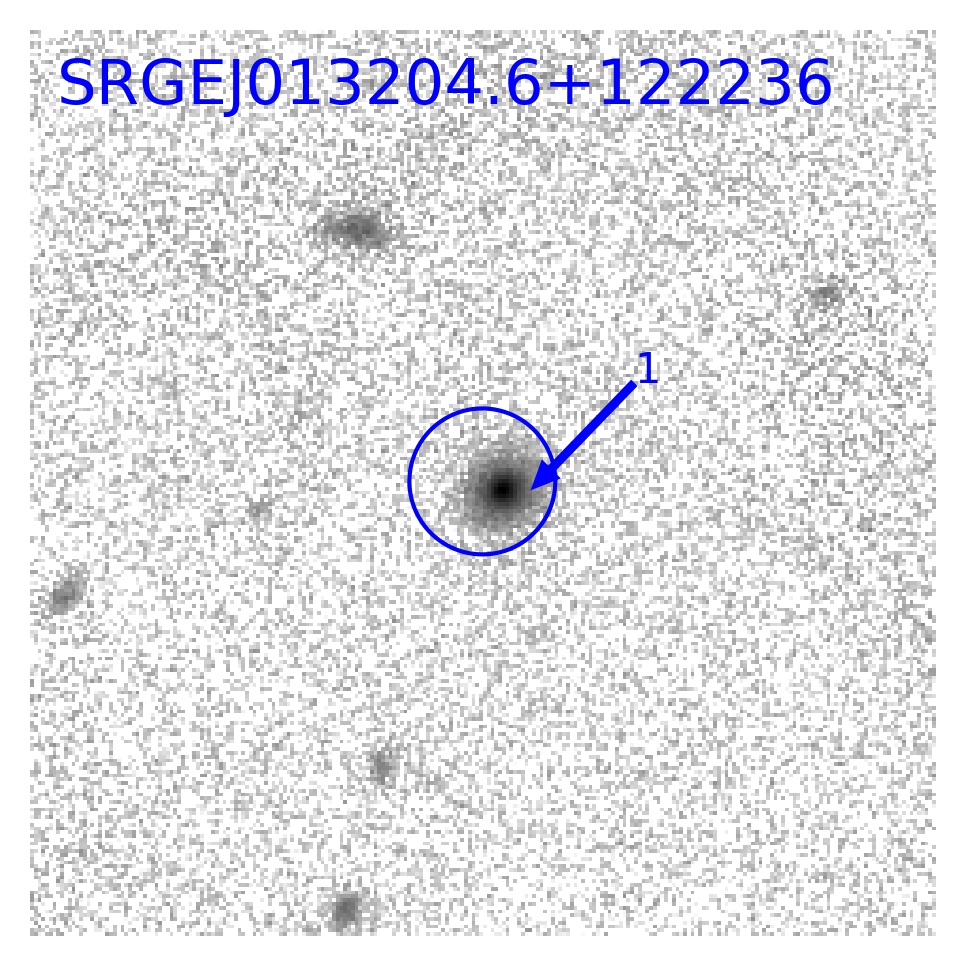}
\end{subfigure}
\begin{subfigure}[t]{0.19\textwidth}
\centering
\includegraphics[width=\linewidth]{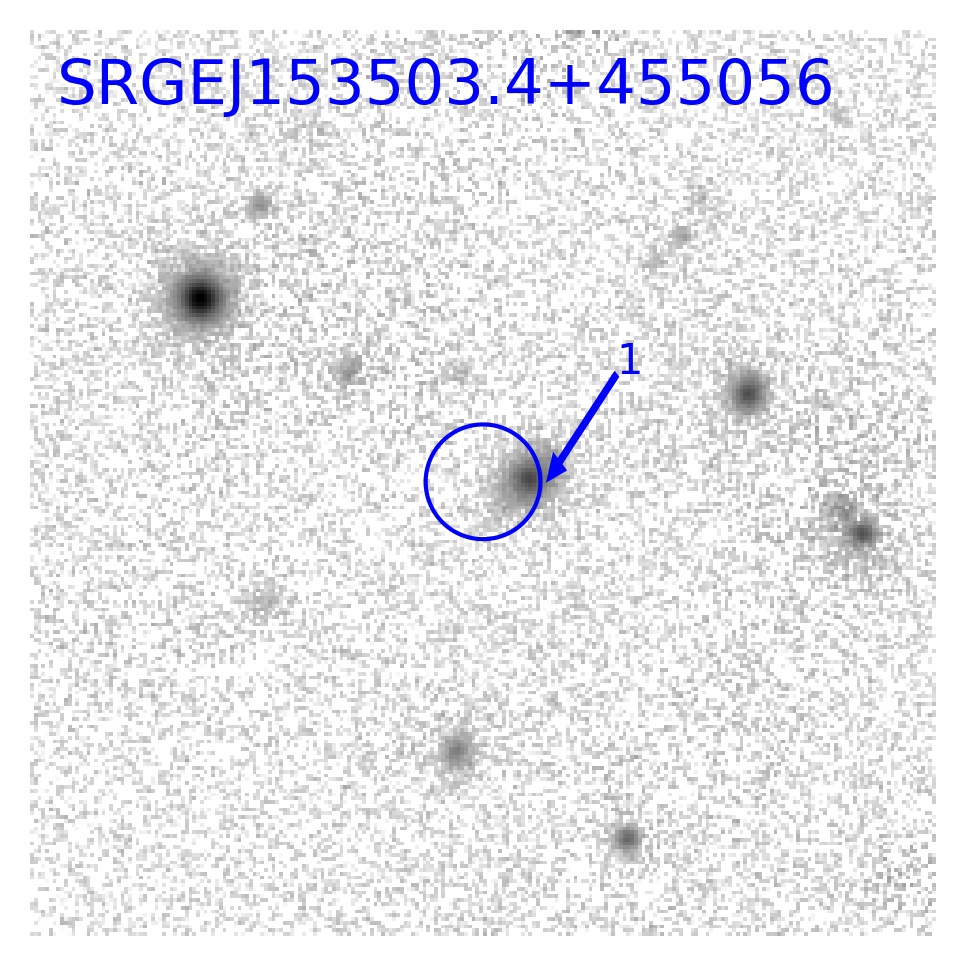}
\end{subfigure}
\begin{subfigure}[t]{0.19\textwidth}
\centering
\includegraphics[width=\linewidth]{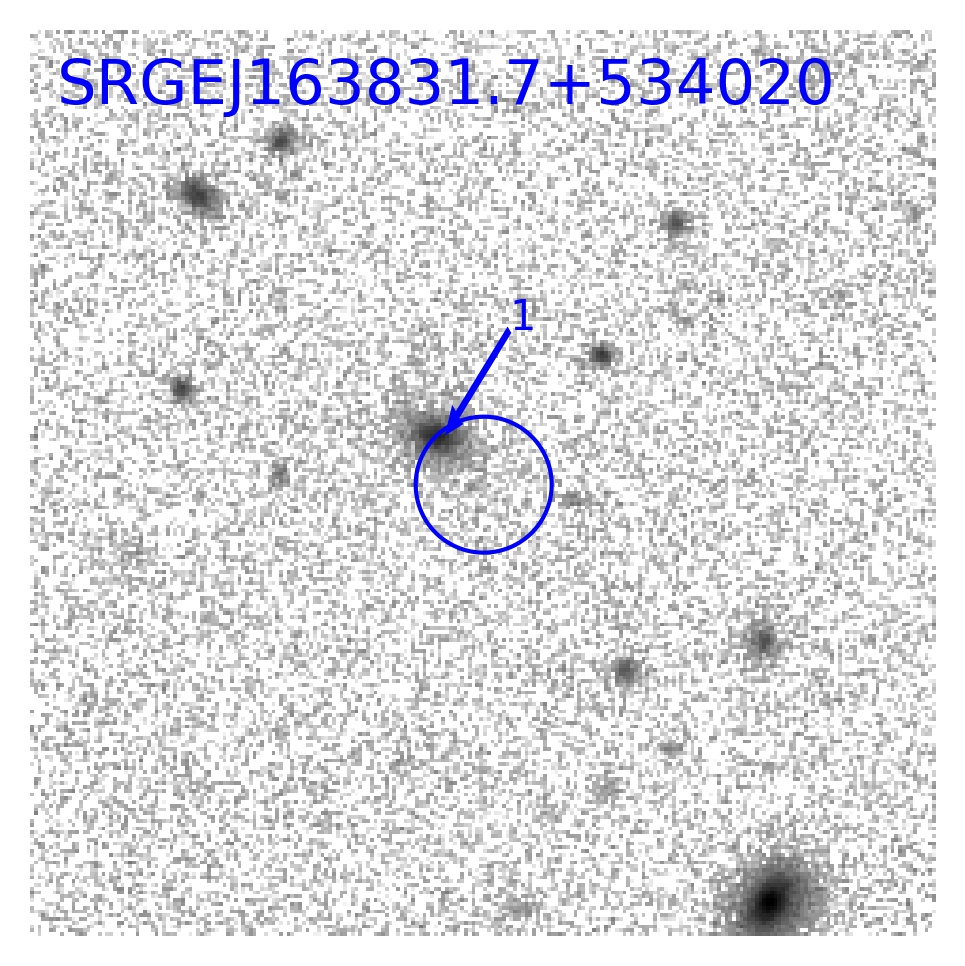}
\end{subfigure}
\begin{subfigure}[t]{0.19\textwidth}
\centering
\includegraphics[width=\linewidth]{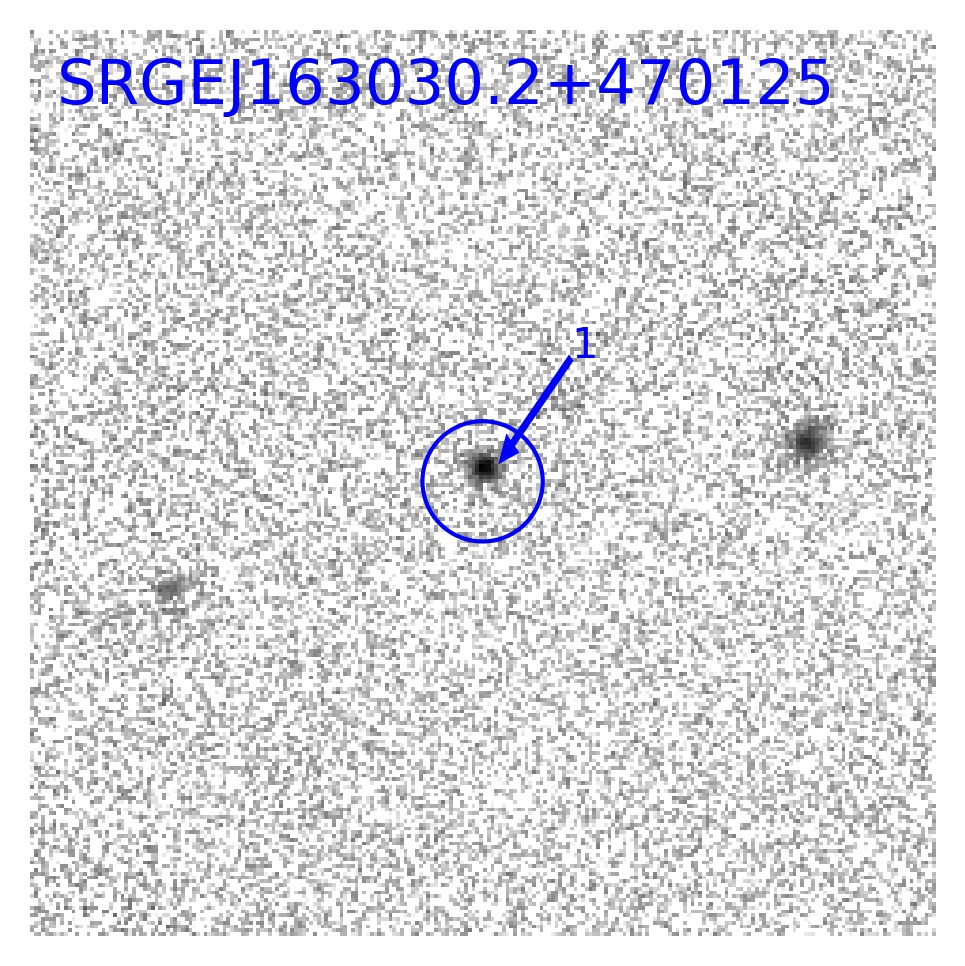}
\end{subfigure}
\begin{subfigure}[t]{0.19\textwidth}
\centering
\includegraphics[width=\linewidth]{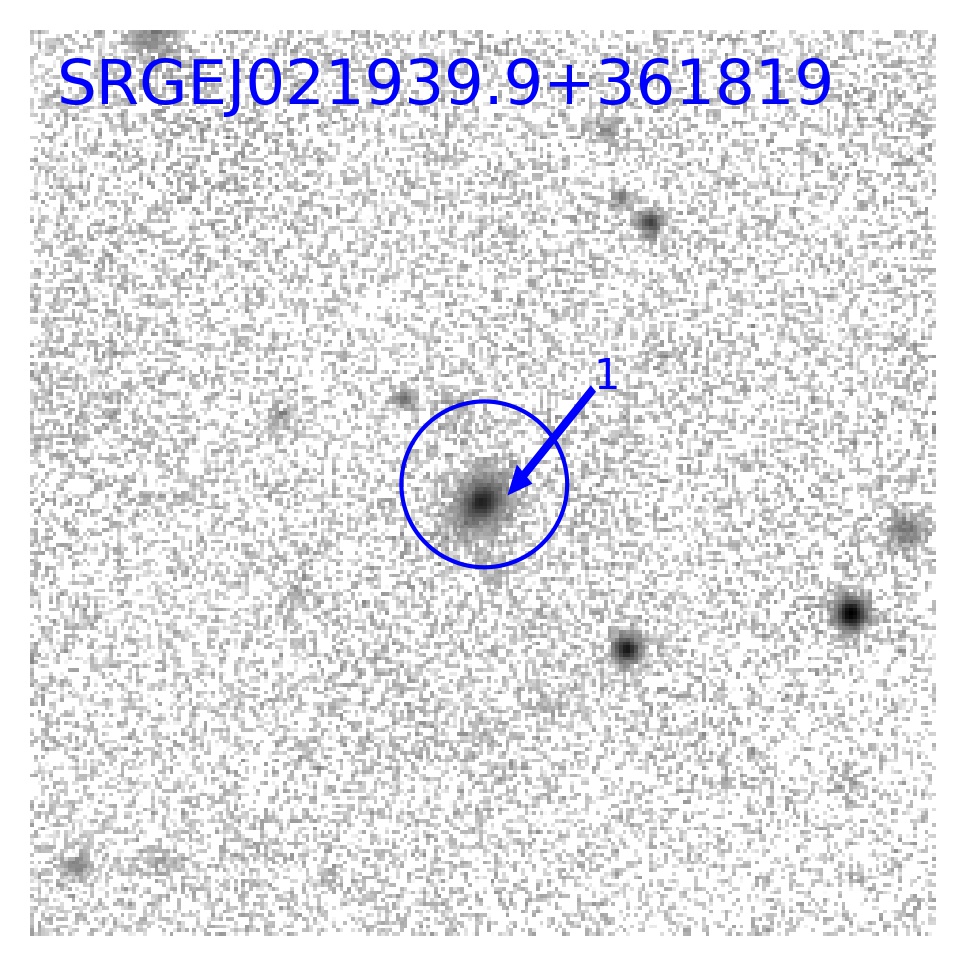}
\end{subfigure}
\begin{subfigure}[t]{0.19\textwidth}
\centering
\includegraphics[width=\linewidth]{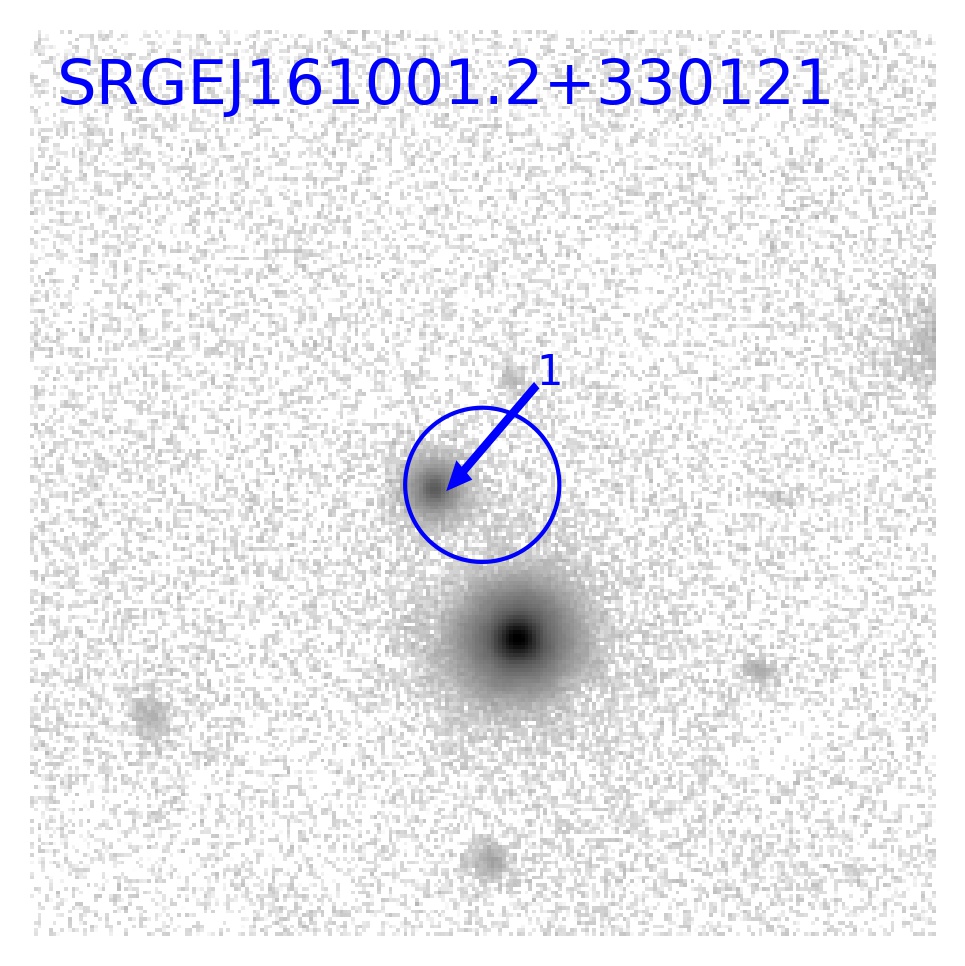}
\end{subfigure}
\begin{subfigure}[t]{0.19\textwidth}
\centering
\includegraphics[width=\linewidth]{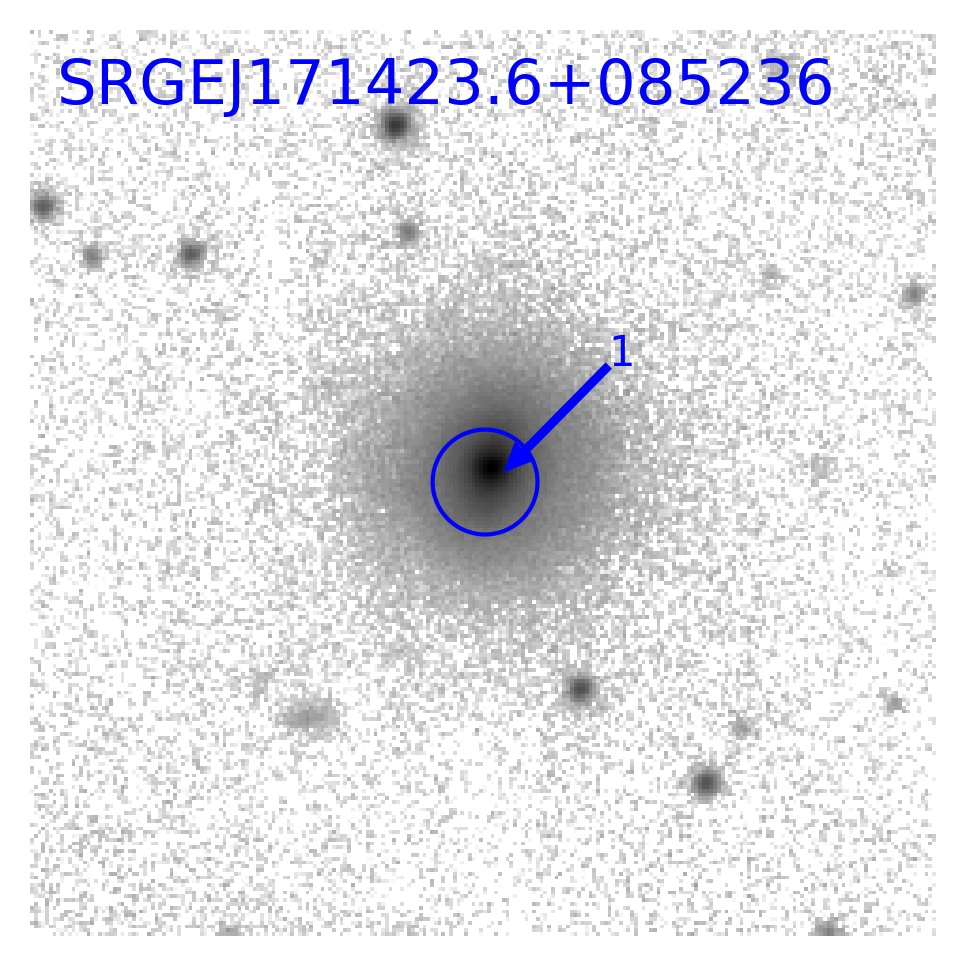}
\end{subfigure}
\begin{subfigure}[t]{0.19\textwidth}
\centering
\includegraphics[width=\linewidth]{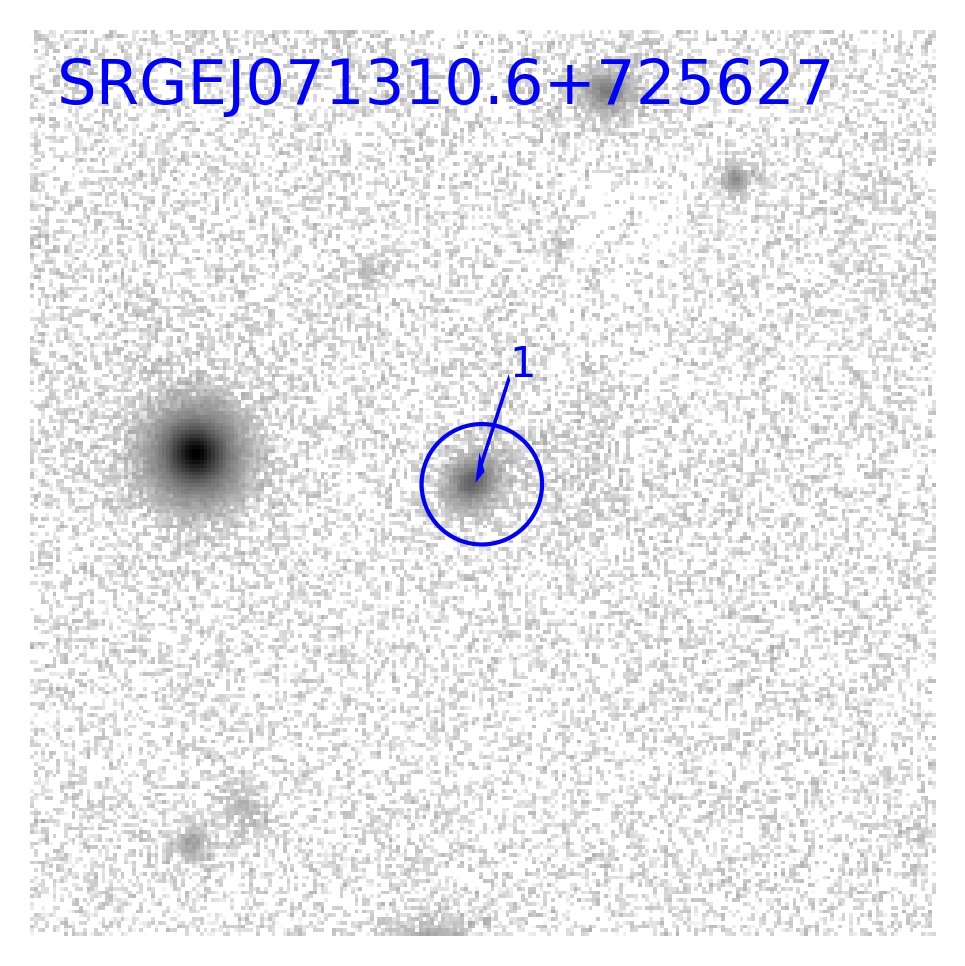}
\end{subfigure}
\begin{subfigure}[t]{0.19\textwidth}
\centering
\includegraphics[width=\linewidth]{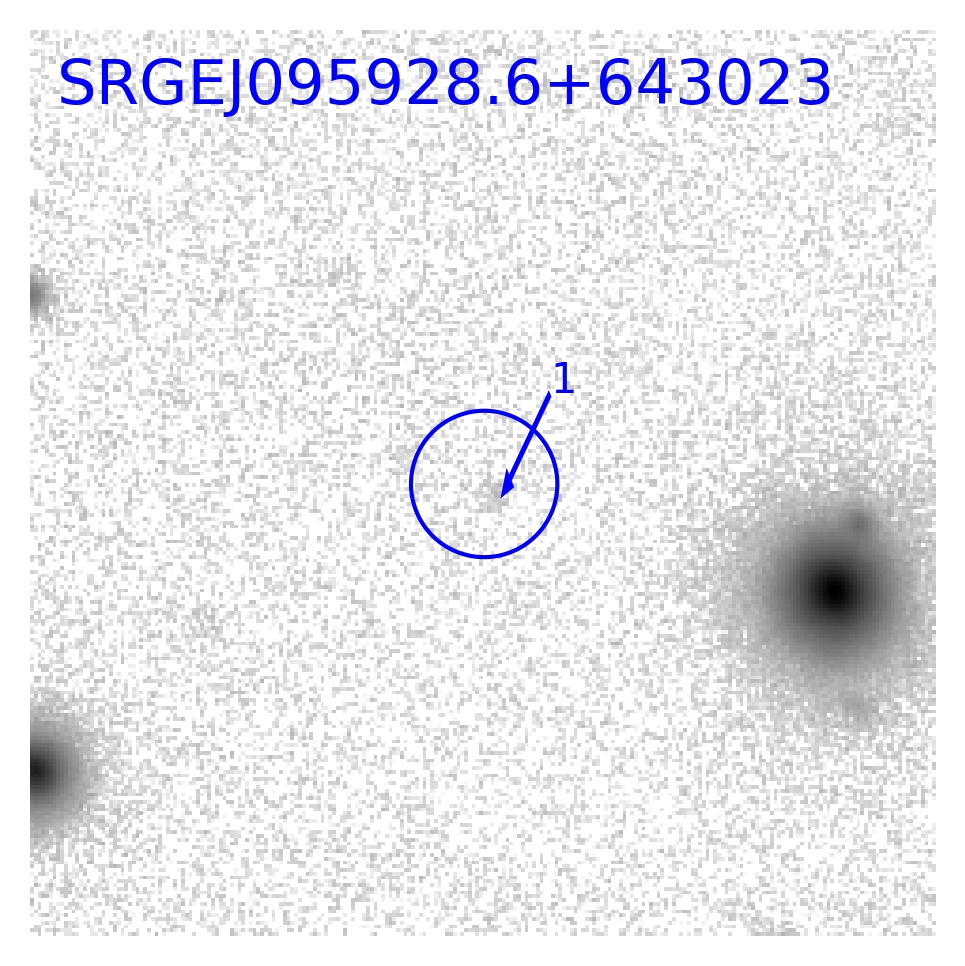}
\end{subfigure}
\begin{subfigure}[t]{0.19\textwidth}
\centering
\includegraphics[width=\linewidth]{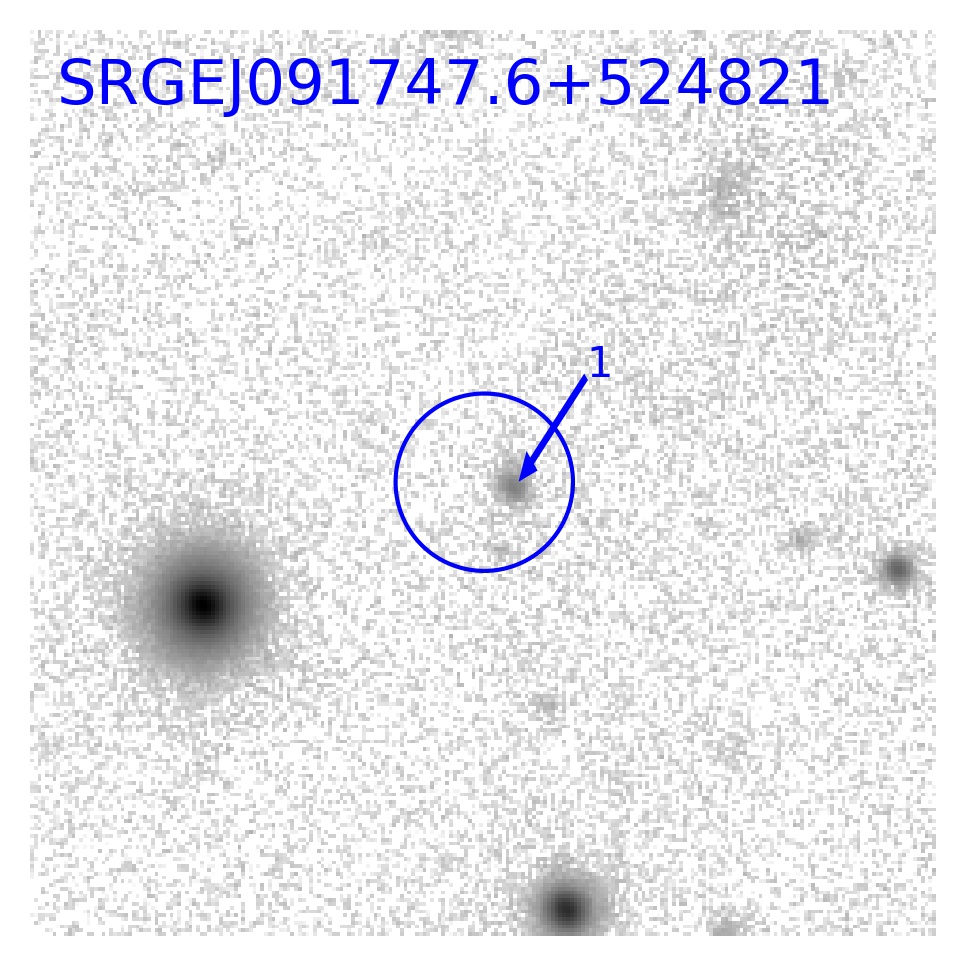}
\end{subfigure}
\begin{subfigure}[t]{0.19\textwidth}
\centering
\includegraphics[width=\linewidth]{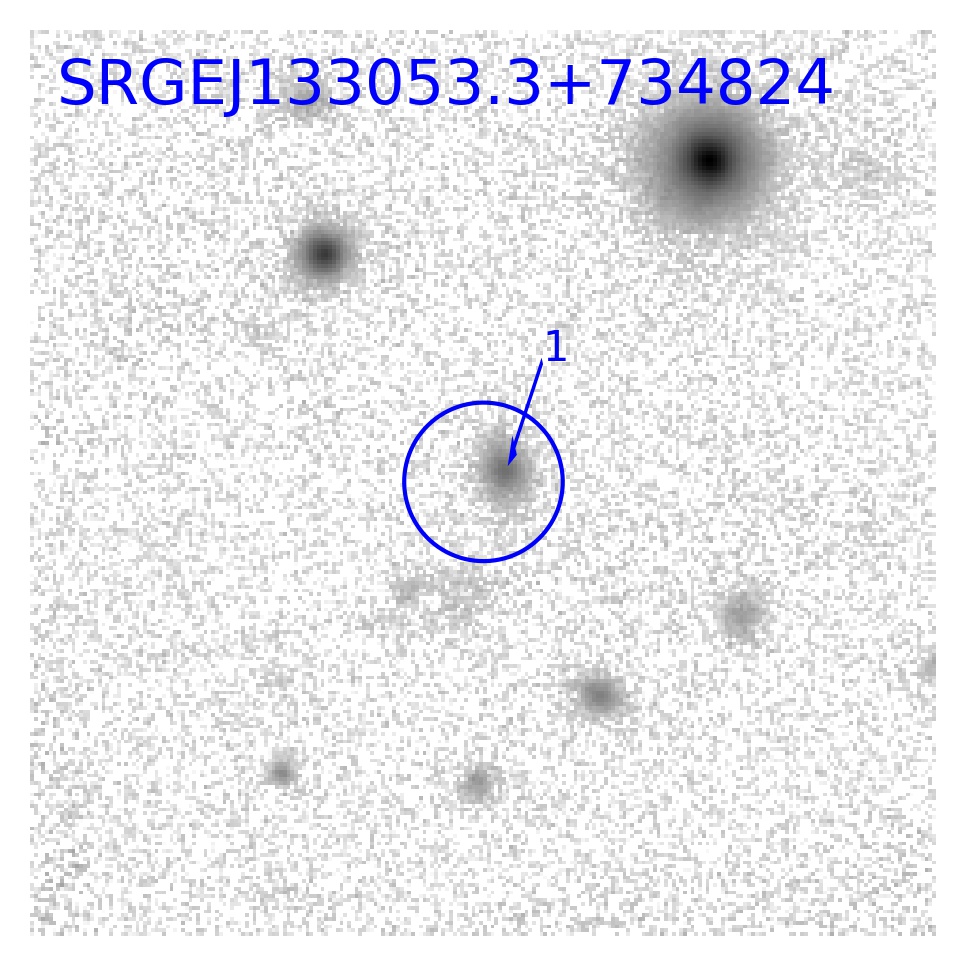}
\end{subfigure}
\begin{subfigure}[t]{0.19\textwidth}
\centering
\includegraphics[width=\linewidth]{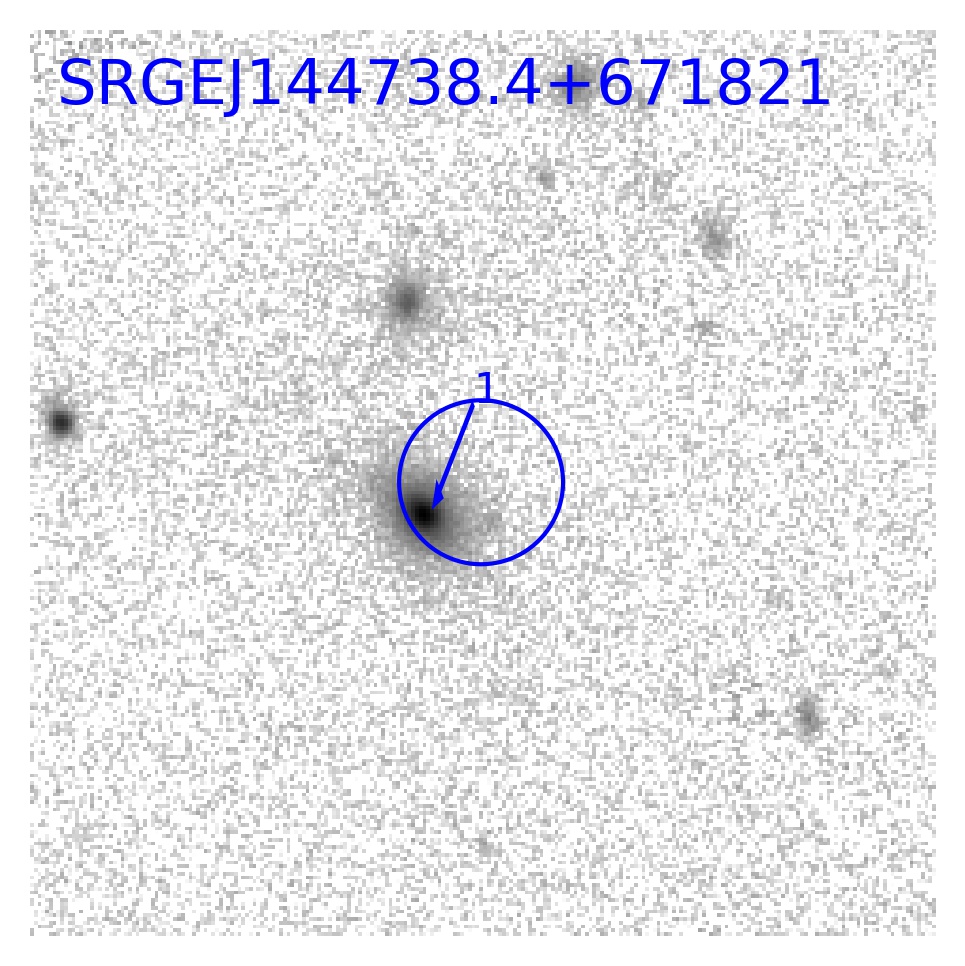}
\end{subfigure}
\caption{Pan-STARRS i-band 1'$\times$1' images around the TDEs. In each panel, the circle shows the \erosita\ localization region (with the size given in Table~\ref{tab:sample}), while the arrow shows the object for which optical spectroscopy was performed.
    \label{fig:charts}
    }
\end{figure*}

\begin{table*}
  \caption{Log of photometric observations.} 
  \label{tab:photometry}
  \begin{tabular}{lllllll}
  \hline
  \multicolumn{1}{c}{Object (SRGE)} &
  \multicolumn{1}{c}{Date} &
  \multicolumn{1}{c}{Telescope} &
  \multicolumn{1}{c}{Exposure (s), $gri$} &
  \multicolumn{1}{c}{m$_{g}$} & 
  \multicolumn{1}{c}{m$_{r}$} & 
  \multicolumn{1}{c}{m$_{i}$} \\
  \hline
  J135514.8+311605 & 2020 July 21 & RTT150 & 10$\times$60, 10$\times$60, 10$\times$60 & 20.49$\pm$0.07 & 19.60$\pm$0.04 & 19.13$\pm$0.05 \\ 
  J153503.4+455056 & 2020 July 23--27 & CMO & 4$\times$200, 8$\times$200+4$\times$300, 200+3$\times$300 & 20.06$\pm$0.05 & 19.02$\pm$0.03 & 18.55$\pm$0.05 \\  
& 2020 Aug. 5 & CMO & 3$\times$200, 3$\times$200+2$\times$300, 9$\times$200 & 20.01$\pm$0.05 & 19.02$\pm$0.02 & 18.54$\pm$0.04 \\
& 2020 Sept. 17 & CMO & 3$\times$300, 4$\times$300, 3$\times$300 & 20.09$\pm$0.02 & 19.02$\pm$0.01 & 18.49$\pm$0.01 \\
  J163030.2+470125 & 2020 Oct. 17 & BTA & 4$\times$40, 4$\times$40, 4$\times$40& 21.45$\pm$0.08 & 20.42$\pm$0.06 & 20.04$\pm$0.06 \\
  J161001.2+330121 & 2021 Mar. 12 & RTT150 & 2$\times$300, 3$\times$300, 2$\times$300 & 19.28$\pm$0.01 & 18.51$\pm$0.01 & 18.11$\pm$0.01 \\
  J171423.6+085236 & 2020 Oct. 18 & CMO RC600 & 2$\times$200, 13$\times$200, 2$\times$200 & 16.90$\pm$0.03 & 16.02$\pm$0.02 & 15.57$\pm$0.06 \\
                   & 2020 Oct. 21 & CMO RC600 &  8$\times$240, 3$\times$240, 3$\times$240 & 16.86$\pm$0.02 & 16.00$\pm$0.03 & 15.59$\pm$0.02 \\
  J095928.6+643023 & 2021 Mar. 10 & AZT-33IK & 18$\times$90, 18$\times$90, 18$\times$90 & 23.34$\pm$0.22 & 22.31$\pm$0.11 & 22.23$\pm$0.13 \\
  \hline
  \end{tabular}
\end{table*}

\begin{table}
  \caption{Log of spectroscopic observations.} 
  \label{tab:spectroscopy}
  \begin{tabular}{llll}
  \hline
  Object (SRGE) &
  Date &
  Telescope &
  Exp. (s)\\
  \hline
  J135514.8+311605 &  2020 Dec. 17 & BTA & 4$\times$900  \\ 
  & 2021 Jun. 7 & Keck-I & 435 \\
  J013204.6+122236 & 2021 Jul. 6 & Keck-I & 600  \\
  J153503.4+455056 & 2020 Sep. 16--20 & CMO & 21$\times$1200  \\
   & 2021 May 13  & Keck-I & 570 \\
  J163831.7+534020 & 2021 Apr. 14 & Keck-I & 1250\\
  J163030.2+470125 & 2020 Oct. 17 & BTA & 4$\times$900 \\ 
   &2021 Jun. 7  & Keck-I &  1010  \\
  J021939.9+361819 & 2021 Jul. 6 & Keck-I & 750  \\
  J161001.2+330121 &2021 Apr. 14  & Keck-I &300 \\
  J171423.6+085236 &  2020 Oct. 18, 21 & CMO & 2$\times$1200  \\
   & 2021 Jun. 7 & Keck-I & 280 \\
  J071310.6+725627 & 2020 Nov. 11--12 & AZT-33IK & 7$\times$600  \\   
   & 2020 Nov. 20 & Keck-I & 900 \\
  J095928.6+643023 & 2021 May 13 & Keck-I &2500   \\
  J091747.6+524821 & 2021 Apr. 14 & Keck-I & 600  \\
  J133053.3+734824 & 2021 Apr. 14 & Keck-I & 400  \\
  J144738.4+671821 & 2021 Mar. 11 & AZT-33IK &  5$\times$600\\
  & 2021 Apr. 14 & Keck-I & 300  \\
  \hline
  \end{tabular}
\end{table}

\section{Optical/infared properties}
\label{s:opt}

Figure~\ref{fig:charts} shows optical images around the studied objects from the Panoramic Survey Telescope and Rapid Response System DR1 (Pan-STARRS, PS1) \citep{Flewelling2020, Waters2020}. There is a single potential optical counterpart within each \erosita\ localization region. All of these candidates appear to be extended and thus can be TDE host galaxies. Their optical positions are provided in Table~\ref{tab:properties} below.

We carried out spectroscopy and photometry of the candidate optical counterparts of the \erosita\ transients using a number of telescopes and instruments, namely: the CCD-photometer (CMO RC600, \citealt{Berdnikov_2020}) on the RC600 60-cm telescope of the Caucasus Mountain Observatory of the Sternberg Astronomical Institute (CMO SAI MSU, Russia), the ADAM low and medium resolution spectrograph \citep{Afanasiev_2016,Burenin_2016} and the Andor iKon-M imaging camera on the AZT-33IK 1.6-meter telescope \citep{kamus02} of the Sayan Observatory (Russia), the T\"{U}BITAK Faint Object Spectrograph and Camera\footnote{https://tug.tubitak.gov.tr/en/teleskoplar/rtt150-telescope-0} (TFOSC) on the Russian-Turkish 1.5-meter Telescope (RTT150) of the T\"{U}BITAK National Observatory (Turkey), the Transient Double-Beam Spectrograph (TDS, \citealt{Potanin_2020}) and the NBI CCD-photometer on the 2.5-meter telescope of CMO SAI MSU (Russia), the SCORPIO-2 universal focal reducer \citep{Afanasiev_2011} on the BTA 6-meter telescope of the Special Astrophysical Observatory (Russia), and the Low Resolution Imaging Spectrograph (LRIS, \citealt{Oke1995}) on the Keck-I 10-meter telescope (USA). 

Table~\ref{tab:photometry} presents a log of our photometric follow-up observations and the ($gri$) apparent magnitudes measured during these observations. Table~\ref{tab:spectroscopy} presents a log of our spectroscopic follow-up observations. Further details on the observations and data reduction are presented in Appendix~\ref{subsec:details}.

\subsection{Optical light curves}
\label{s:optlc}

The left panels of Fig.~\ref{fig:optlc} show long-term optical light curves of the TDE host galaxies constructed from our follow-up photometry (Table~\ref{tab:photometry}) and archival photometry provided by Pan-STARRS DR2 (PS2) and the Sloan Digital Sky Survey (SDSS, \citealt{Alam2015}). The right panels show forced differential photometry light curves from ZTF, which cover epochs both before and after the \erosita\ X-ray observations. Since the ZTF coverage at the locations of SRGE\,J153503.4+455056, SRGE\,J171423.6+085236, and SRGE\,J144738.4+671821 is poor, we also show the Asteroid Terrestrial-impact Last Alert System (ATLAS; \citealt{Tonry2018, Smith2020}) forced photometry in the cyan ($c$) and orange ($o$) bands for these objects. 


\begin{figure*}
\begin{subfigure}[t]{\textwidth}
\centering
\includegraphics[width=0.35\linewidth]{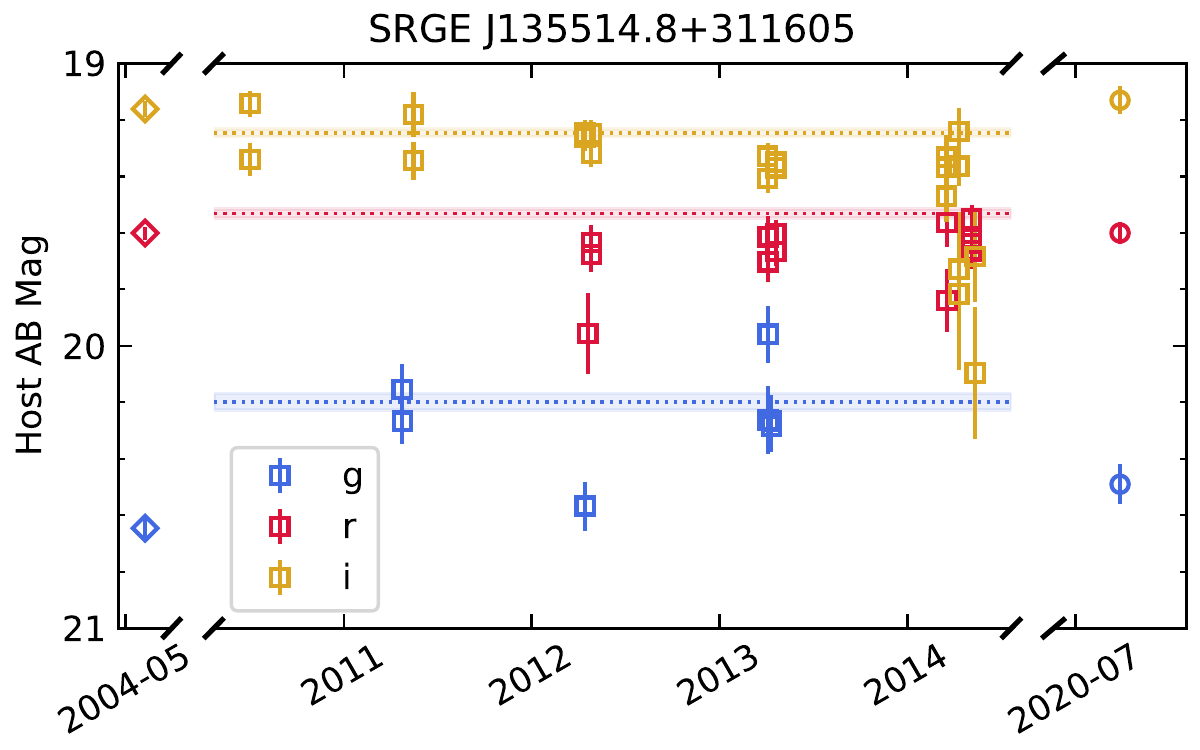}
\includegraphics[width=0.45\linewidth]{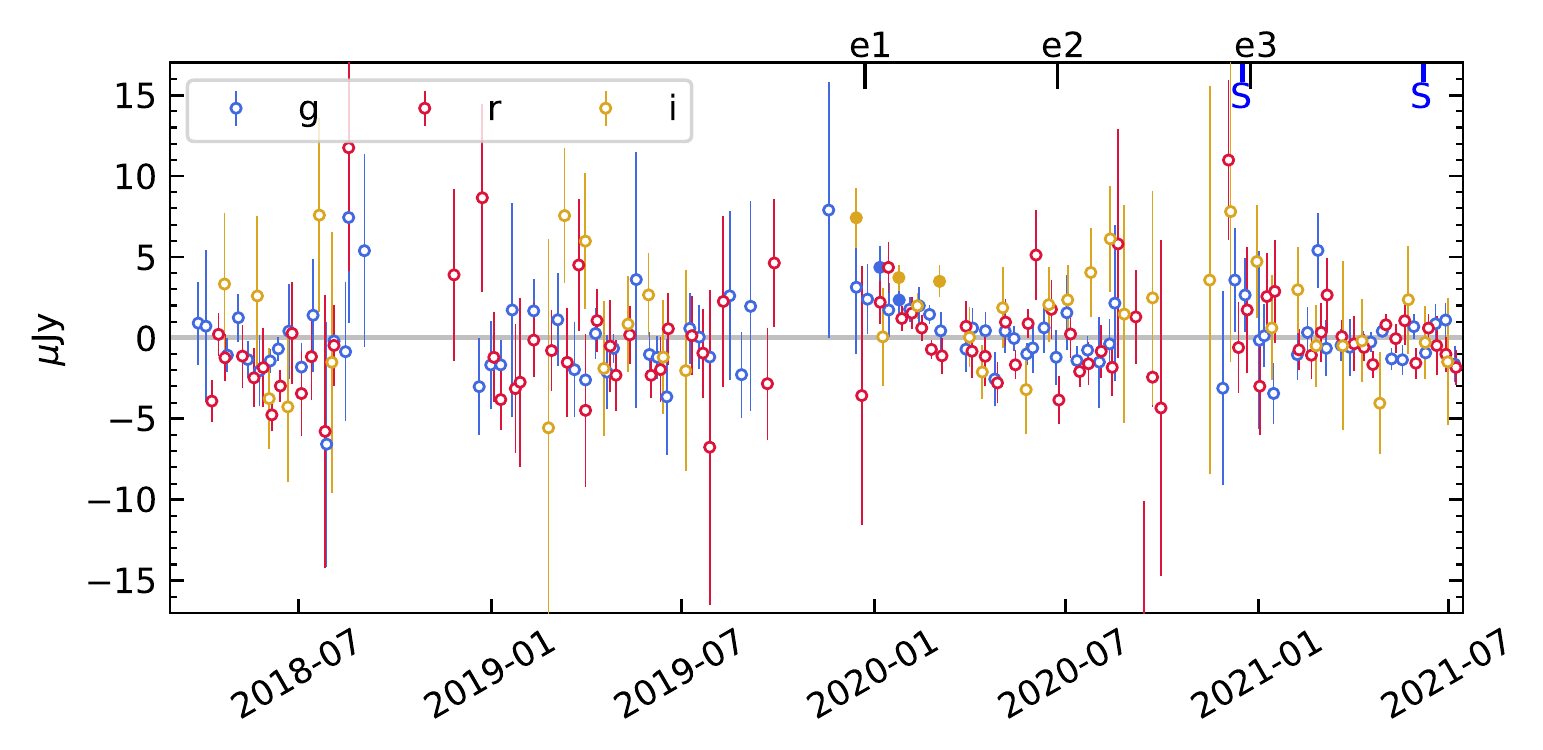}
\end{subfigure}
\begin{subfigure}[t]{\textwidth}
\centering
\includegraphics[width=0.35\linewidth]{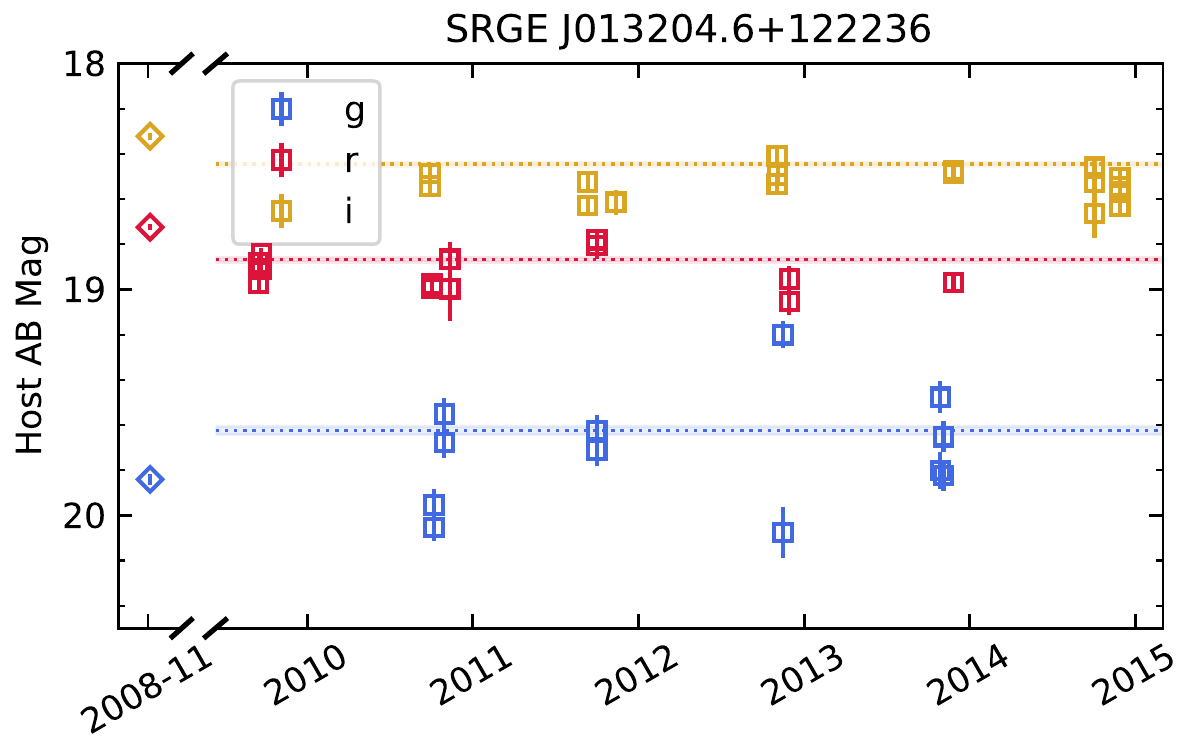}
\includegraphics[width=0.45\linewidth]{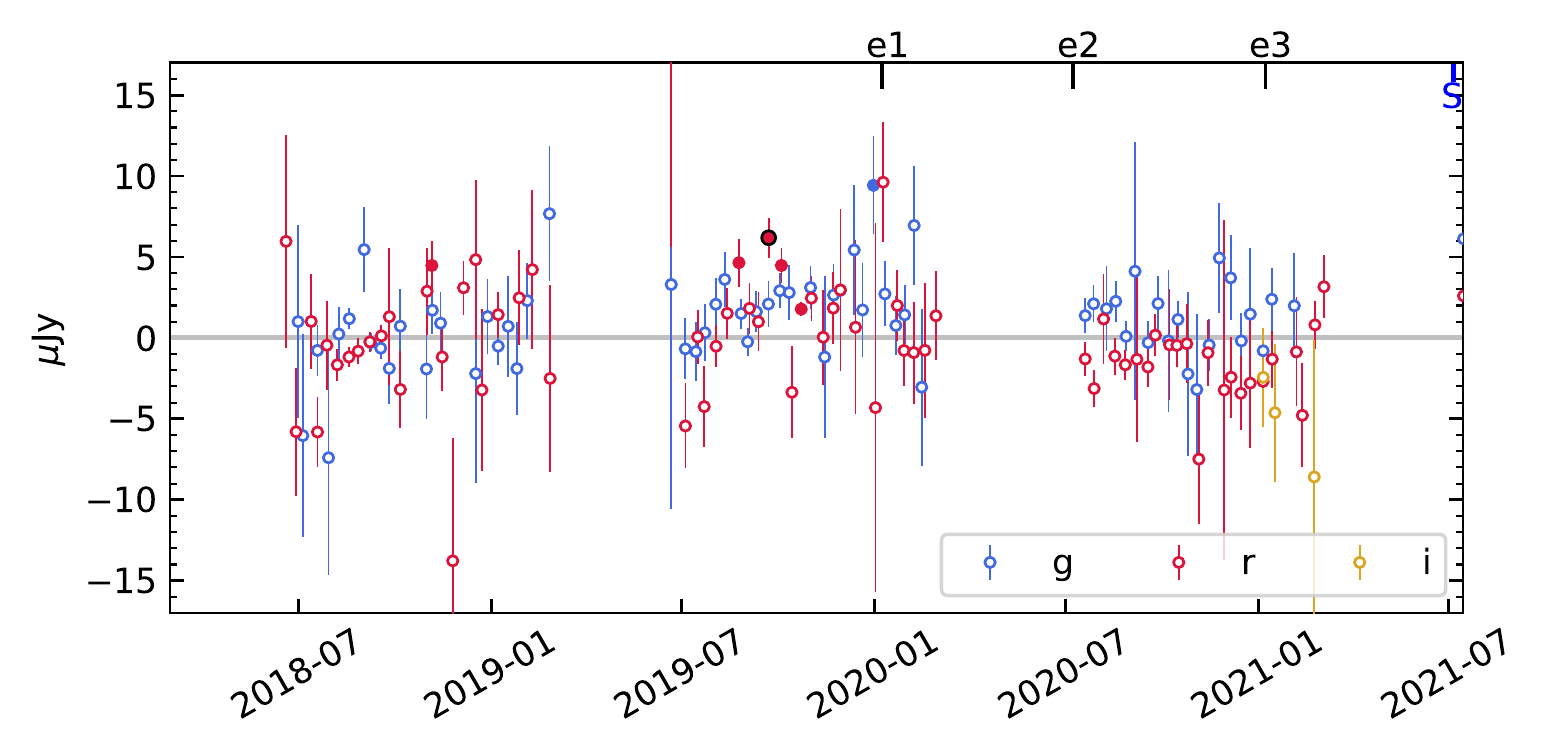}
\end{subfigure}
\begin{subfigure}[t]{\textwidth}
\centering
\includegraphics[width=0.35\linewidth]{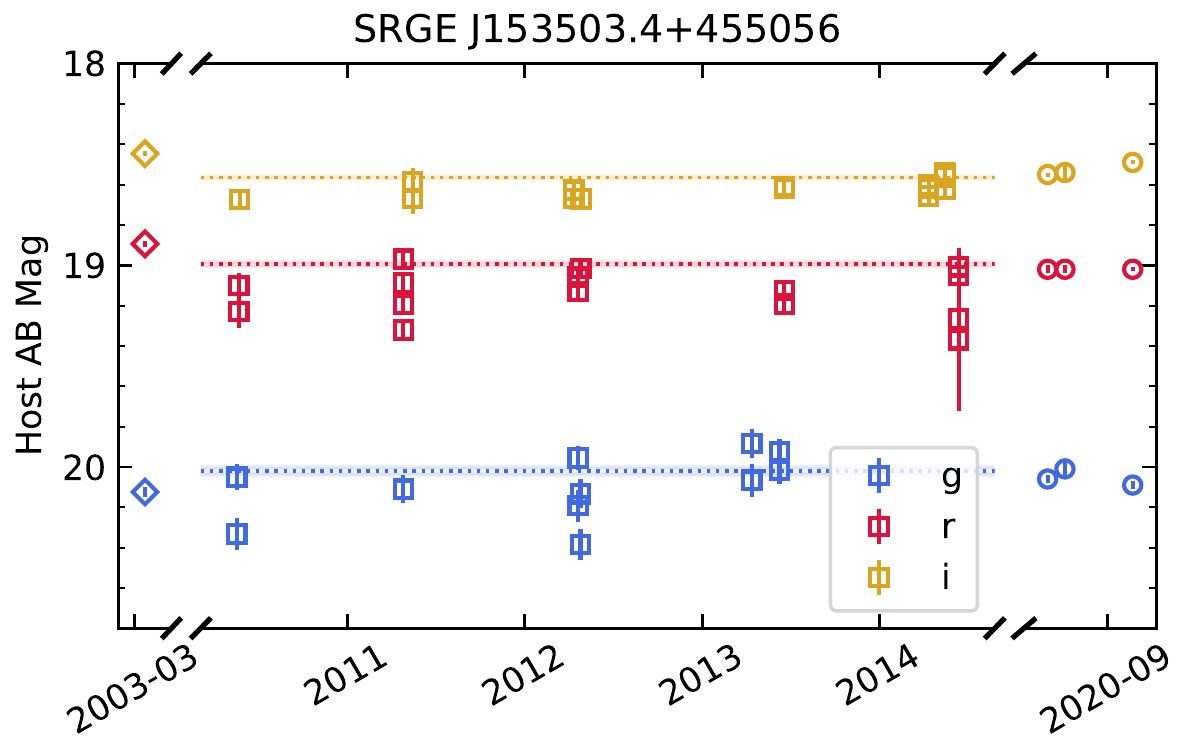}
\includegraphics[width=0.45\linewidth]{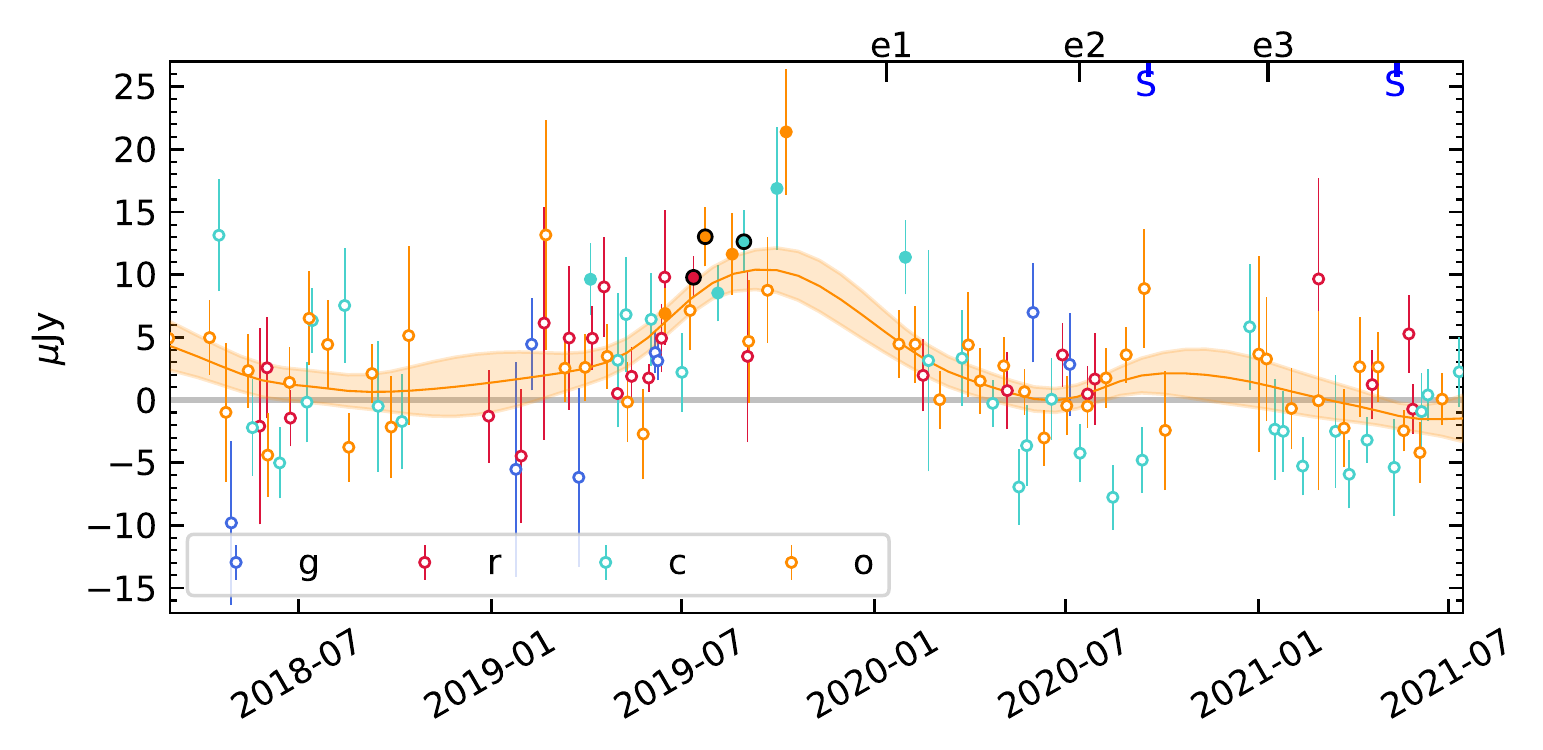}
\end{subfigure}
\begin{subfigure}[t]{\textwidth}
\centering
\includegraphics[width=0.35\linewidth]{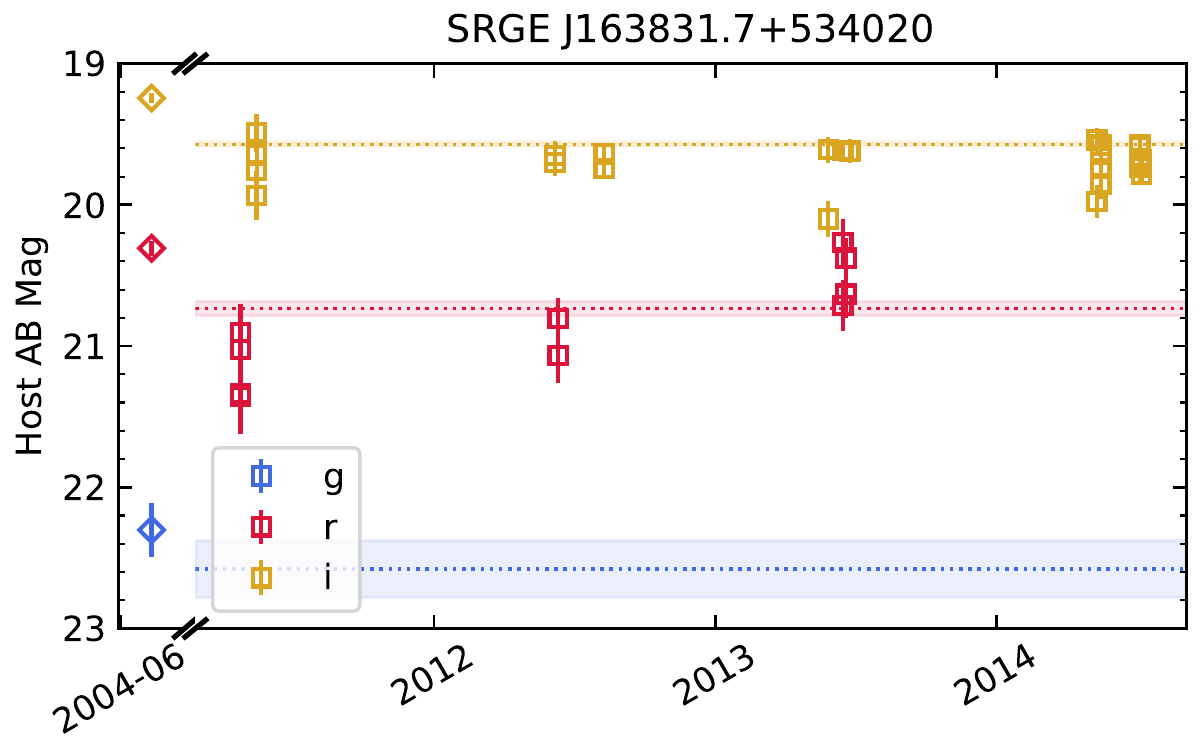}
\includegraphics[width=0.45\linewidth]{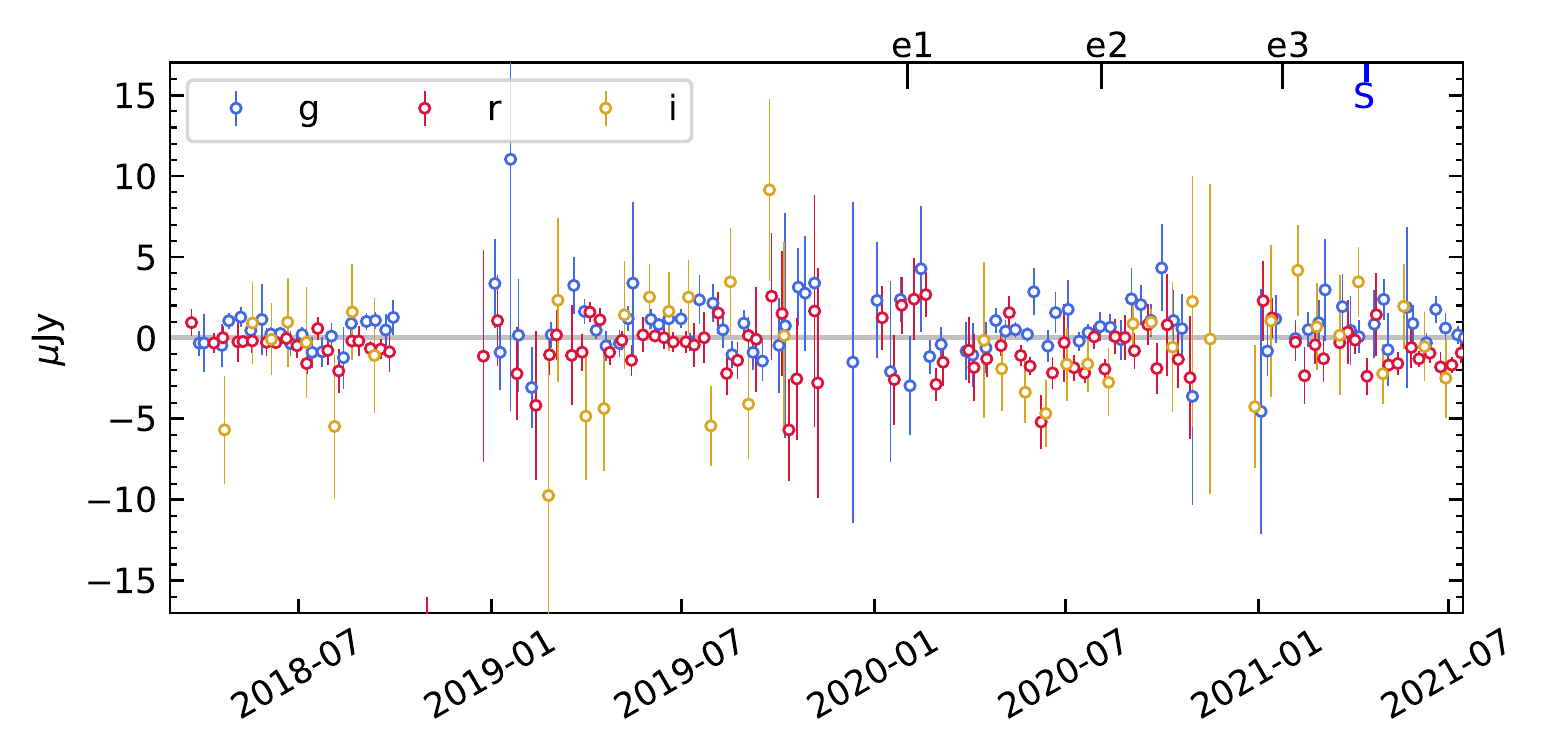}
\end{subfigure}
\begin{subfigure}[t]{\textwidth}
\centering
\includegraphics[width=0.35\linewidth]{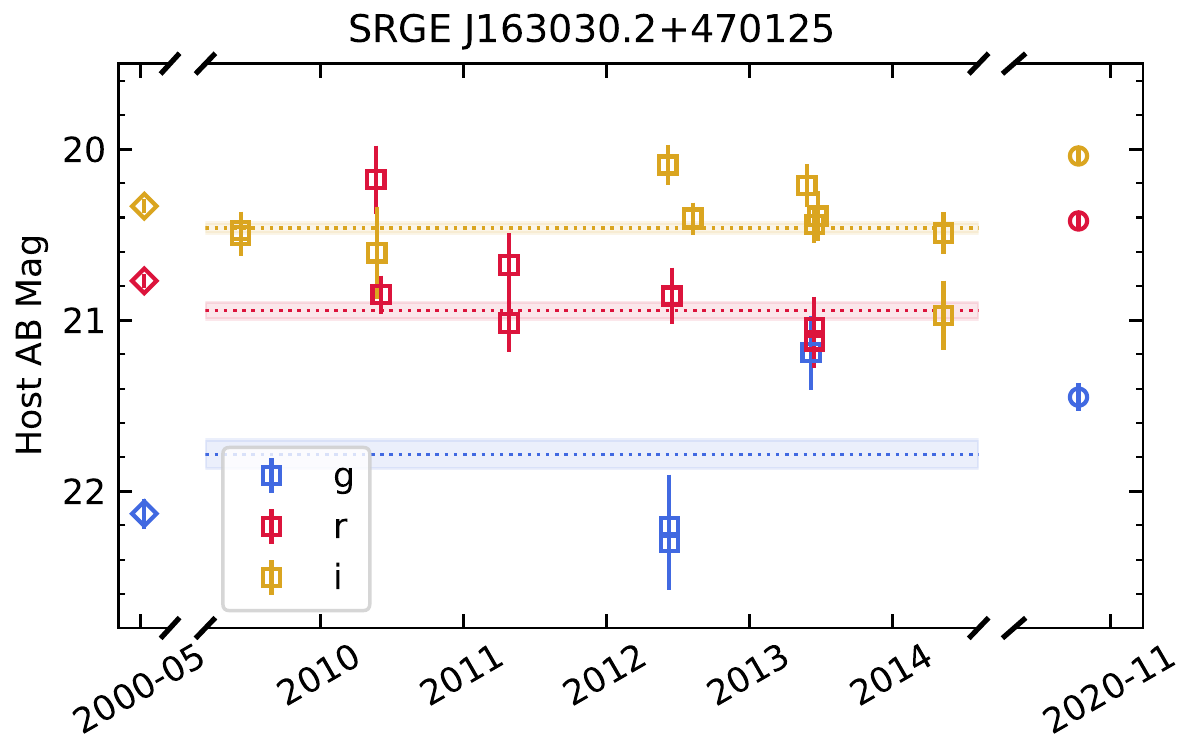}
\includegraphics[width=0.45\linewidth]{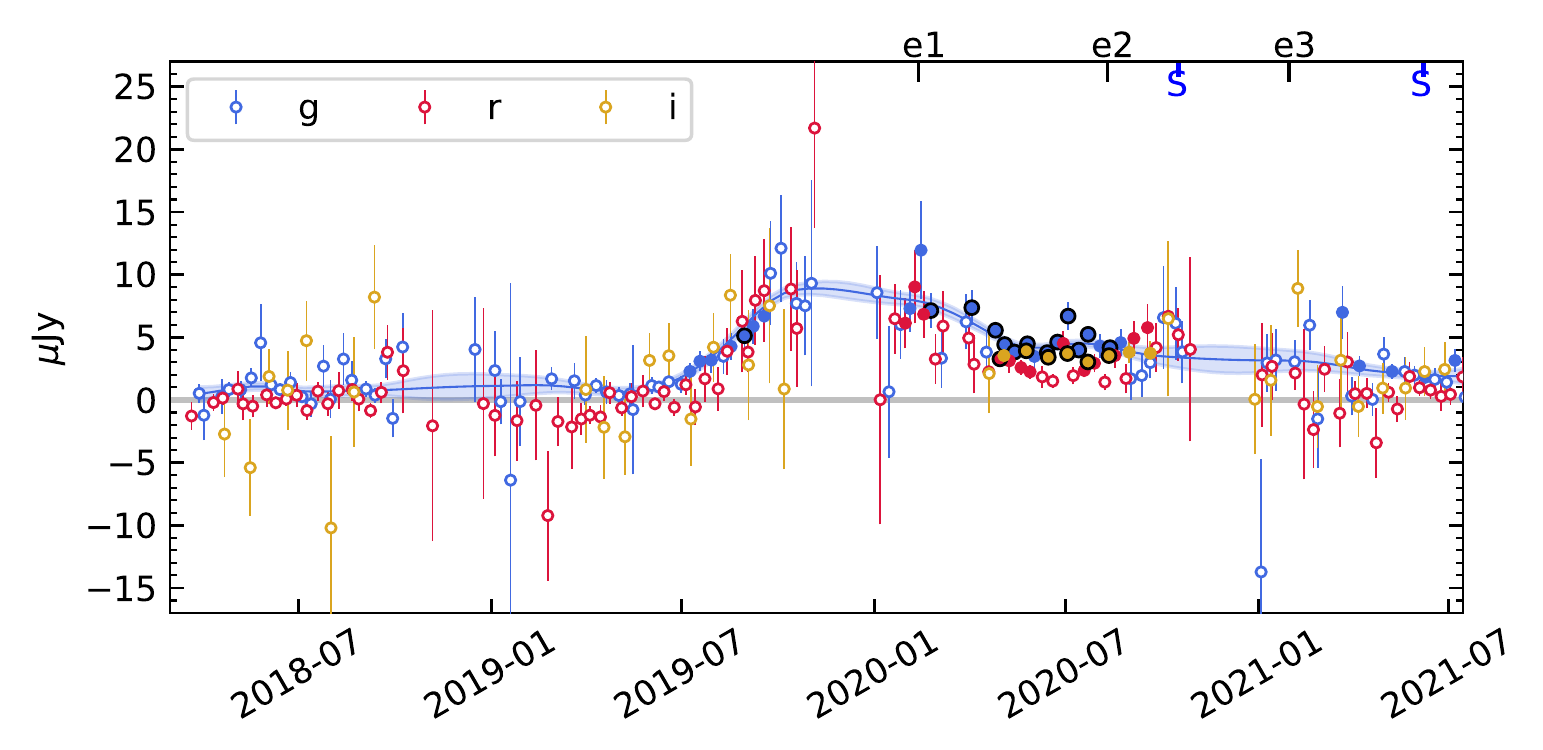}
\end{subfigure}
\caption{\textit{Left}: Optical light curves of \srg\ TDE host galaxies, showing SDSS (diamonds), Pan-STARRS DR2 (PS2, squares), and photomeetry obtained by us (circles, Table~\ref{tab:photometry}). The horizontal dotted lines mark the PS2 ($gri$) Kron magnitudes from stacked images. \textit{Right}: ZTF and ATLAS differential photometry performed at the centroids of the host galaxies. $<3\sigma$ data points are shown in hollow circles, $>3\sigma$ data points are shown in solid circles, and $>5\sigma$ data points are further highlighted using the black edge color. Along the upper axis, epochs of the \srg/\erosita\ visits (`e1', `e2', `e3') and spectroscopic observations (`S') are marked. For visual clarity, the ZTF-$g$ and ZTF-$r$ photometry are binned by 10\,days, and the ZTF-$i$ and ATLAS photometry are binned by 20\,days. For the four TDEs with optical flares, we show Gaussian process model fits following procedures described in Appendix~B.4 of \citet{Yao2020}. The models are fitted to a single band where the photometric uncertainty is the smallest or the temporal coverage is the highest. 
\label{fig:optlc}}
\end{figure*}

\addtocounter{figure}{-1}
\begin{figure*}
\centering
\begin{subfigure}[t]{\textwidth}
\centering
\includegraphics[width=0.35\linewidth]{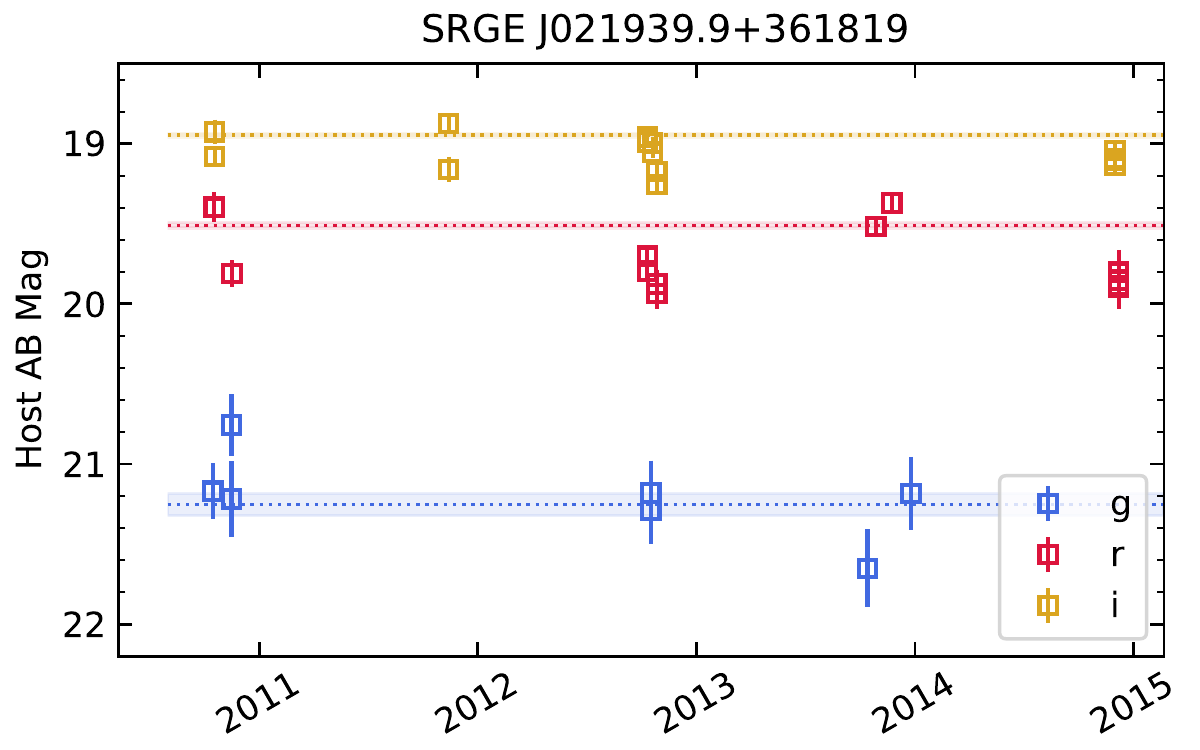}
\includegraphics[width=0.45\linewidth]{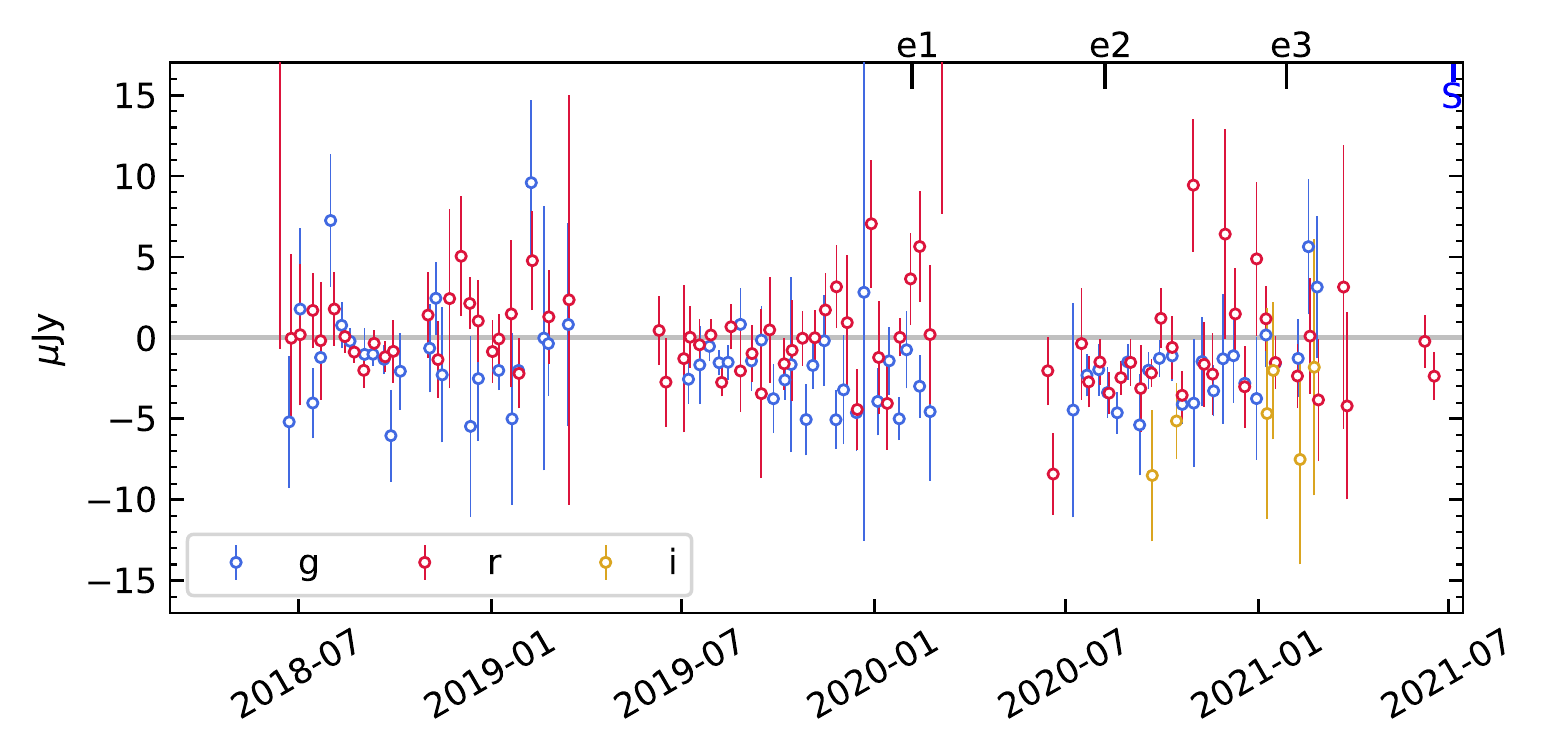}
\end{subfigure}
\begin{subfigure}[t]{\textwidth}
\centering
\includegraphics[width=0.35\linewidth]{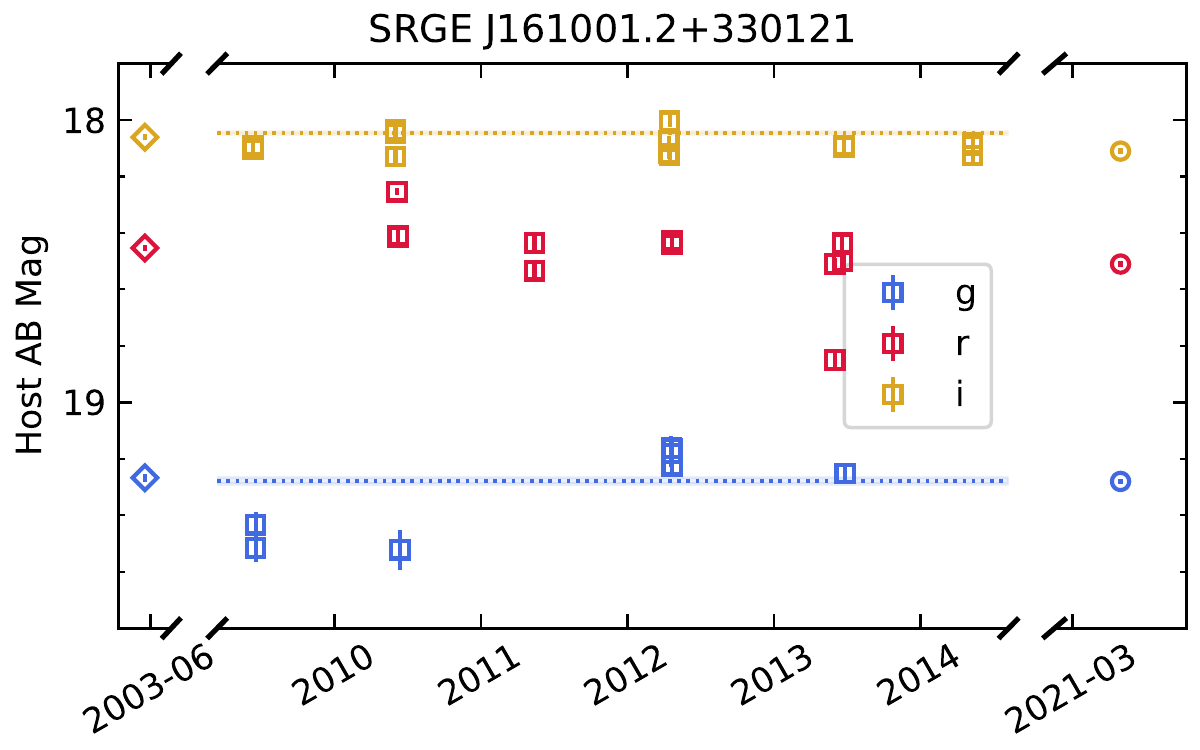}
\includegraphics[width=0.45\linewidth]{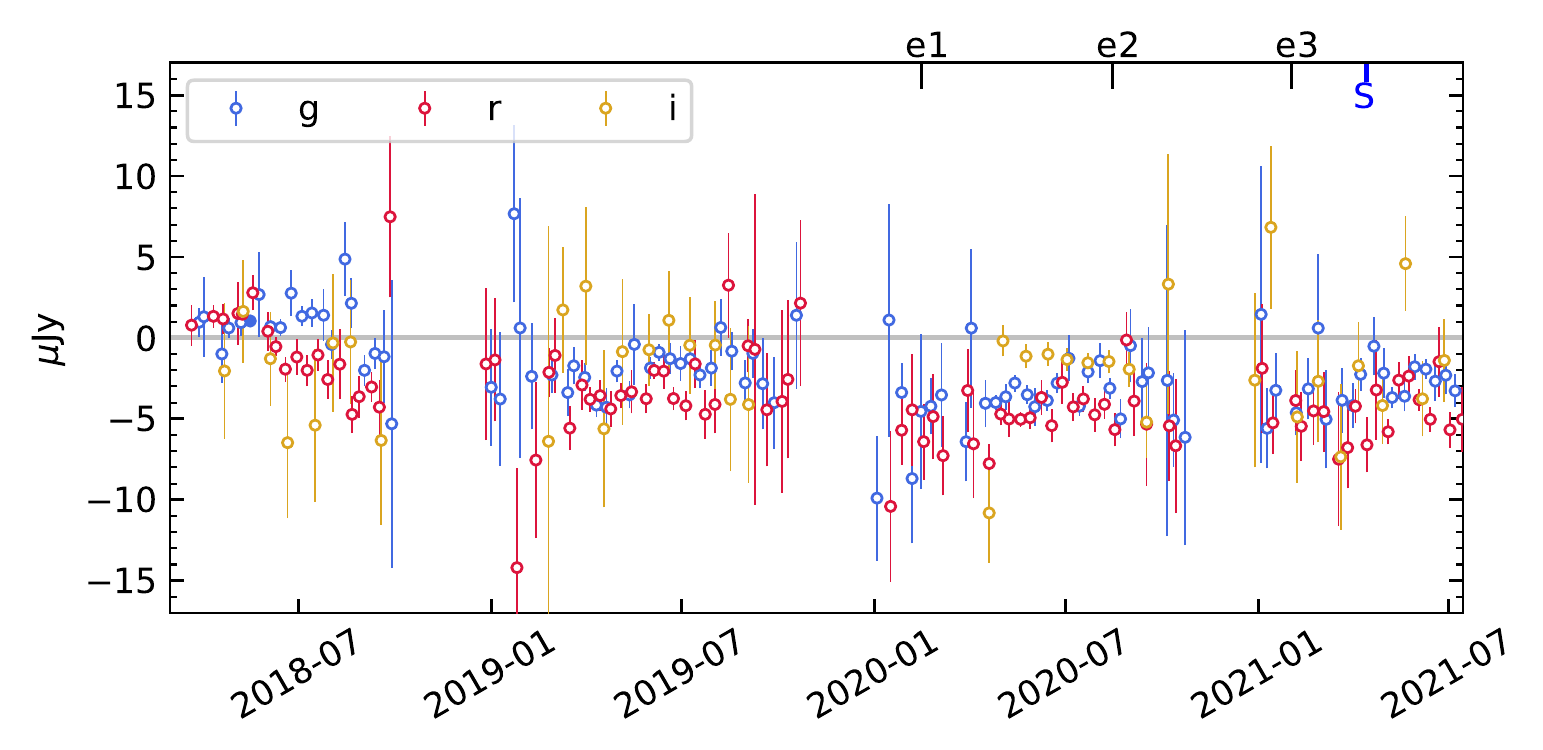}
\end{subfigure}
\begin{subfigure}[t]{\textwidth}
\centering
\includegraphics[width=0.35\linewidth]{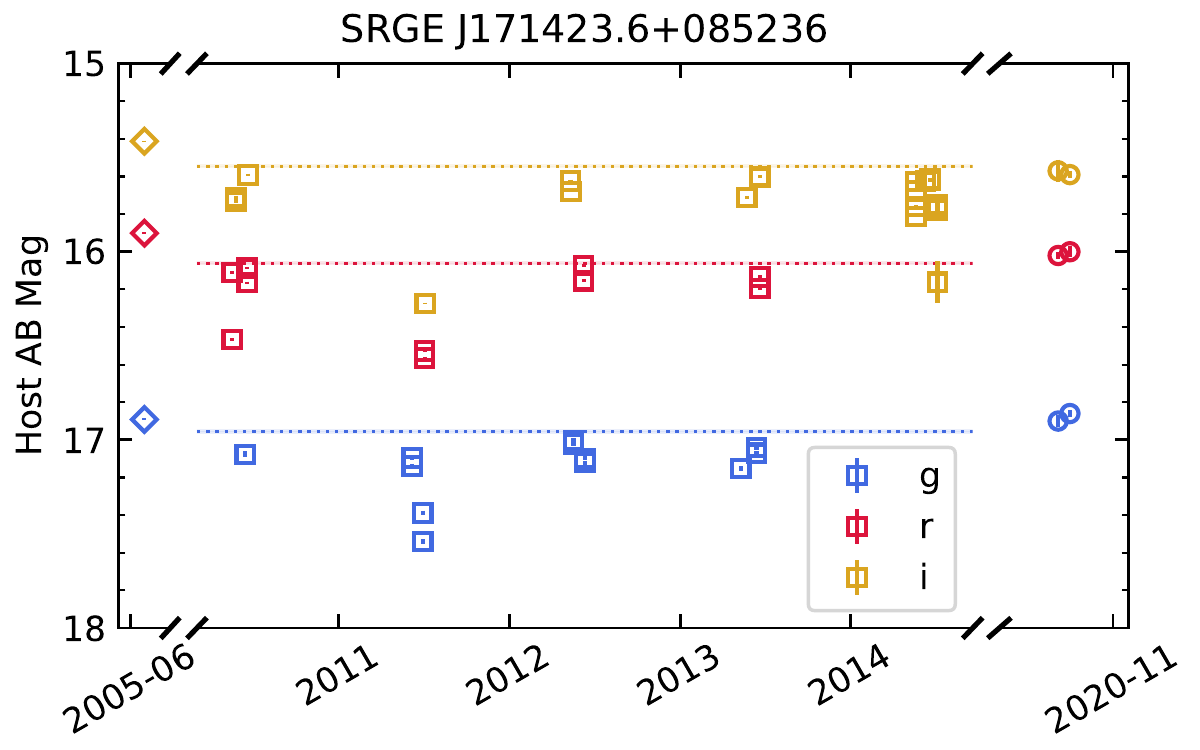}
\includegraphics[width=0.45\linewidth]{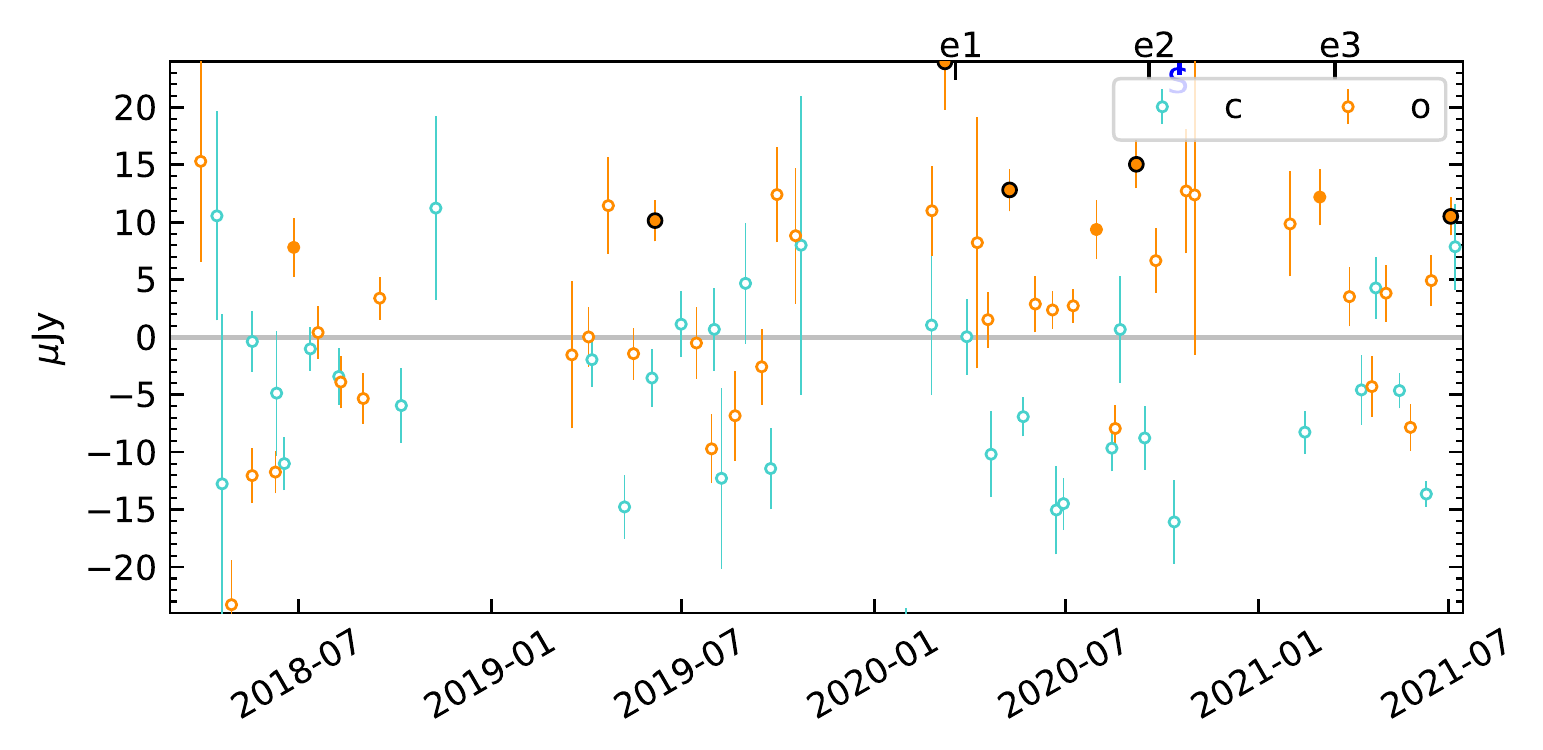}
\end{subfigure}
\begin{subfigure}[t]{\textwidth}
\centering
\includegraphics[width=0.35\linewidth]{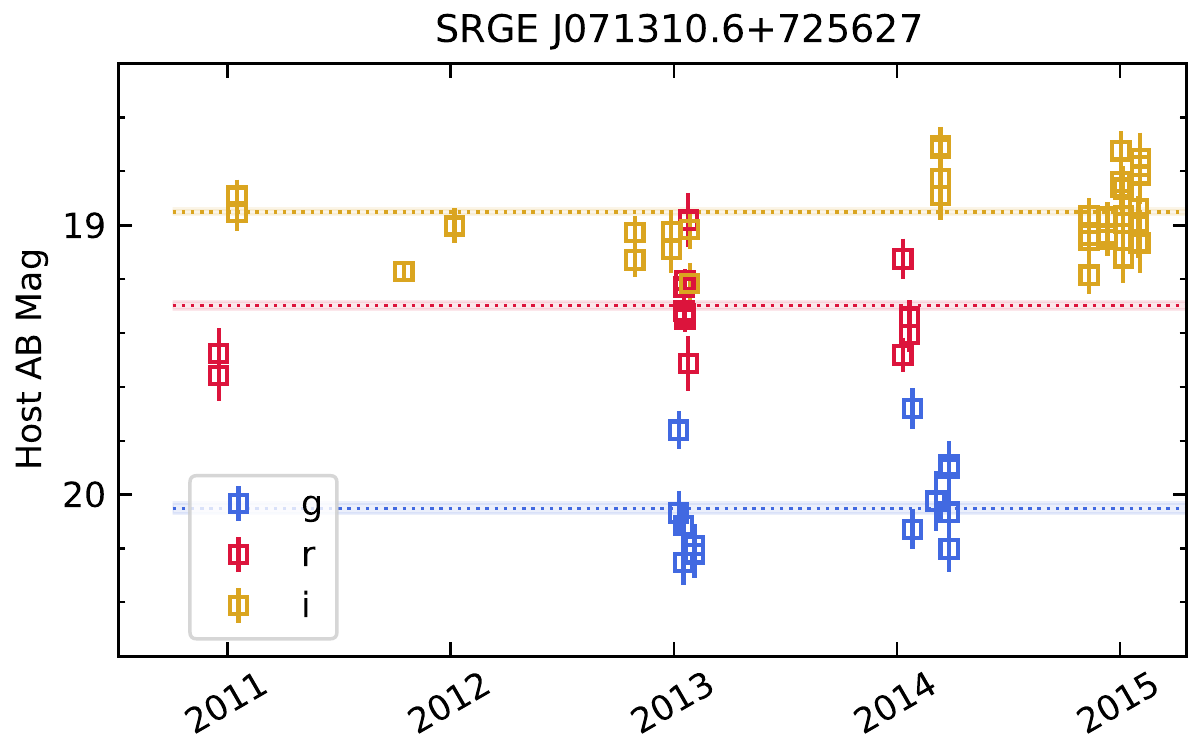}
\includegraphics[width=0.45\linewidth]{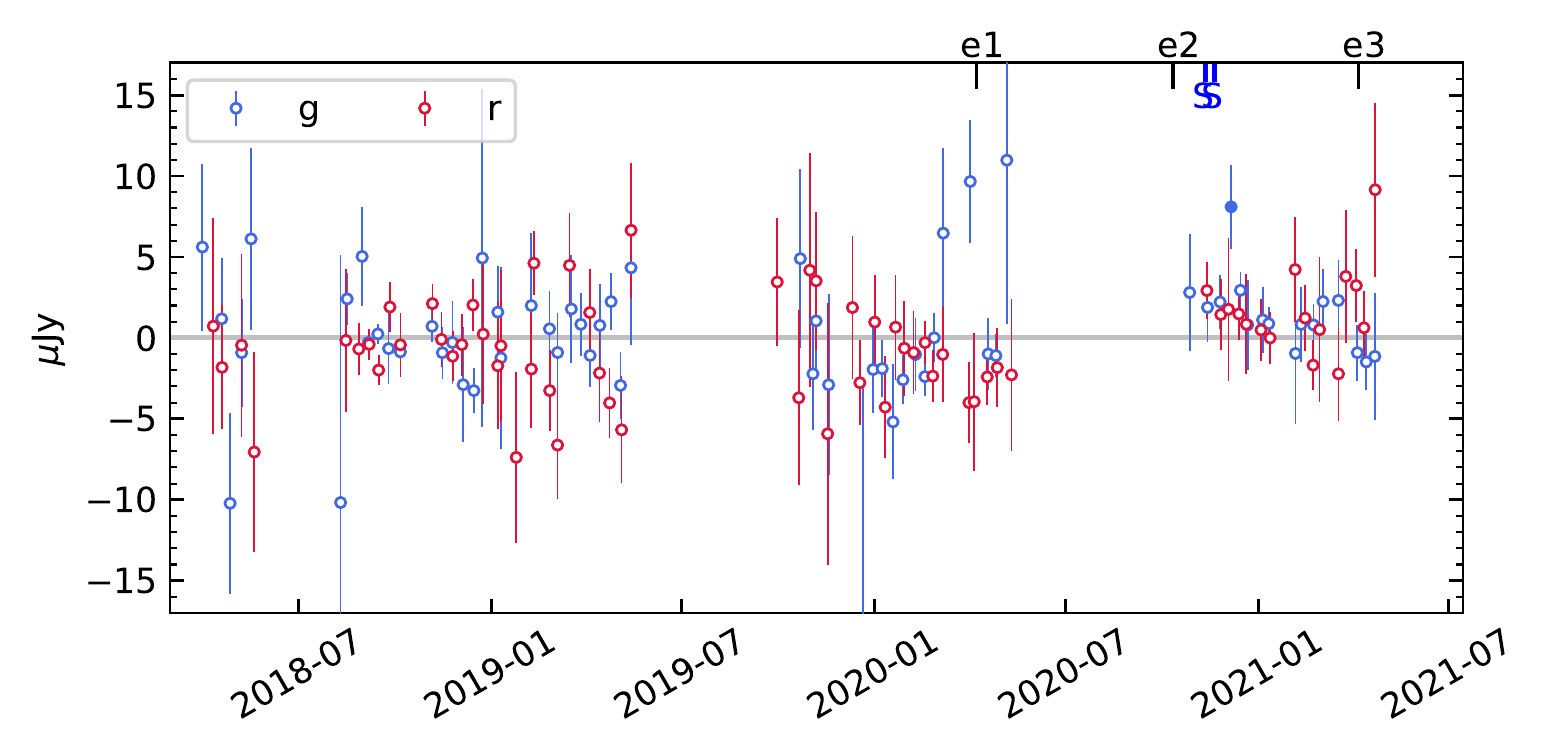}
\end{subfigure}
\begin{subfigure}[t]{\textwidth}
\centering
\includegraphics[width=0.35\linewidth]{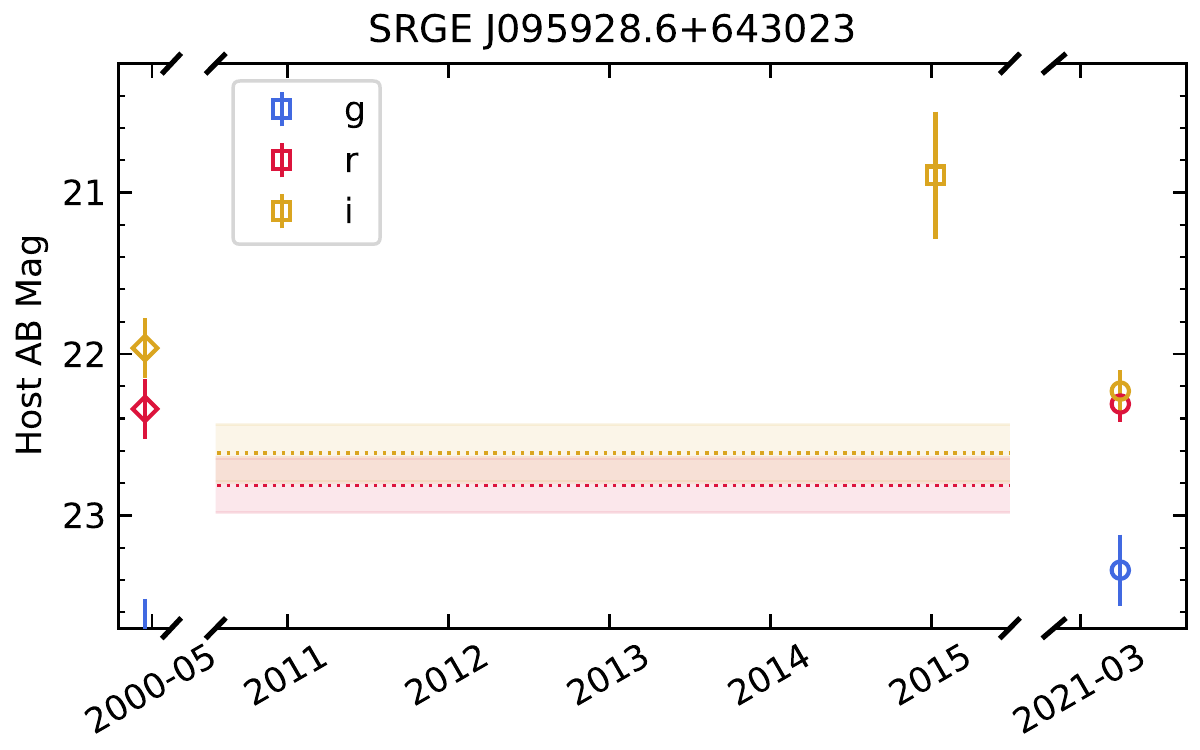}
\includegraphics[width=0.45\linewidth]{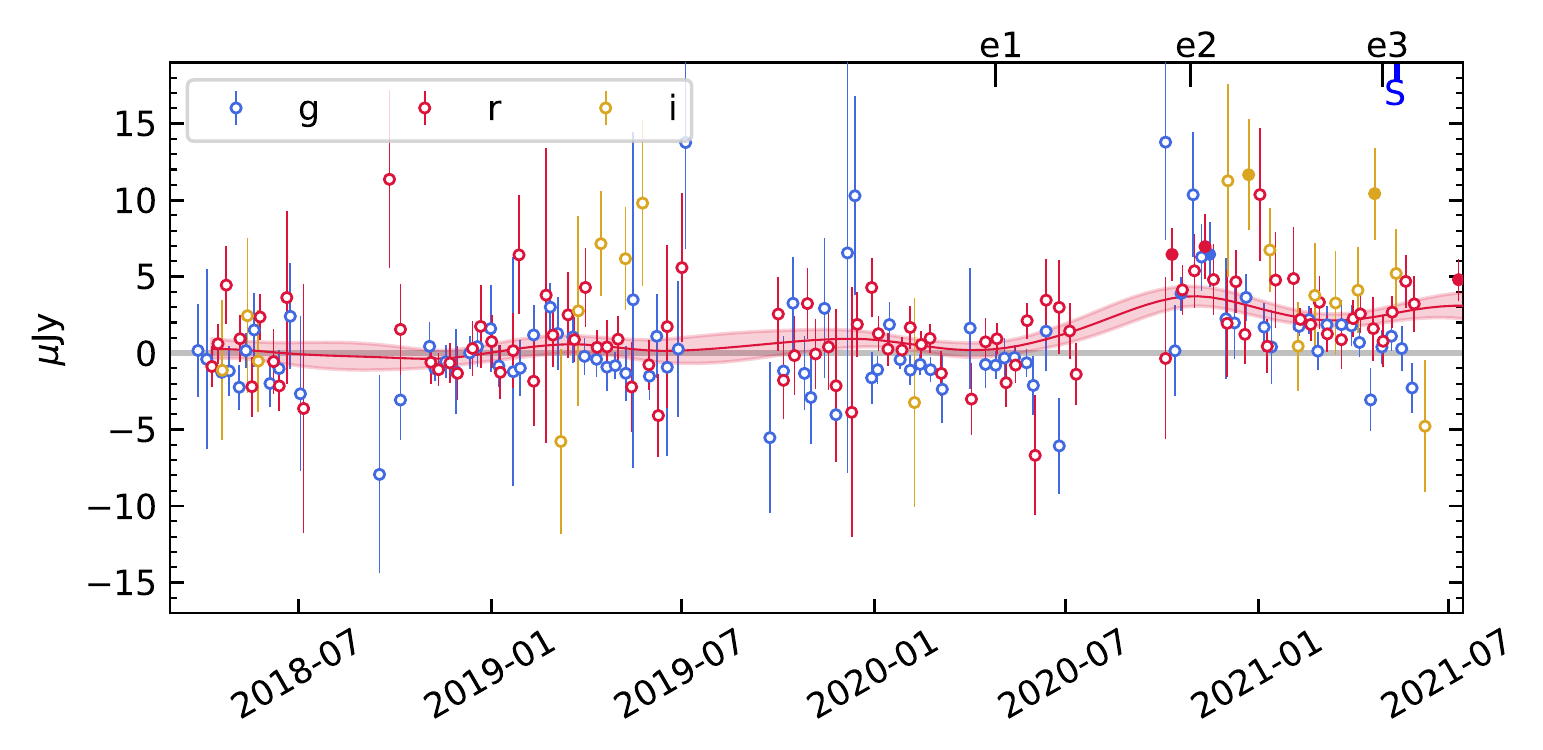}
\end{subfigure}
\caption{Continuation.}
\end{figure*}

\addtocounter{figure}{-1}
\begin{figure*}
\begin{subfigure}[t]{\textwidth}
\centering
\includegraphics[width=0.35\linewidth]{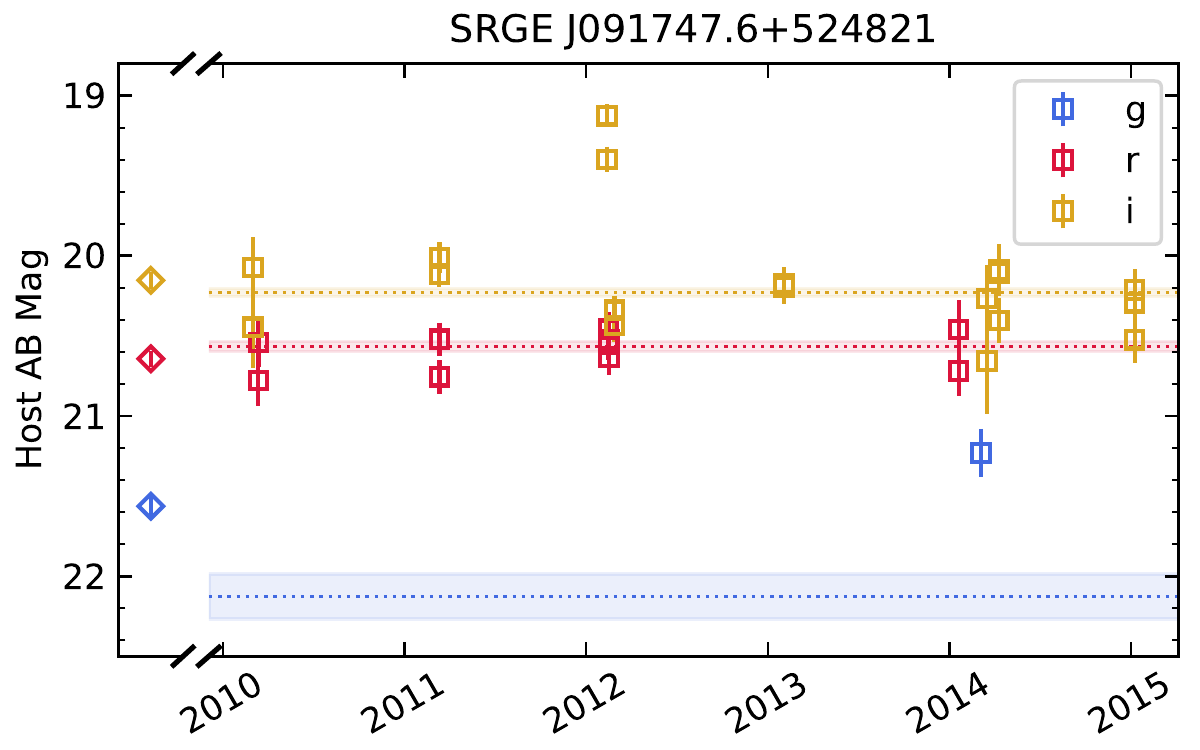}
\includegraphics[width=0.45\linewidth]{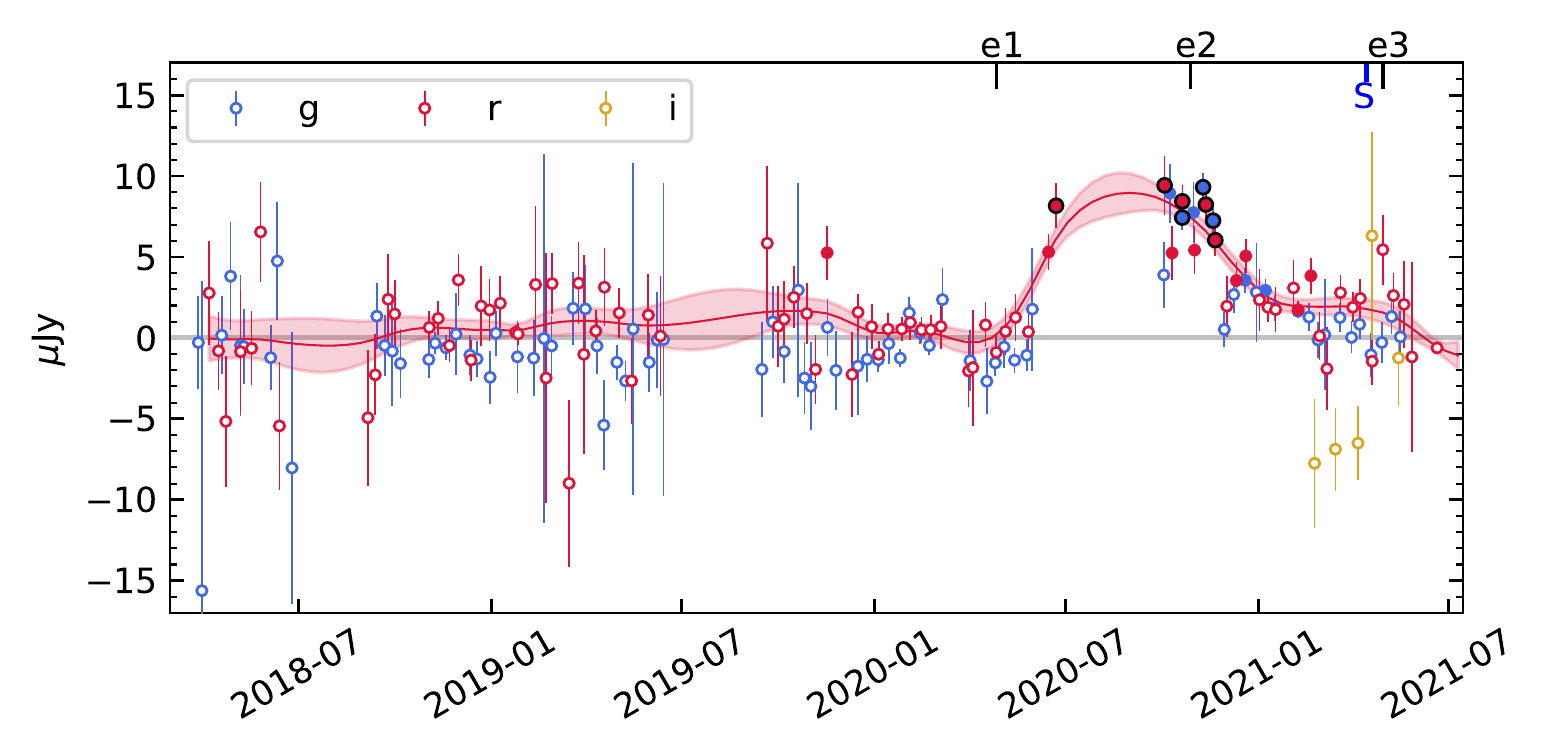}
\end{subfigure}
\begin{subfigure}[t]{\textwidth}
\centering
\includegraphics[width=0.35\linewidth]{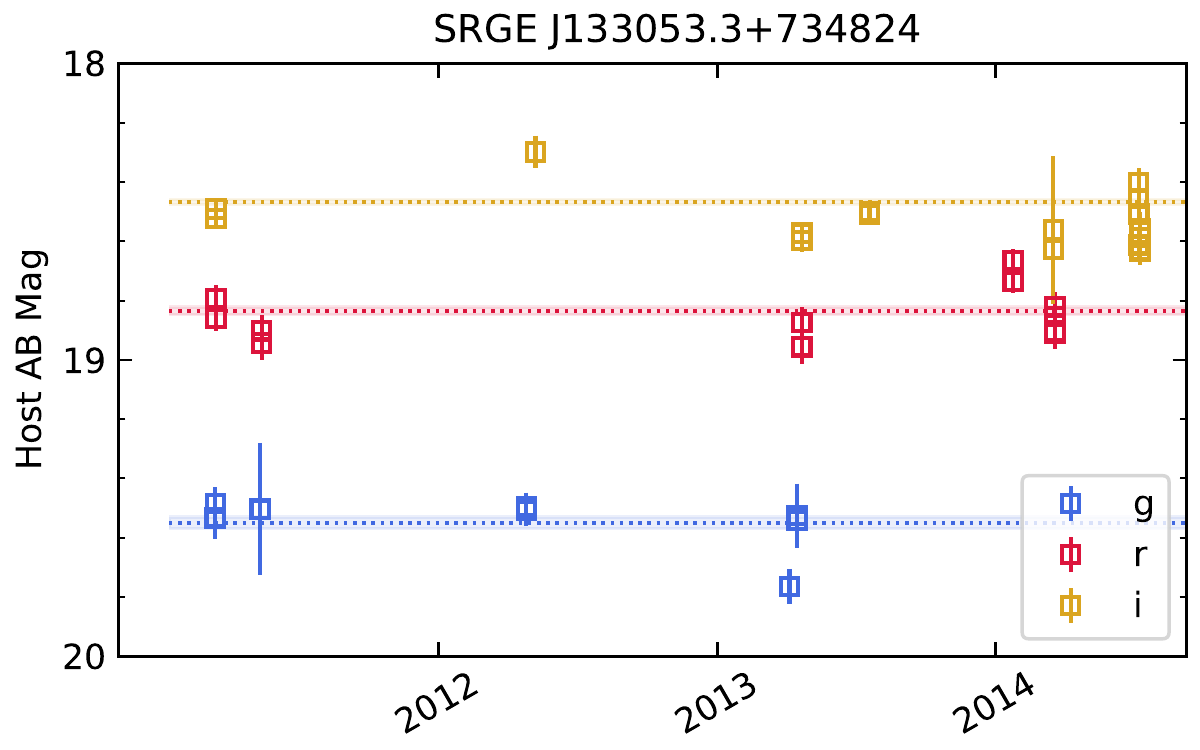}
\includegraphics[width=0.45\linewidth]{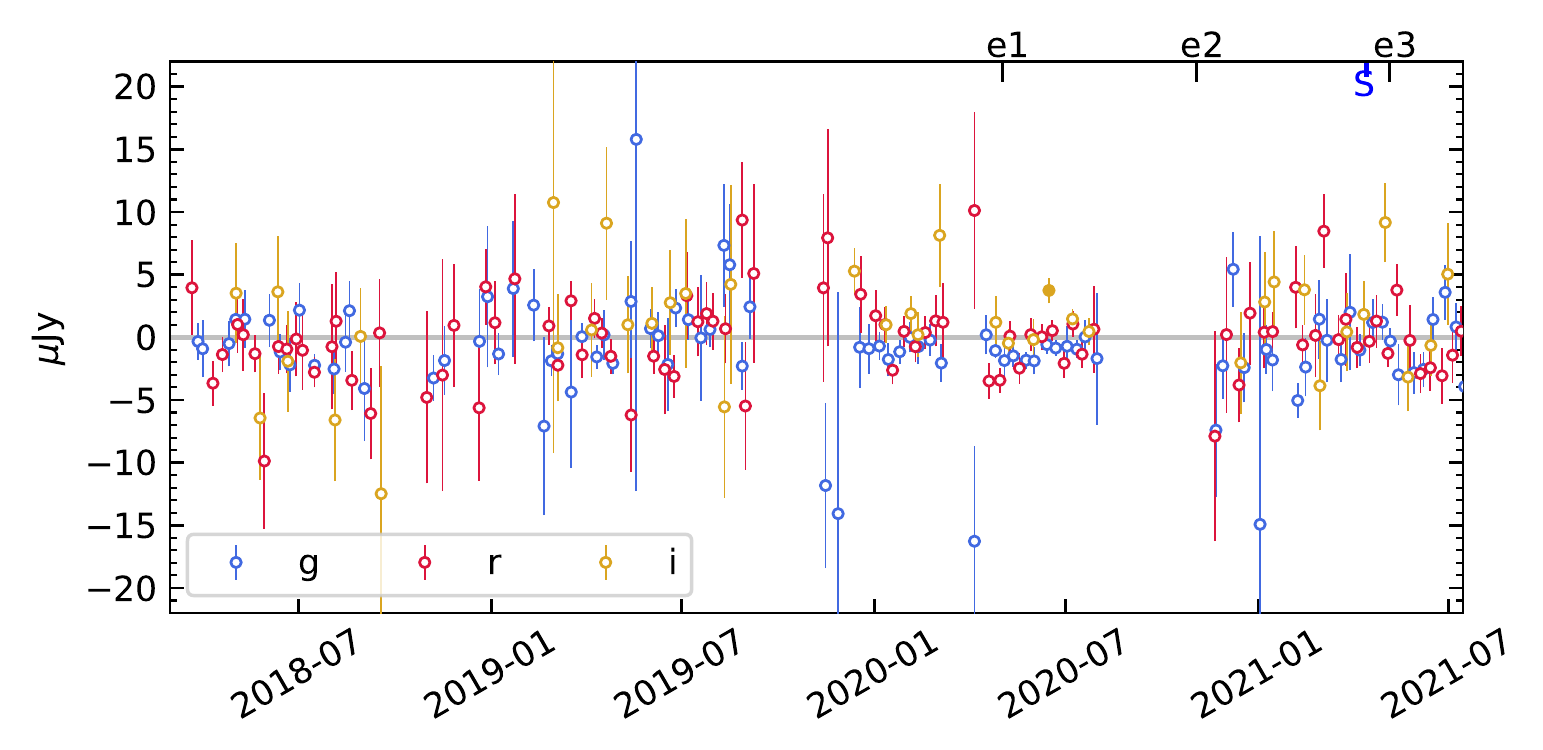}
\end{subfigure}
\begin{subfigure}[t]{\textwidth}
\centering
\includegraphics[width=0.35\linewidth]{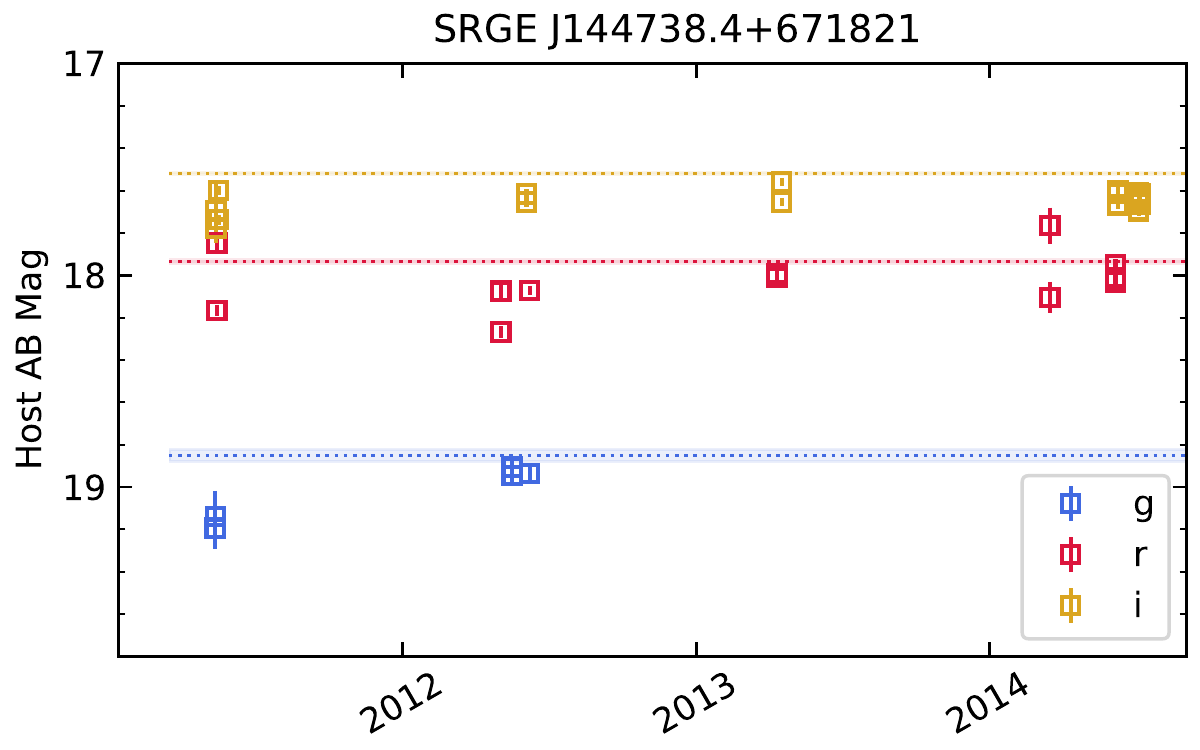}
\includegraphics[width=0.45\linewidth]{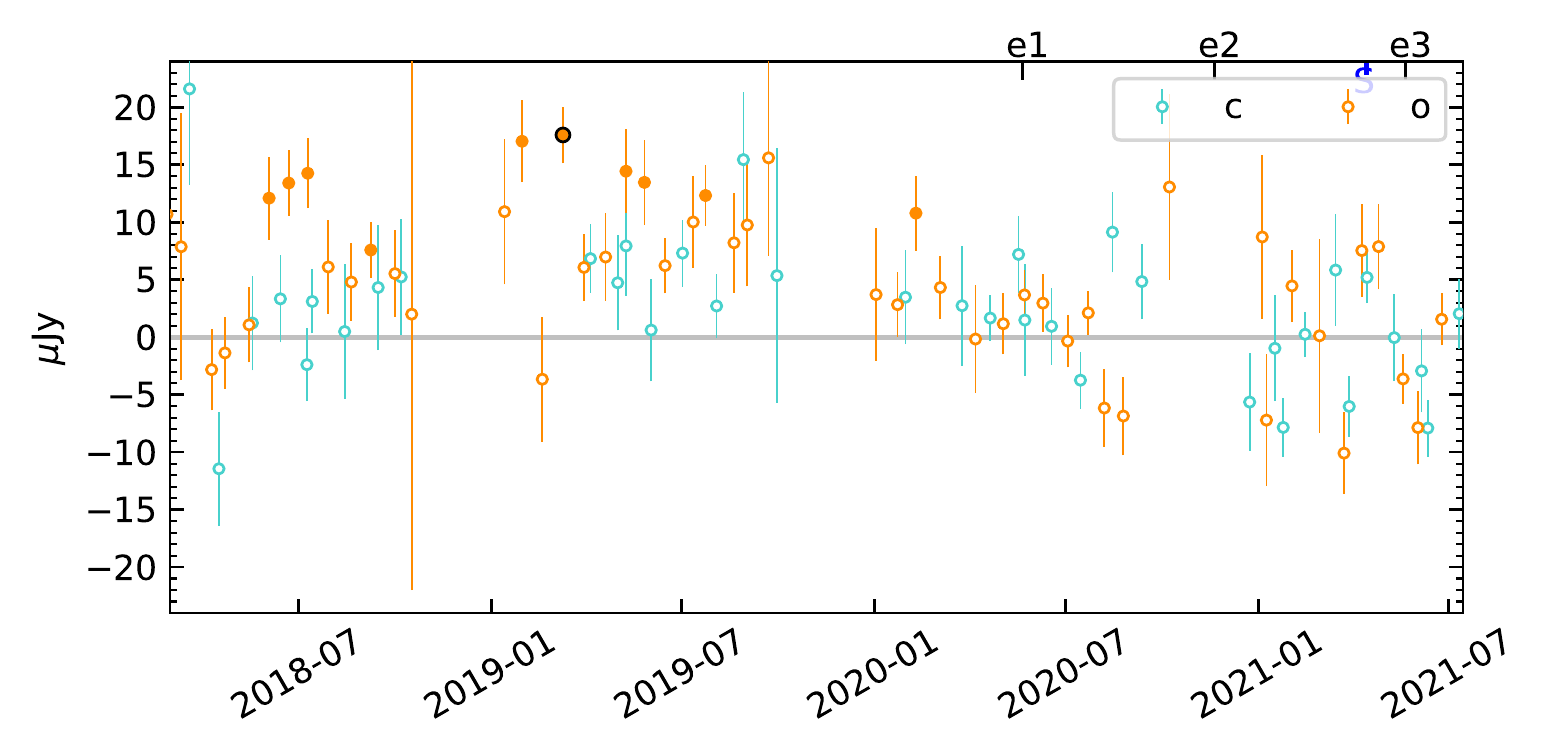}
\end{subfigure}
\caption{Continuation.}
\end{figure*}

Three, or possibly four, TDEs exhibit optical flares. For SRGE\,J153503.4+455056, we observe an optical brightening starting around July 2019 and ending before July 2020, i.e., just before its discovery in X-rays by \erosita\ on 13--16 July 2020, while during the previous visit in January 2020 \erosita\ did not detect this source. For SRGE\,J163030.2+470125, we observe a clear optical brightening starting in July 2019 and lasting at least until the end of 2020, whereas \erosita\ discovered this transient in X-rays in August 2020 and did not detect it in February 2020 and at the end of January 2021. For SRGE\,J091747.6+524821, we observe an optical brightening between June 2020 and March 2021, while \erosita\ discovered the X-ray transient in October 2020 and did not detect it in April 2020 and April 2021. A fourth object, SRGE\,J095928.6+643023, might also exhibit a weak optical flare around the eRASS2 visit in October--November 2020.

Therefore, for two events (SRGE\,J153503.4+455056 and SRGE\,J163030.2+470125) there is evidence that the optical activity started before the X-ray activity, whereas the data for another two transients (SRGE\,J095928.6+643023 and SRGE\,J091747.6+524821) are consistent with the X-ray and optical activity occurring nearly concurrently. 

The other nine events do not show evidence of optical flares. For SRGE\,J171423.6+085236 and SRGE\,J144738.4+671821, we observe moderate variability in the difference photometry light curve from July 2015 to July 2021 (ATLAS data before March 2018 is not shown in Fig.~\ref{fig:optlc} for visual clarity). An amplitude of $<20\,\mu$Jy in difference photometry corresponds to $<0.07$\,mag for SRGE\,J171423.6+085236 and $<0.16$\,mag for SRGE\,J144738.4+671821. Therefore, the scatter shown in Fig.~\ref{fig:optlc} is probably due to imperfect image subtraction at the bright galaxy nuclei, rather than variability associated with the stellar tidal disruption. Some other events (e.g., SRGE\,J135514.8+311605, SRGE\,J013204.6+122236, SRGE\,J071310.6+725627) show $>3\sigma$ data points in the binned differential photometry light curves. However, those data are consistent with random fluctuations. 

\begin{table*}
  \caption{Constraints on the optical luminosity.} 
  \label{tab:optlim}
  \begin{tabular}{lccccccc}
  \hline
  Object 
 & Optical flare peak
 & \multicolumn{3}{c}{$L_{\rm bb}^{\dagger}$, $10^{43}\,{\rm erg\,s^{-1}}$}
 & \multicolumn{3}{c}{$L_{g}^{\dagger}$, $10^{43}\,{\rm erg\,s^{-1}}$} \\
(SRGE) & AB mag & $ 1.3\times 10^4$\,K  & $ 2.5\times 10^4$\,K & $4.0\times 10^4$\,K
 & $1.3\times 10^4$\,K & $2.5\times 10^4$\,K & $4.0\times 10^4$\,K\\
\hline
J135514.8+311605 & $g>21.83$
& $<0.66$ & $<1.92$ & $<5.67$
& $<0.39$ & $<0.33$ & $<0.31$ \\ 
J013204.6+122236 & $g>21.51$
& $<0.40$ & $<1.24$ & $<3.75$
& $<0.22$ & $<0.20$ & $<0.19$ \\ 
J153503.4+455056 & $21.33<o<20.58$
& 1.56--3.12 & 6.01--11.99 & 19.52--38.95
& 0.94--1.87 & 1.05--2.09 & 1.09--2.17 \\ 
J163831.7+534020 & $g>21.84$, $r>21.86$
& $<4.49$ & $<10.24$ & $<26.67$
& $<3.46$ & $<2.29$ & $<1.91$ \\ 
J163030.2+470125 & $21.47 <g<20.71$
& 1.92--3.86 & 5.13--10.32 & 14.66--29.52 
& 1.21--2.43 & 0.94--1.89 & 0.86--1.73 \\  
J021939.9+361819 & $g>21.43$
& $<3.34$ & $<8.21$ & $<22.78$ 
& $<2.26$ & $<1.61$ & $<1.43$ \\ 
J161001.2+330121 & $g>21.68$
& $<0.34$ & $<1.06$ & $<3.18$
& $<0.19$ & $<0.17$ & $<0.16$ \\ 
J171423.6+085236 & $c>20.04$
& $<0.13$ & $<0.47$ & $<1.51$
& $<0.07$ & $<0.07$ & $<0.07$ \\ 
J071310.6+725627 & $g>21.41$
& $<0.28$ & $<0.89$ & $<2.69$
& $<0.15$ & $<0.14$ & $<0.13$\\ 
J095928.6+643023 & $22.40<r<21.64$
& 1.80--3.62 & 5.80--11.67 & 17.73--35.71
& 1.27--2.56 & 1.19--2.41 & 1.17--2.35 \\ 
J091747.6+524821 & $21.49<r<20.74$ 
& 0.90--1.79 & 3.42--6.83 & 11.11--22.17
& 0.52--1.04 & 0.58--1.15 & 0.60--1.19 \\ 
J133053.3+734824 & $g>21.73$
& $<0.42$ & $<1.29$ & $<3.86$ 
& $<0.24$ & $<0.21$ & $<0.20$ \\ 
J144738.4+671821 & $c>20.41$
& $<1.04$ & $<3.51$ & $<10.90$
& $<0.57$ & $<0.56$ & $<0.56$ \\ 
\hline
  \end{tabular}
  
Notes. Numbers presented in this table are corrected for Galactic extinction. $^\dagger$: Assuming the optical emission can be described by a blackbody with typical temperatures of $1.3\times 10^4$\,K, $2.5\times 10^4$\,K, or $4.0\times 10^4$\,K, $L_{\rm bb}$ is the blackbody luminosity, and $L_g$ is the rest-frame $g$-band luminosity.
\end{table*}

\begin{table*}
    \centering
    \caption{Host properties (line fluxes). \label{tab:host_line}}
    \begin{tabular}{ccccccccc}
    \hline 
        Object (SRGE) & $f_{\rm H\alpha}$ & $f_{\rm H\beta}$ & $f_{\rm [NII]6583}$ & $f_{\rm [SII]6717+6731}$ & $f_{\rm [OII]3726+3729}$ & $f_{\rm [OIII]5007}$  & $L_{\rm [OIII]5007}$ & BPT class\\
    \hline
J013204.6+122236 &
-- &
-- &
-- &
$10.7 \pm 2.4$ &
-- &
$3.5 \pm 1.1$ &
$18.7 \pm 6.0$ &
-- \\
J153503.4+455056 &
$24.9 \pm 1.2$ &
-- &
$14.2 \pm 1.0$ &
-- &
$10.8 \pm 2.7$ &
$4.4 \pm 1.4$ &
$72.2 \pm 23.5$ &
-- \\
J163831.7+534020 &
-- &
$6.3 \pm 0.8$ &
-- &
-- &
$39.6 \pm 6.3$ &
$6.9 \pm 1.1$ &
$987.3 \pm 154.1$ &
-- \\
J161001.2+330121 &
$30.4 \pm 0.9$ &
$7.7 \pm 1.1$ &
$19.1 \pm 0.9$ &
$13.9 \pm 2.2$ &
$24.7 \pm 2.4$ &
-- &
-- &
-- \\
J171423.6+085236 &
$42.3 \pm 1.7$ &
$12.1 \pm 1.3$ &
$38.5 \pm 1.8$ &
$26.0 \pm 3.5$ &
$24.7 \pm 2.4$ &
$28.6 \pm 1.5$ &
$12.6 \pm 0.6$ &
LINER \\
J071310.6+725627 &
$11.8 \pm 0.7$ &
-- &
$6.4 \pm 0.7$ &
$4.3 \pm 1.0$ &
$9.6 \pm 2.5$ &
-- &
-- &
-- \\
J133053.3+734824 &
$50.3 \pm 0.9$ &
$10.4 \pm 1.3$ &
$27.0 \pm 0.8$ &
$15.5 \pm 1.6$ &
$22.6 \pm 2.0$ &
$3.9 \pm 1.1$ &
$25.2 \pm 6.8$ &
Composite \\
J144738.4+671821 &
$10.3 \pm 1.0$ &
-- &
$19.8 \pm 1.3$ &
$14.7 \pm 2.4$ &
$44.4 \pm 4.6$ &
$7.0 \pm 2.4$ &
$29.9 \pm 10.0$ &
LINER/Seyfert \\
    \hline
    \end{tabular}
    
Notes. Observed line fluxes are given in units of $10^{-17}\,{\rm erg\,s^{-1}\,cm^{-2}}$. The luminosity of [OIII]5007 (in units of $10^{38}\,{\rm erg\,s^{-1}}$) is corrected for Galactic extinction. 
\end{table*}

\begin{table}
\caption{Host properties (equivalent width and WISE colors).} 
  \label{tab:host_ew}
  \begin{tabular}{rrrr}
  \hline
  Object (SRGE) & EW(H$\alpha_{\rm em}$) & EW(H$\delta_{\rm A}$) & W1$-$W2\\
  \hline
J135514.8+311605 &
$2.42 \pm 0.62$ &
$0.70$ &
$0.26 \pm 0.12$ \\
J013204.6+122236 &
$0.11 \pm 0.31$ &
$-1.38$ &
$0.11 \pm 0.04$ \\
J153503.4+455056 &
$5.88 \pm 0.22$ &
$0.46$ &
$0.29 \pm 0.06$ \\
J163831.7+534020 &
$\gtrsim 7$ &
$-0.95$ &
$-0.04 \pm 0.05$ \\
J163030.2+470125 &
$1.10 \pm 0.33$ &
$1.21$ &
$0.11 \pm 0.43$ \\
J021939.9+361819 &
$0.52 \pm 0.37$ &
$-1.08$ &
$0.15 \pm 0.08$ \\
J161001.2+330121 &
$6.71 \pm 0.26$ &
$2.60$ &
$0.10 \pm 0.04$ \\
J171423.6+085236 &
$3.29 \pm 0.42$ &
$0.24$ &
$-0.05 \pm 0.04$ \\
J071310.6+725627 &
$4.61 \pm 0.27$ &
$1.40$ &
$-0.01 \pm 0.12$ \\
J095928.6+643023 &
$\gtrsim 0$ &
$3.74$ &
--- \\
J091747.6+524821 &
$0.08 \pm 0.55$ &
$1.20$ &
$0.44 \pm 0.33$ \\
J133053.3+734824 &
$12.11 \pm 0.46$ &
$2.73$ &
$0.16 \pm 0.08$ \\
J144738.4+671821 &
$3.12 \pm 0.24$ &
$-1.91$ &
$0.12 \pm 0.03$ \\
\hline
    \end{tabular}
    
Notes. Equivalent widths are given in units of \AA. The WISE color W1$-$W2 is given in the Vega system.
\end{table}

\subsection{Constraints on optical luminosity}
\label{s:optlum}

The observed optical flare peak magnitudes or upper limits are shown in Table~\ref{tab:optlim}. For the nine events without optical flares, we compute long-term median of $3\sigma$ upper limits using ZTF and ATLAS forced differential photometry. For the four events with optical flares, we calculate maximum of the model fits shown in Fig.~\ref{fig:optlc}. Since the optical data have seasonal gaps, the peak of the optical emission might be missed. Therefore, we consider the actual optical peak to be 1--2 times the model maximum.

Previous studies have shown that the broad-band SED of UV or optically discovered TDEs can be described by blackbody spectra with temperatures ($T_{\rm bb}$) between $\sim 1.3\times 10^4$\,K and $\sim 4.0\times 10^4$\,K \citep{Velzen_2020, Gezari2021}. Assuming three typical values of $T_{\rm bb}$, we report the constraints on optical luminosities in Table~\ref{tab:optlim}. Although the total blackbody luminosity $L_{\rm bb}$ is largely model dependent, the rest-frame $g$-band luminosity $L_g$ only has a weak dependence on $T_{\rm bb}$.

\subsection{WISE detection of a luminous infrared echo in SRGE\,J153503.4+455056}

\begin{figure}
    \centering
    \includegraphics[width=\columnwidth]{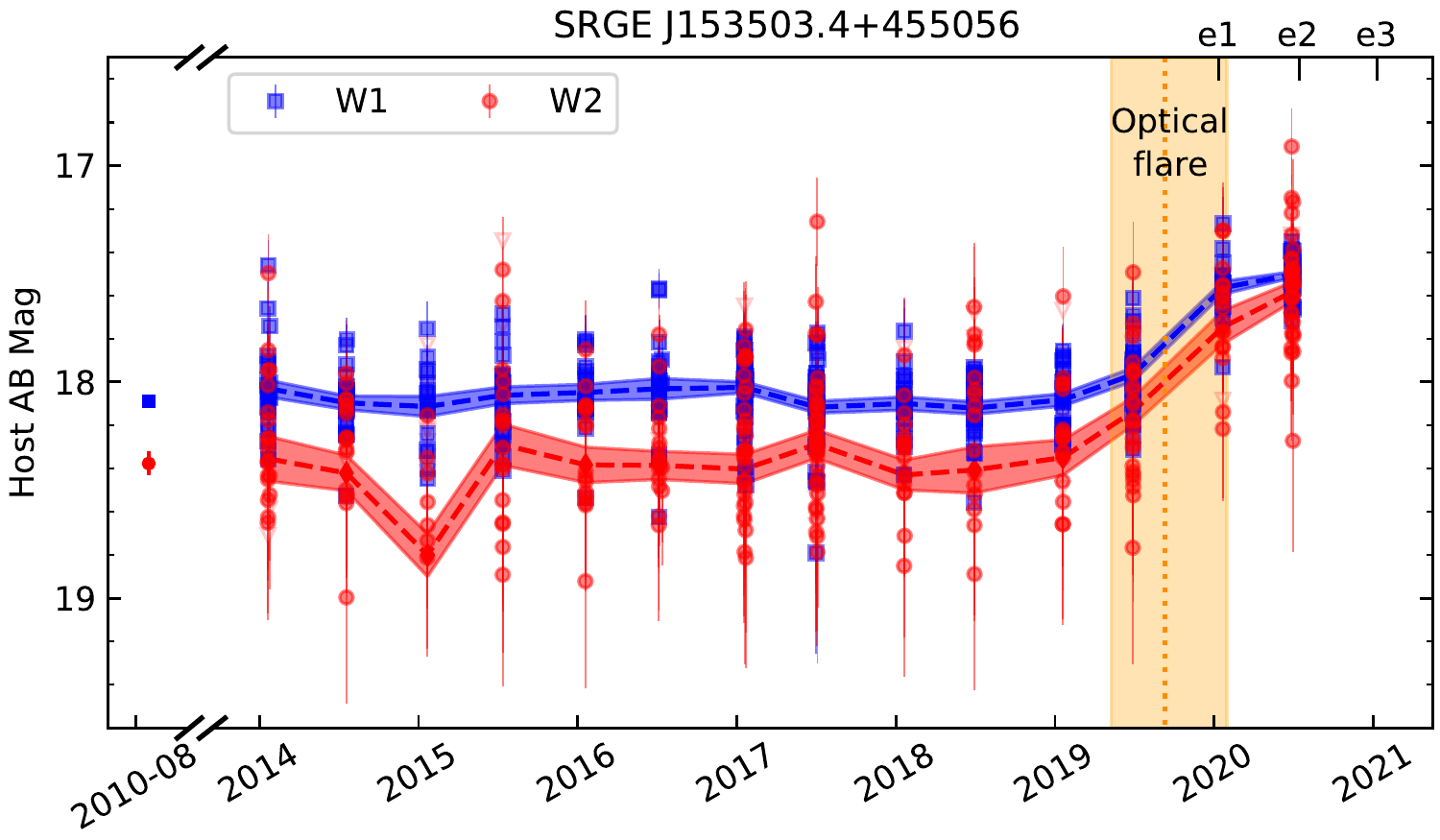}
    \caption{WISE W1 (3.4\,$\mu$m) and W2 (4.6\,$\mu$m) light curves of SRGE\,J153503.4+455056. The blue and red dashed lines mark the median magnitude of each WISE epoch. The vertical dotted line denotes the optical maximum. The vertical orange band marks the e-folding rise and decay timescales of the optical flare around the optical maximum (estimated using the Gaussian process model shown in Fig.~\ref{fig:optlc}). Epochs of the \srg/\erosita\ visits are marked along the upper axis.\label{fig:wiselc}}
\end{figure}

If the circum-nuclear medium of a TDE is dusty, a significant fraction of the UV/optical radiation energy will be absorbed by dust and reprocessed to the infrared \citep{Lu2016}. The resulting IR echoes bear important information on the dust properties at sub-pc scales in quiescent galaxies \citep{vanVelzen2016, vanVelzen2021}. By performing photometry on time-resolved AllWISE/NEOWISE coadds, \citet{Jiang2021} reported the detection of IR echoes in 8 of 23 optically selected TDEs. The ratio of their peak dust luminosity ($L_{\rm dust, peak}\sim10^{41}$--$10^{42}\,{\rm erg\,s^{-1}}$) and their UV/optical luminosity ($L_{\rm bb, peak}\sim 10^{44}\,{\rm erg\,s^{-1}}$) suggests a dust-covering factor $f_c = L_{\rm dust, peak}/L_{\rm bb, peak}$ of $\lesssim 0.01$. 

In order to search for IR echoes in the \srg\ TDE sample, we collected AllWISE and NEOWISE-R data for the TDE hosts. Significant IR brightening was observed in SRGE\,J153503.4+455056 (Fig.~\ref{fig:wiselc}). In the most recent NEOWISE epoch (from 2021 June 26 to July 1), compared with the 2010 baseline, its IR flux has increased by $150\pm7\,\mu$Jy in W1 and $174\pm16\,\mu$Jy in W2. The corresponding dust luminosity and temperature are $L_{\rm dust}=(3.0\pm0.2)\times 10^{43}\,{\rm erg\,s^{-1}}$ and $T_{\rm dust}=1393^{+196}_{-145}$\,K. The derived $L_{\rm dust}$ is a lower limit on $L_{\rm dust,peak}$, but is already at the same level of the most luminous dust echo detected in optically selected TDEs \citep{Jiang2021}. We have constrained the peak UV/optical luminosity of SRGE\,J153503.4+455056 to be $L_{\rm bb, peak}< 4\times 10^{44}\,{\rm erg\,s^{-1}}$ (Table~\ref{tab:optlim}). Taken together, we infer a dust covering factor of $f_c > 0.1$, which is greater than typical values seen in optically selected TDEs. 

Future detailed studies on the IR properties of \srg/\erosita\ selected TDEs should reveal if the dust environment of X-ray selected TDEs is statistically different from that of optically selected events.

\subsection{Optical spectra}
\label{s:optspec}


\begin{figure*}
\begin{subfigure}[t]{0.48\textwidth}
\centering
\includegraphics[width=\linewidth]{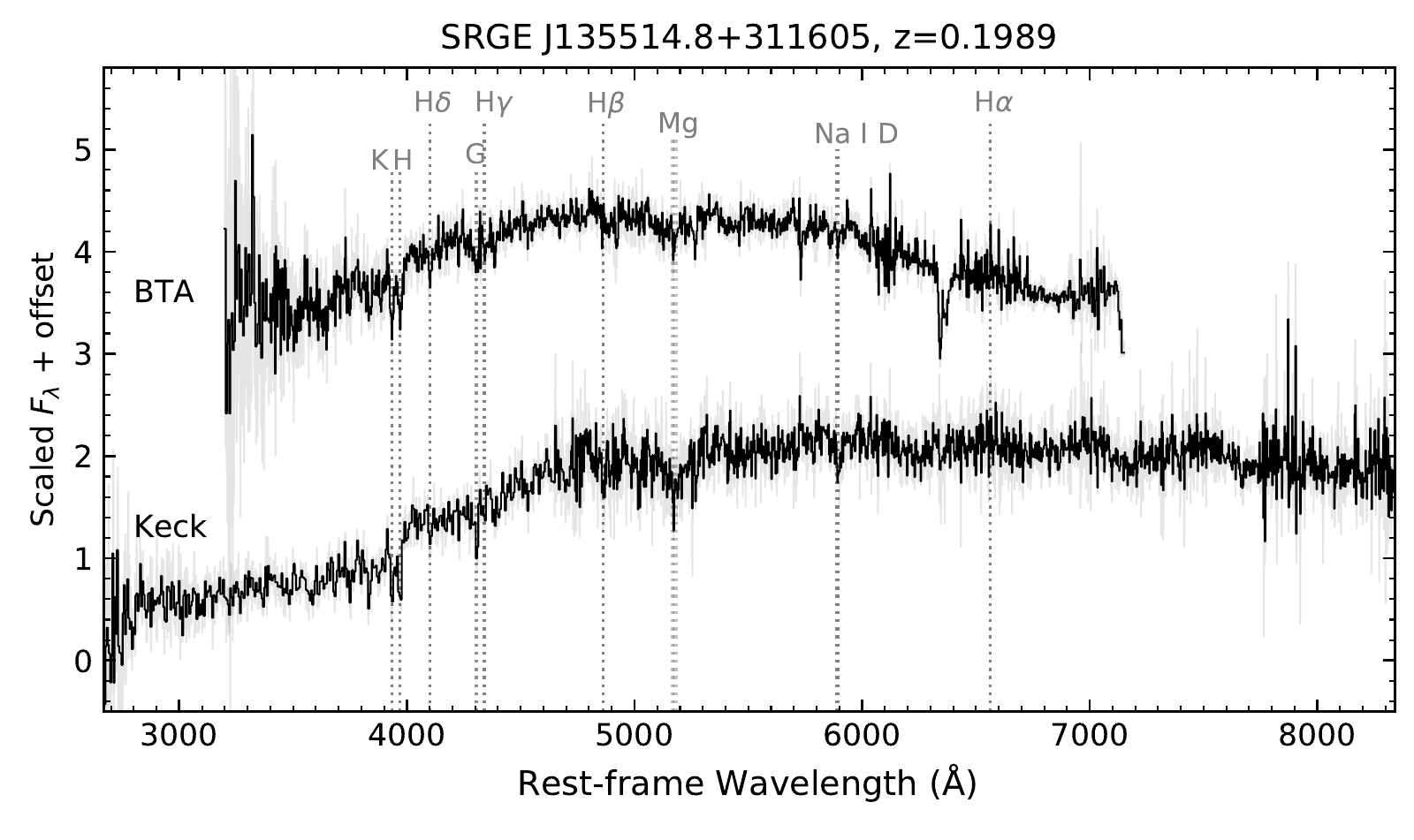}
\end{subfigure}
\begin{subfigure}[t]{0.48\textwidth}
\centering
\includegraphics[width=\linewidth]{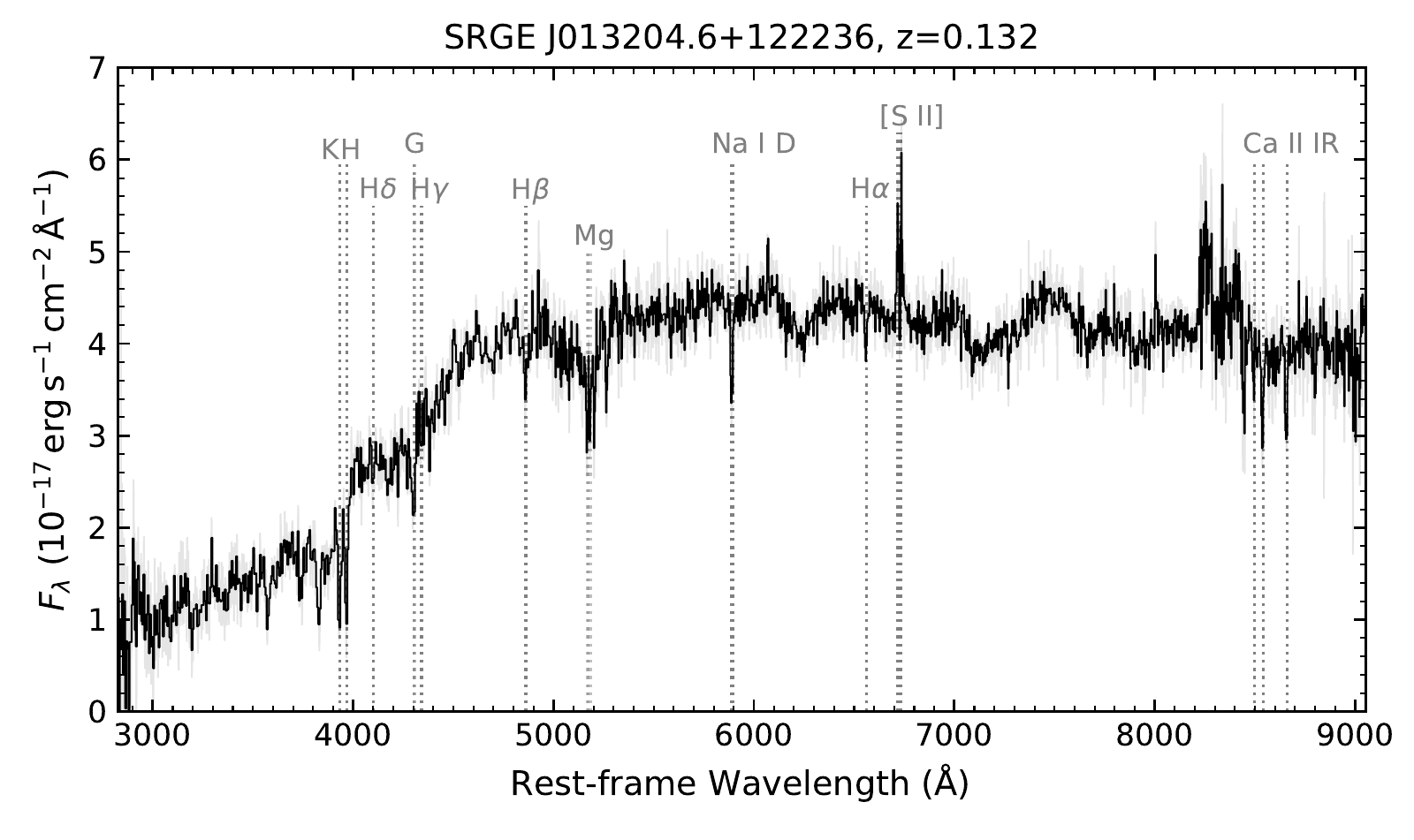}
\end{subfigure}
\begin{subfigure}[t]{0.48\textwidth}
\centering
\includegraphics[width=\linewidth]{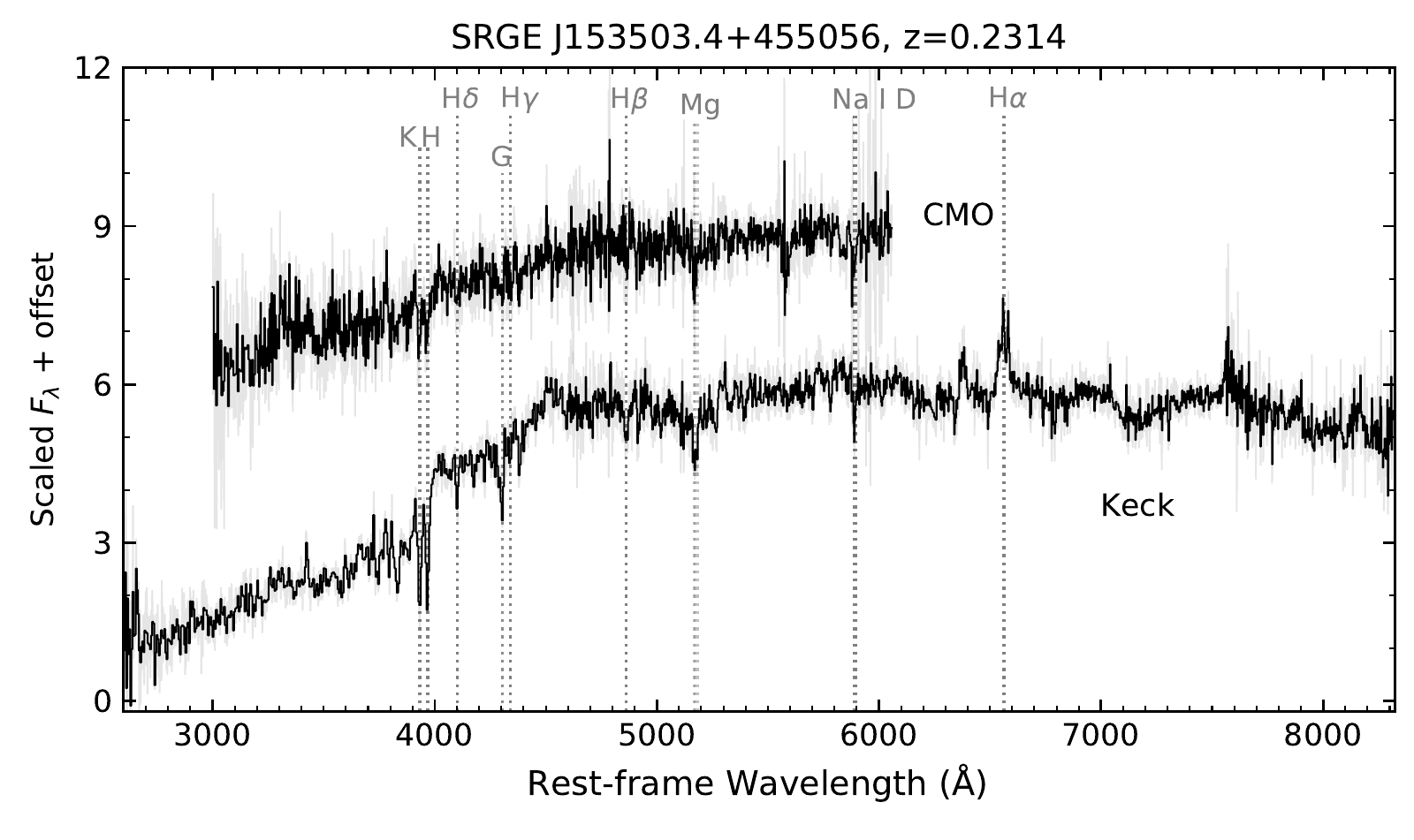}
\end{subfigure}
\begin{subfigure}[t]{0.48\textwidth}
\centering
\includegraphics[width=\linewidth]{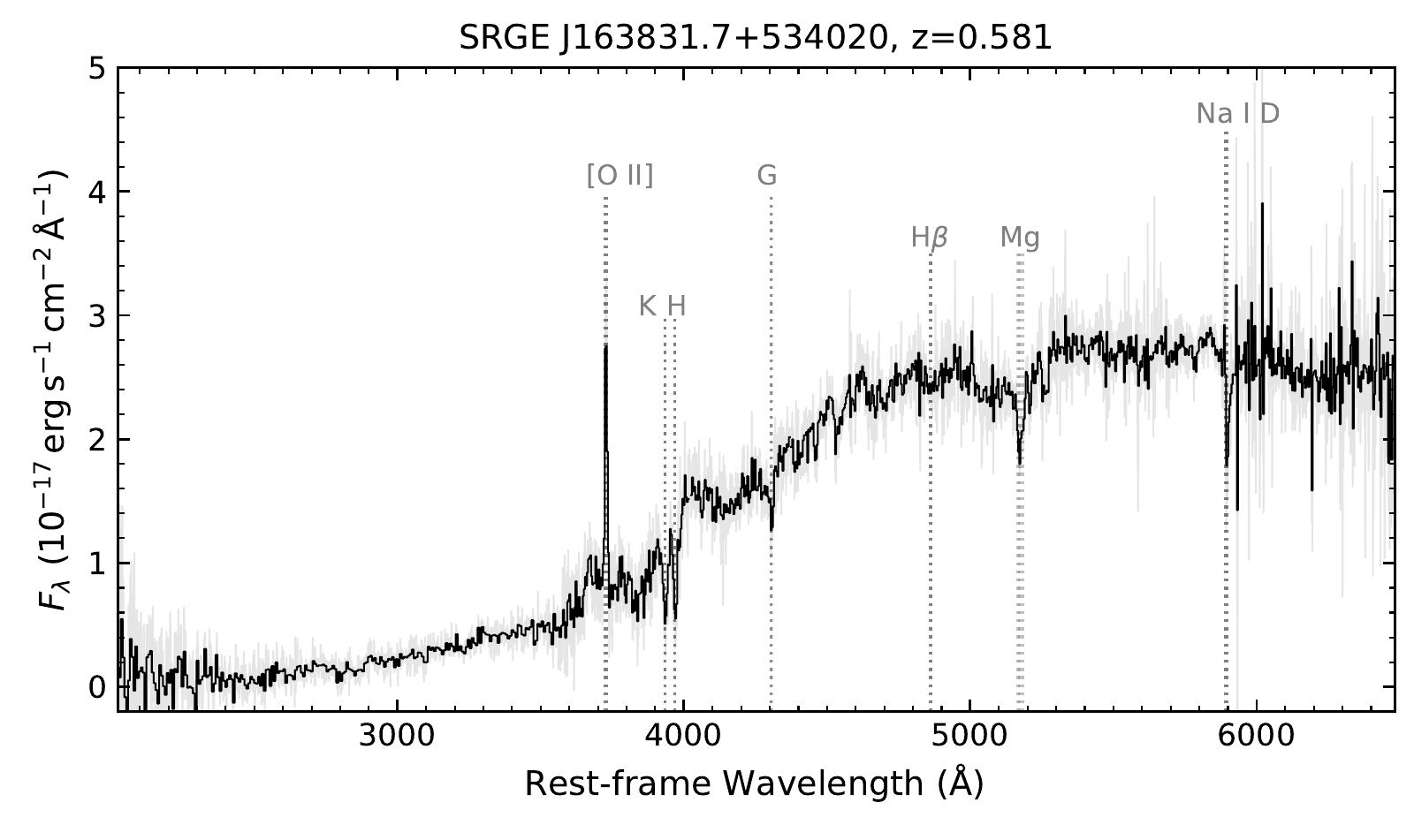}
\end{subfigure}
\begin{subfigure}[t]{0.48\textwidth}
\centering
\includegraphics[width=\linewidth]{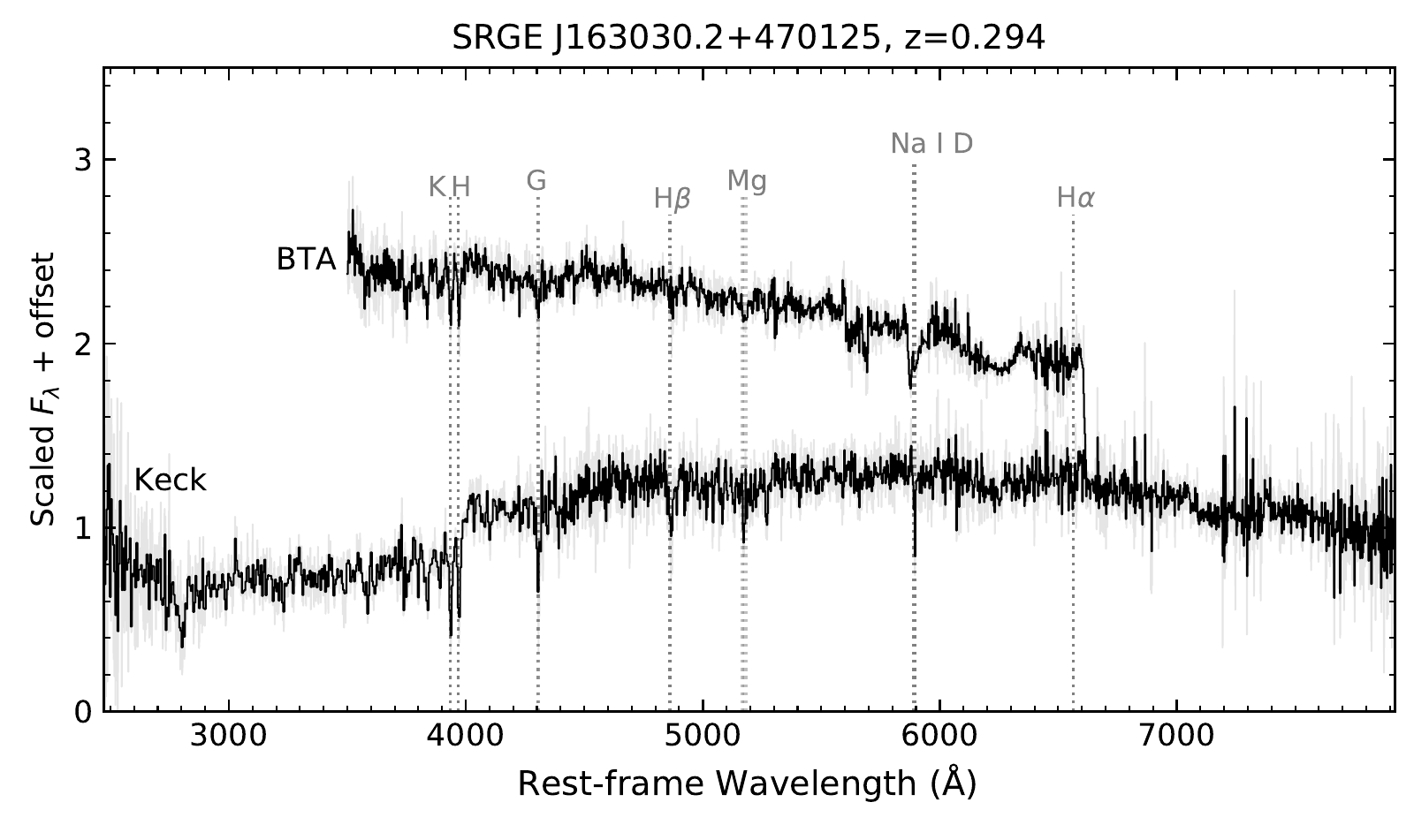}
\end{subfigure}
\begin{subfigure}[t]{0.48\textwidth}
\centering
\includegraphics[width=\linewidth]{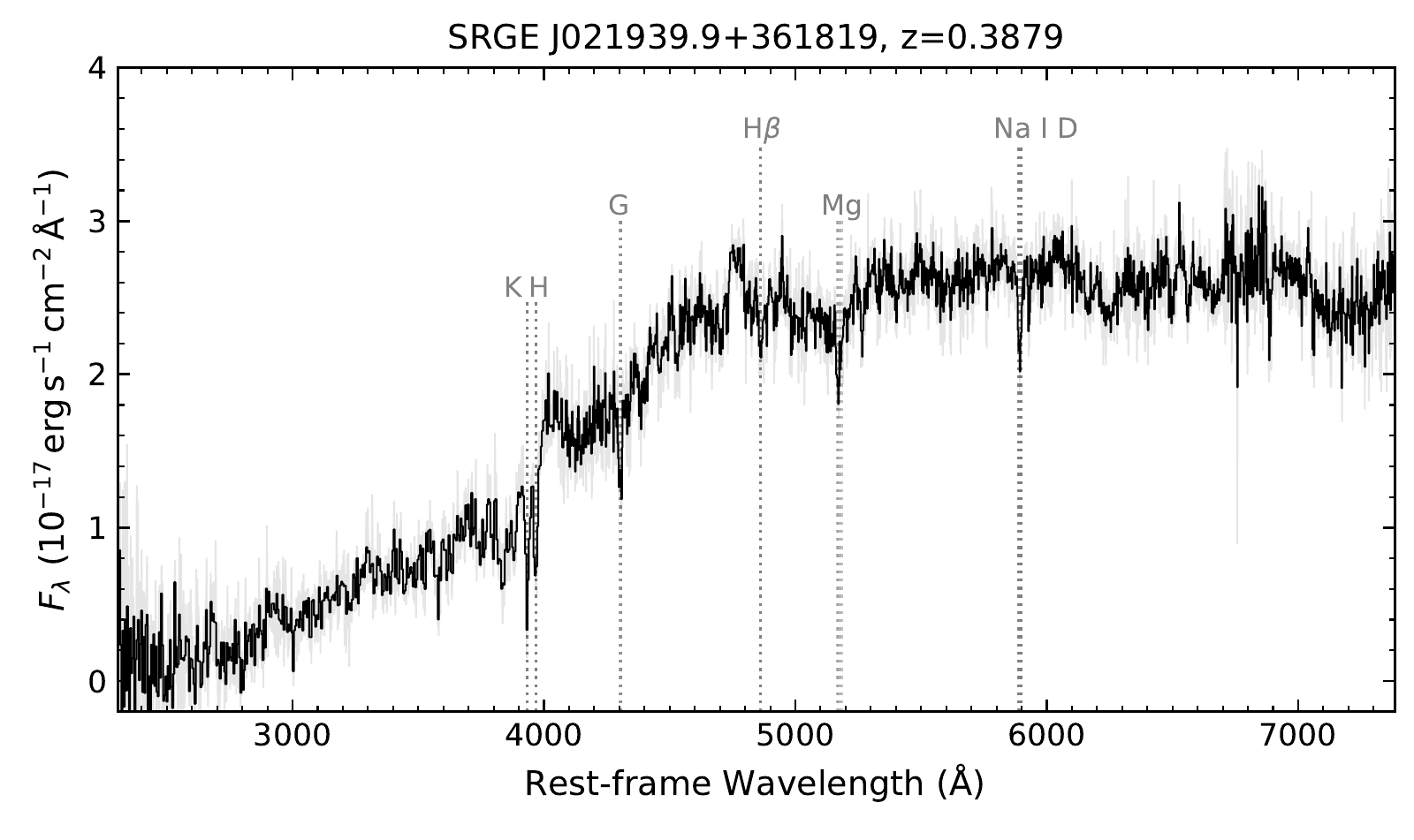}
\end{subfigure}
\begin{subfigure}[t]{0.48\textwidth}
\centering
\includegraphics[width=\linewidth]{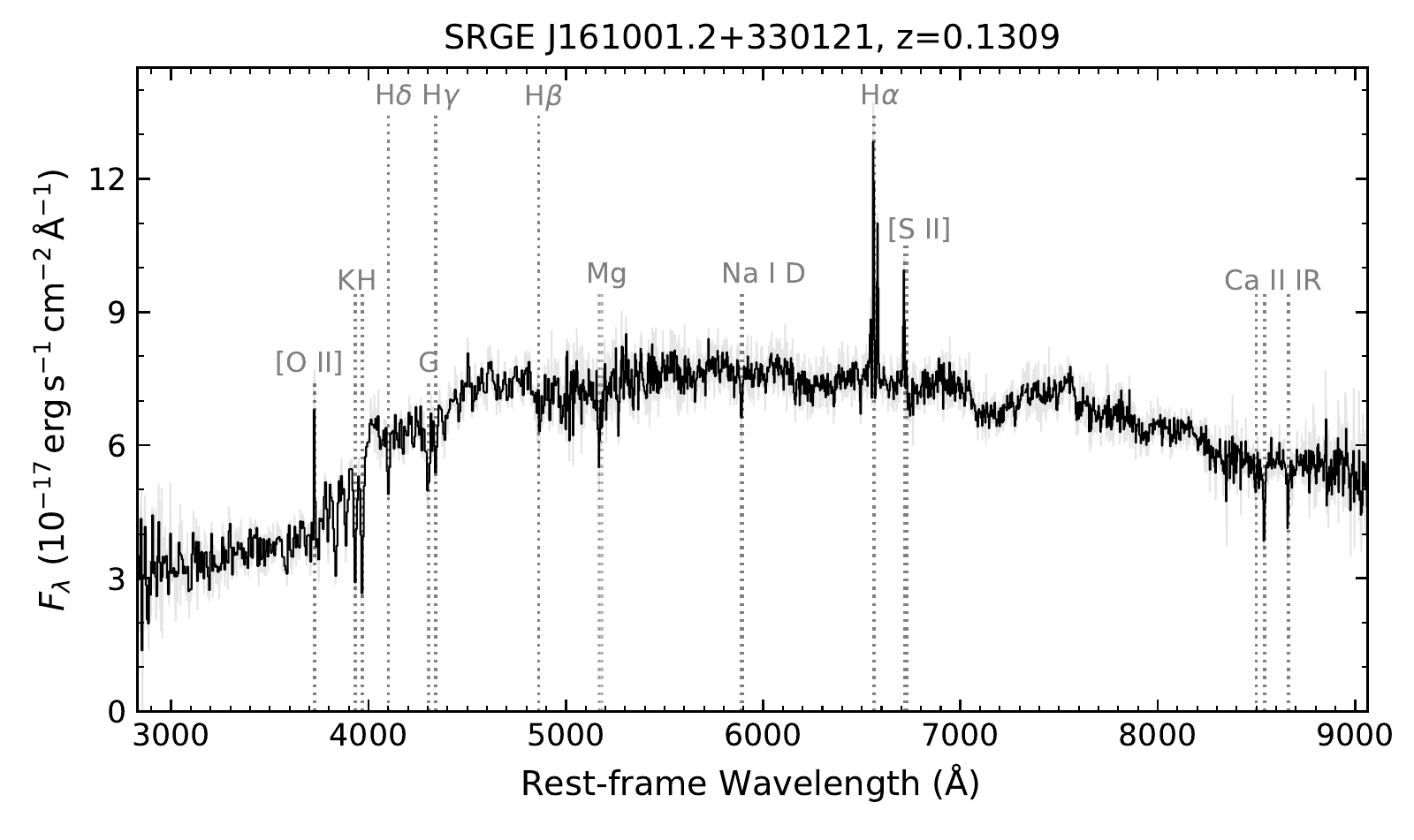}
\end{subfigure}
\begin{subfigure}[t]{0.48\textwidth}
\centering
\includegraphics[width=\linewidth]{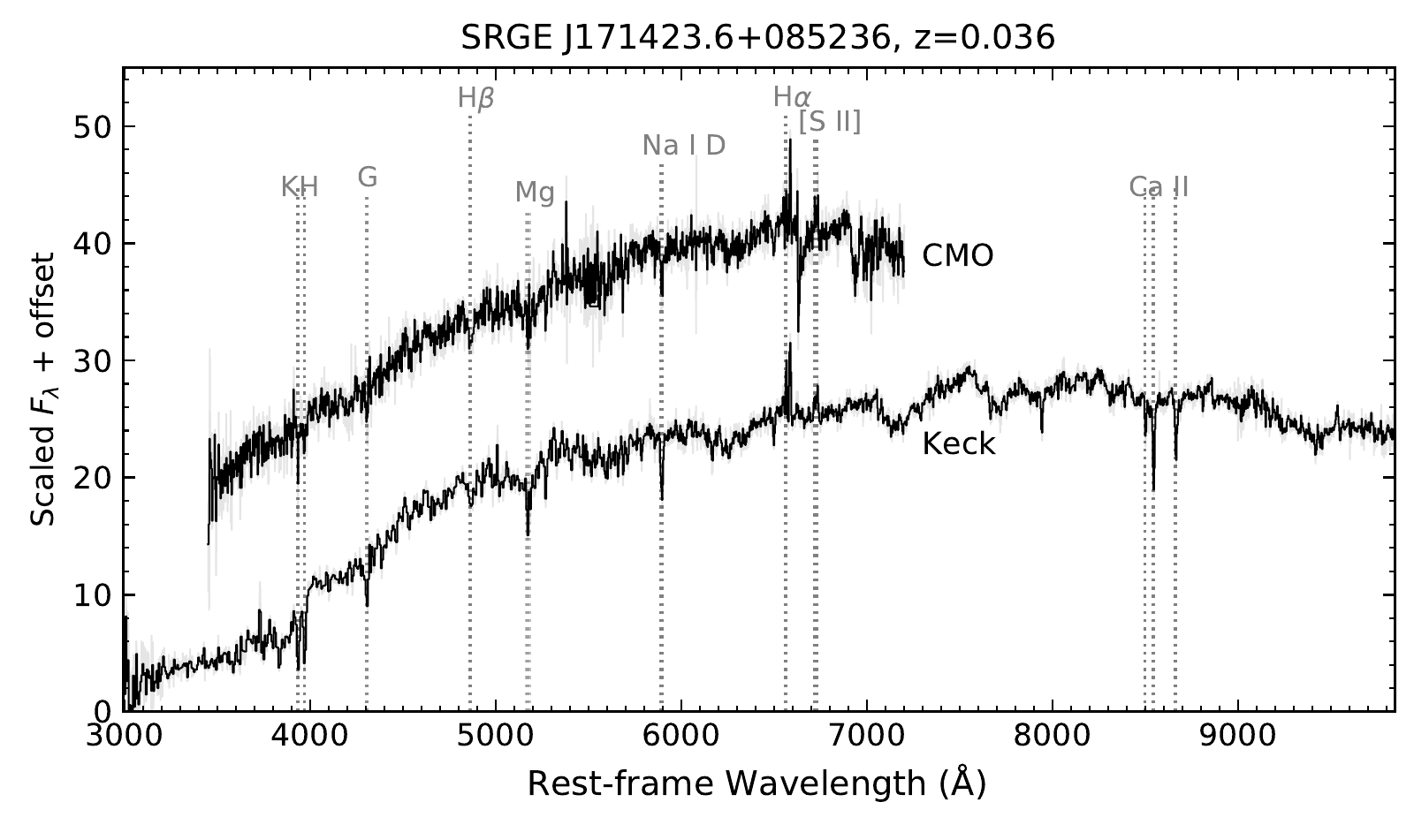}
\end{subfigure}
\caption{Optical spectra of \srg/\erosita\ TDEs and/or their host galaxies.
\label{fig:optspec}}
\end{figure*}

\addtocounter{figure}{-1}
\begin{figure*}
\centering
\begin{subfigure}[t]{0.48\textwidth}
\centering
\includegraphics[width=\linewidth]{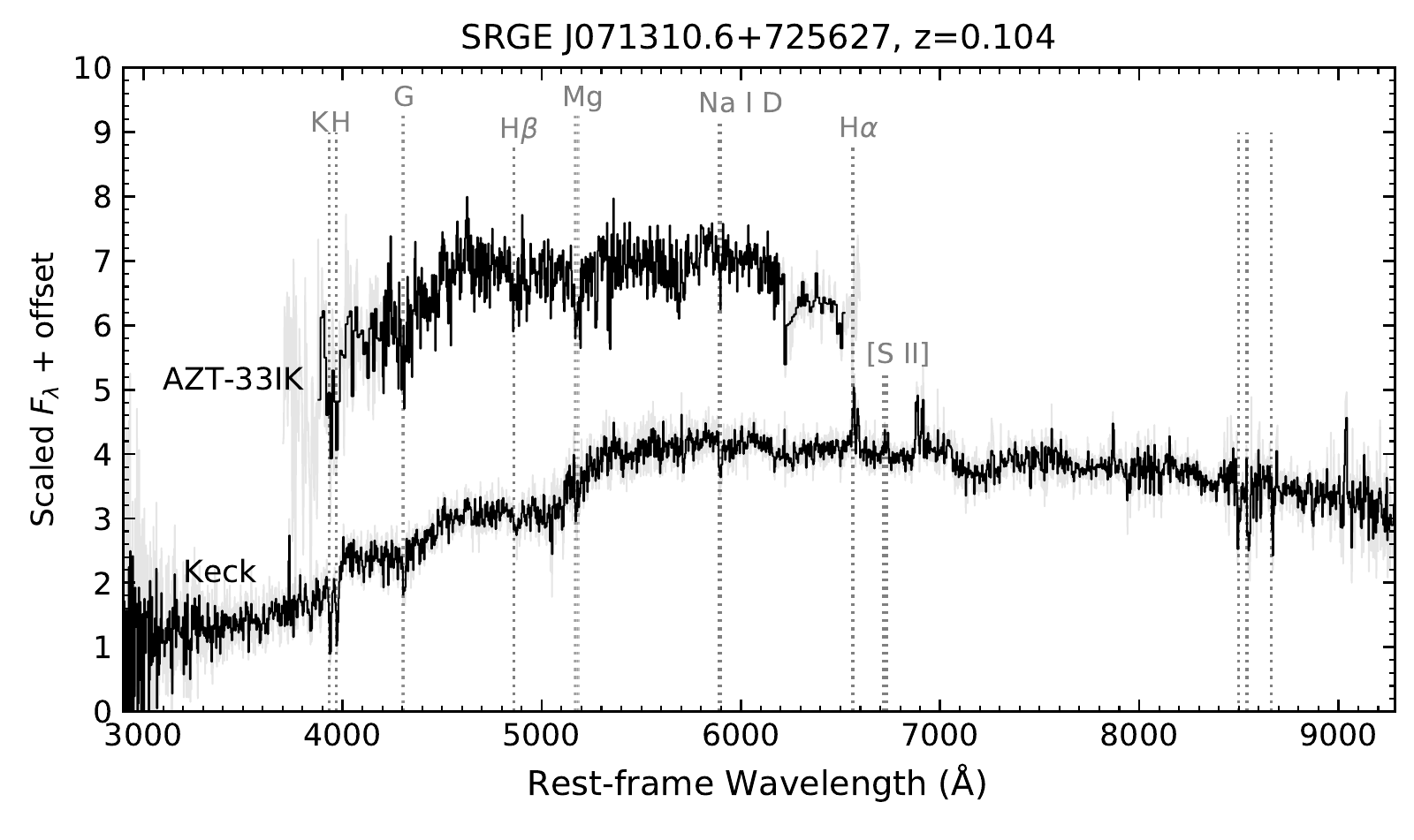}
\end{subfigure}
\begin{subfigure}[t]{0.48\textwidth}
\centering
\includegraphics[width=\linewidth]{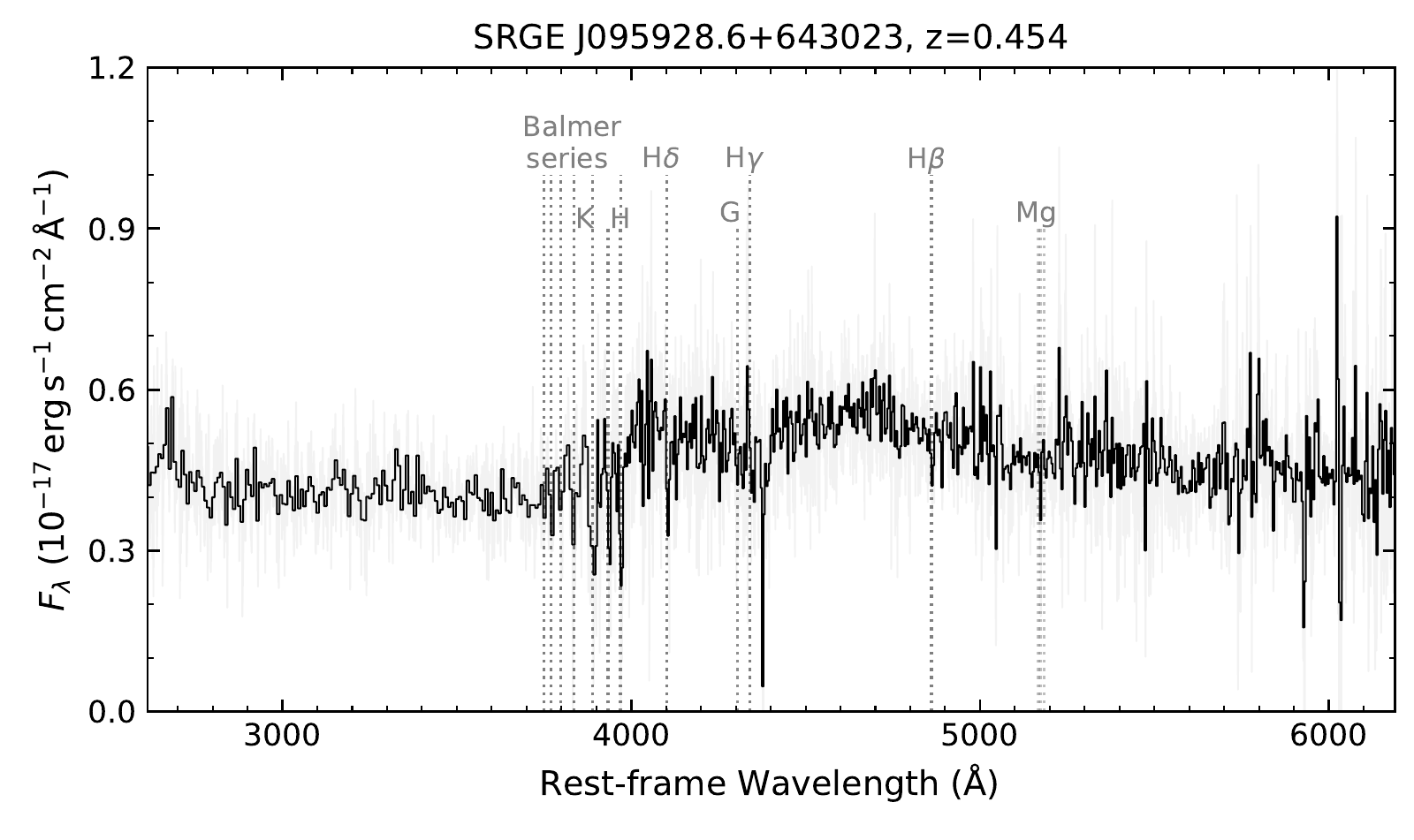}
\end{subfigure}
\begin{subfigure}[t]{0.48\textwidth}
\centering
\includegraphics[width=\linewidth]{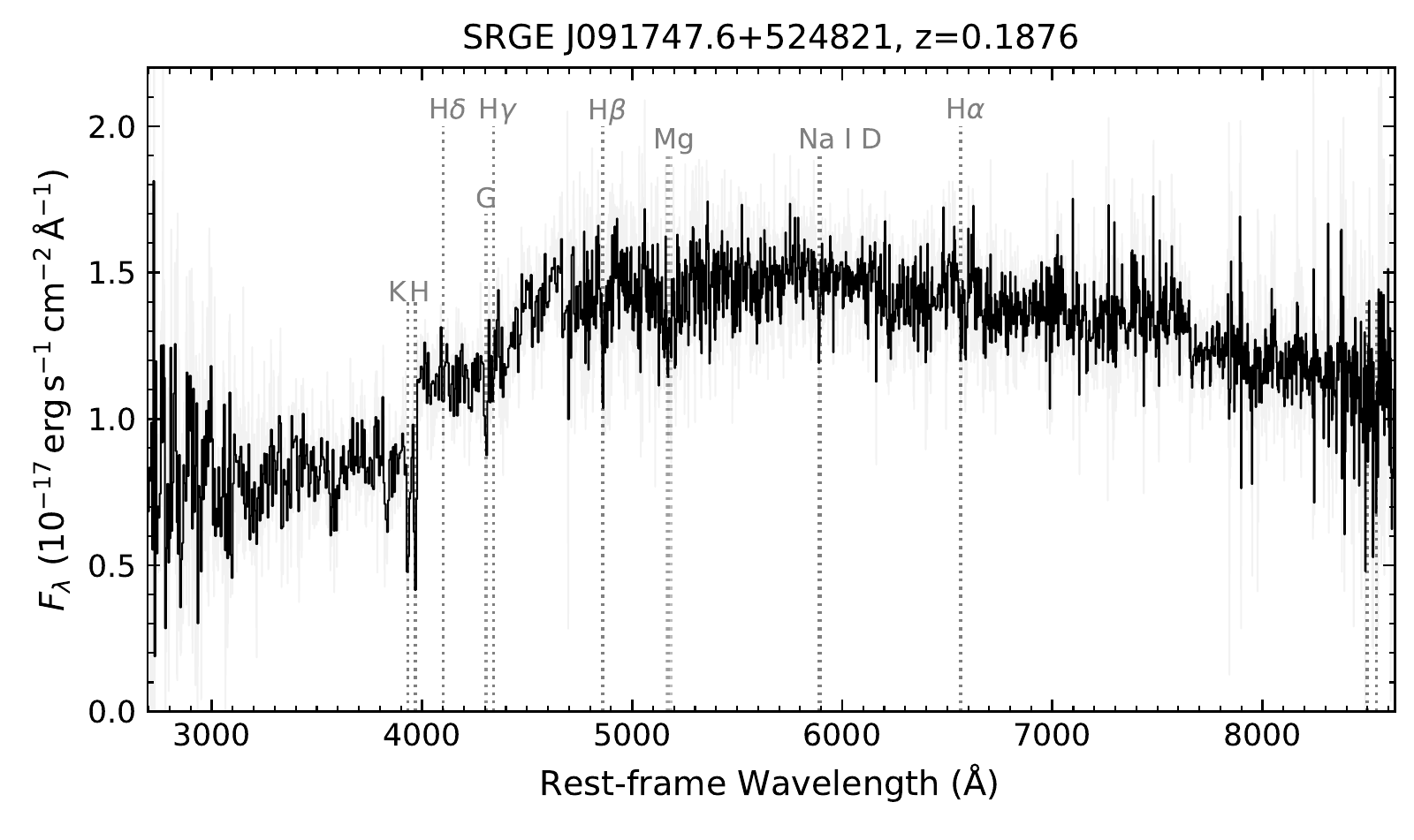}
\end{subfigure}
\begin{subfigure}[t]{0.48\textwidth}
\centering
\includegraphics[width=\linewidth]{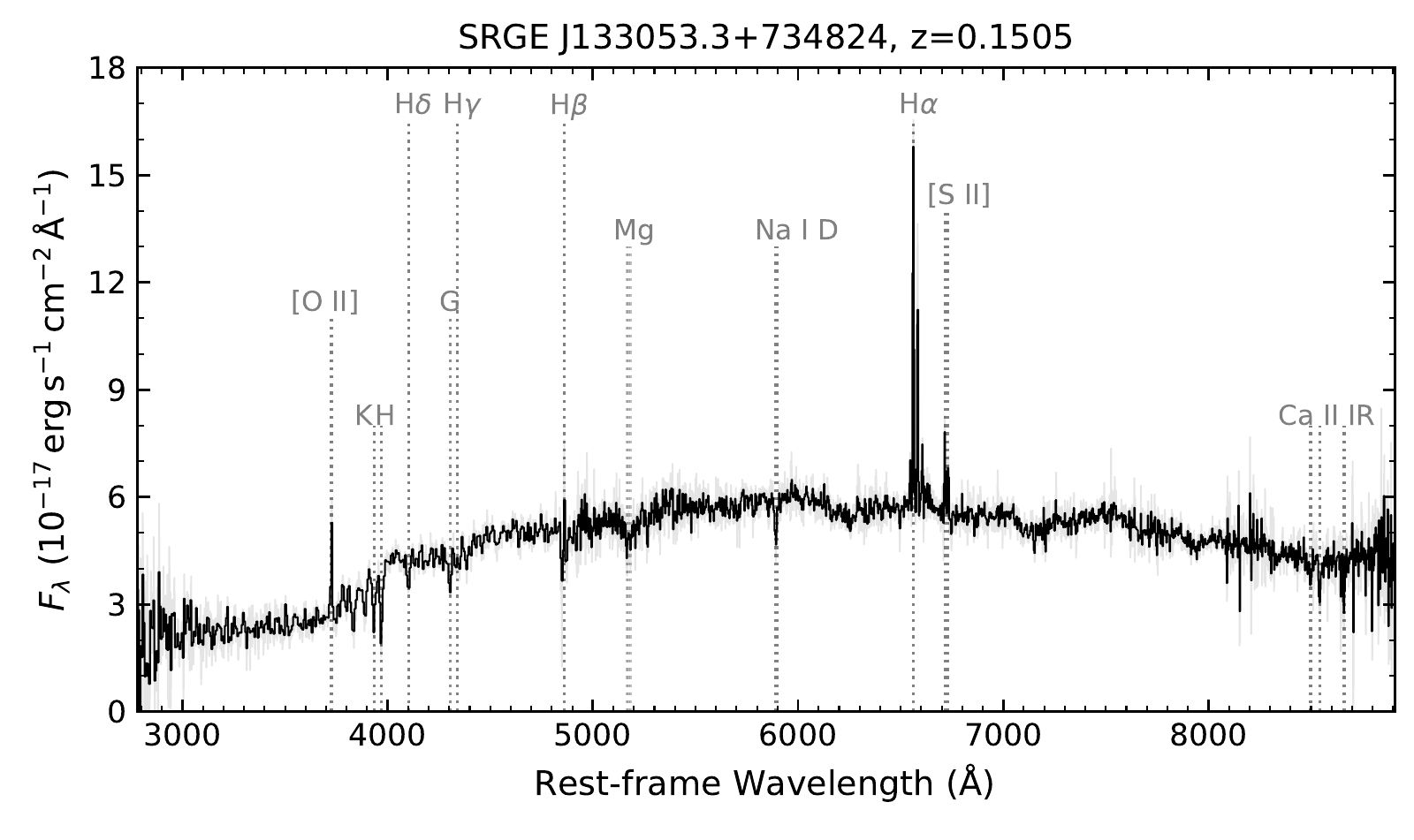}
\end{subfigure}
\begin{subfigure}[t]{0.48\textwidth}
\centering
\includegraphics[width=\linewidth]{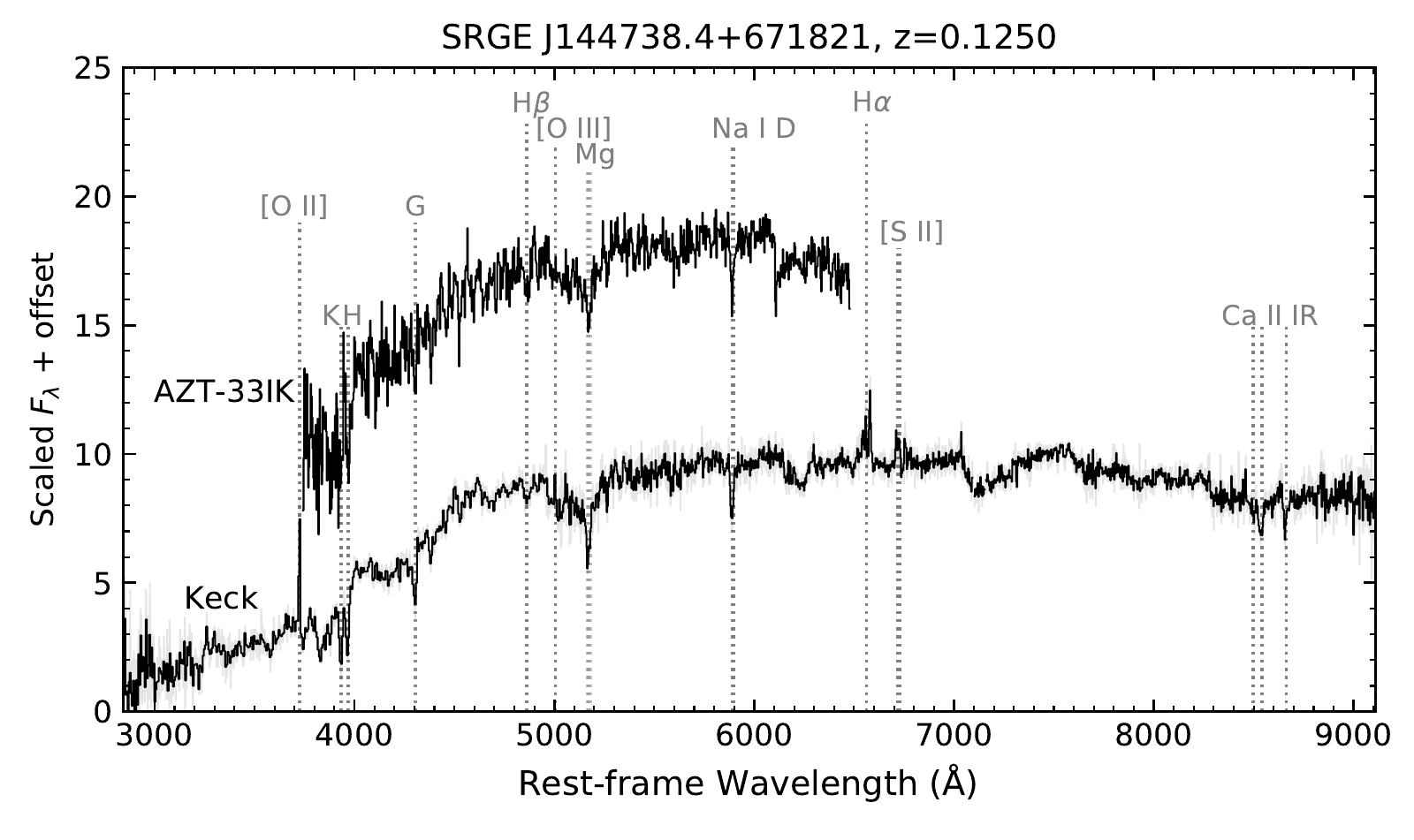}
\end{subfigure}
\caption{Continuation.}
\end{figure*}

Figure~\ref{fig:optspec} shows the optical spectra obtained during our follow-up program. For some objects, only a single-epoch spectrum is available, while for others we obtained a couple of spectra with an interval of several months. The achieved spectral quality is sufficient for a reliable measurement of the redshift in all cases.

Only in the spectrum of SRGE\,J163030.2+470125 do we see a clear indication of optical emission associated with the presumed stellar tidal disruption on top of the host galaxy light. Specifically, the first spectrum taken in October 2020, i.e. nearly 2 months after the \erosita\ discovery, exhibits a blue continuum, while this blue excess is not observed anymore in a second spectrum obtained 8 months later (in June 2021). This spectral evolution is consistent with the brightening seen in the optical light curve between the middle of 2019 and the end of 2020. 

The other 3 objects demonstrating optical flares in their light curves (as was discussed in \S\ref{s:optlc}), namely SRGE\,J153503.4+455056, SRGE\,J095928.6+643023, and SRGE\,J091747.6+524821, do not show any signatures in their optical spectra that could be attributed to a TDE. This is, however, fully consistent with the fact that we conducted spectroscopy for these events {\it after} the flares faded away. 

To characterize the emission content of the host galaxies, we fit the observed Keck spectra with stellar population models using \texttt{ppxf} \citep{Cappellari2017}. We use the MILES spectra library \citep{Vazdekis2015}, and emission lines of Balmer series, [OII], [SII], [OIII], [OI], and [NII]. Among the 13 hosts, emission lines are confidently detected in 8 objects. Table~\ref{tab:host_line} presents their line fluxes and classes on the Baldwin, Phillips, \& Terlevich (BPT) diagnostic diagram \citep{Baldwin1981, Veilleux1987, Kewley_2006}.

In order to explore the star formation history of \srg-selected TDE hosts (discussed in \S\ref{subsec:host_properties} below), we measure the equivalent width (EW) of the Lick H$\delta_{\rm A}$ index in the \texttt{ppxf} best-fit stellar continuum. This index was defined to capture stellar absorption from A stars \citep{Worthey1997}. We then measure the EW of the H$\alpha$ emission using the following formula 
\begin{equation}
    {\rm EW(H}\alpha_{\rm em}) = -[{\rm EW(H}\alpha) - {\rm EW(H}\alpha_{\rm abs})],
\end{equation}
where ${\rm EW(H}\alpha)$ is the total EW of H$\alpha$ in the observed spectrum, and ${\rm EW(H}\alpha_{\rm abs})$ is the EW of the absorption line in the \texttt{ppxf} best-fit stellar continuum. The H$\alpha$ line of SRGE\,J163831.7+534020 and SRGE\,J095928.6+643023 are redshifted out of the LRIS bandpass. For these two objects, we assume that the emission line flux of H$\alpha$ is $\gtrsim 3 \times$ the emission line flux of H$\beta$. For SRGE\,J095928.6+643023, the H$\beta$ emission line is not confidently detected in the observed spectrum, and therefore ${\rm EW(H}\alpha_{\rm em})\gtrsim \rm 0\,\AA$. For SRGE\,J163831.7+534020, $f_{\rm H\alpha}\gtrsim 1.9\times 10^{-16}\,{\rm erg\,s^{-1}\,cm^{-2}}$. By inspecting its observed spectrum (Fig.~\ref{fig:optspec}), we take the H$\alpha$ continuum flux to be $\sim 2.7\times 10^{-17}\,{\rm erg\,s^{-1}\,cm^{-2}\,\AA^{-1}}$. Therefore, ${\rm EW(H}\alpha_{\rm em})\gtrsim 7 \rm\,\AA$. Table~\ref{tab:host_ew} presents the measurements of EW(H$\alpha_{\rm em}$) and EW(H$\delta_{\rm A}$), as well as the corresponding WISE W1$-$W2 colors. 

Most of the emission lines observed in the spectra are consistent with being due to star formation in the host galaxies. However, we see clear signatures of AGN activity (namely, high [NII]$\lambda6583$/H$\alpha$, [SII]$\lambda\lambda6717,6731$/H$\alpha$, and [OIII]$\lambda5007$/H$\beta$ ratios), for a few objects, in particular SRGE\,J171423.6+085236, SRGE\,J133053.3+734824, and SRGE\,J144738.4+671821, which are classified according to the BPT diagram as ``LINER'', ``Composite'' (i.e. [HII]/LINER), and ``LINER'' or ``Seyfert'', respectively. Furthermore, although the [OIII]$\lambda5007$ and H$\beta$ emission lines are very weak or absent in some of the spectra, we can try to use the equivalent width of the H$\alpha$ emission line as a proxy of the [OIII]$\lambda5007$/H$\beta$ ratio on the BPT diagram \citep{cid_2010} together with the [NII]$\lambda6583$/H$\alpha$ ratio. Based on the values of EW(H$\alpha_{\rm em})$ from Table~\ref{tab:host_ew}, we infer that SRGE\,J153503.4+455056, SRGE\,J161001.2+330121, and SRGE\,J071310.6+725627 might also be LINERs. 

Despite this evidence of AGN activity, it is unlikely that it is responsible for the X-ray transient phenomena revealed by \srg/\erosita\ in these objects. First of all, $W1-W2<0.3$ for all of the suspected AGN (see Tables~\ref{tab:host_line} and \ref{tab:host_ew}), which indicates that the mid-infrared emission associated with an active nucleus is overwhelmed by that from the surrounding galaxy, hence the suspected AGN cannot be luminous. 

\begin{figure}
    \centering
    \includegraphics[width=\columnwidth]{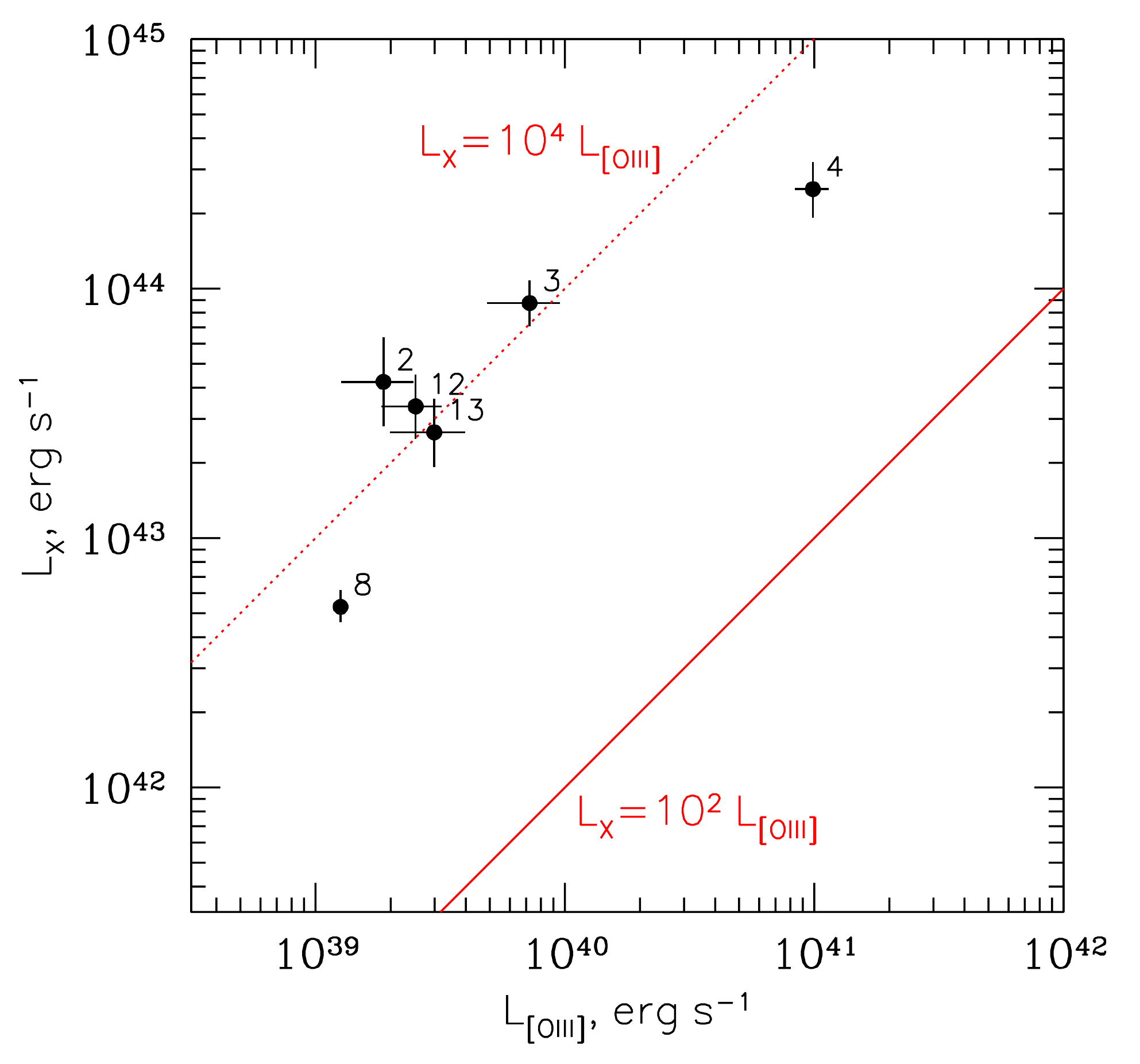}
    \caption{X-ray (0.2--6\,keV) luminosities of the \srg\ TDEs (labeled by their internal numbers in Table~\ref{tab:sample}) vs. the [OIII]$\lambda5007$ luminosity of their host galaxies. Only objects exhibiting a significant [OIII]$\lambda5007$ emission line in the spectrum are shown. The solid line shows a typical relation between $\lx$ and $\loiii$ in AGN \citep{Heckman_2005}, while the dotted line indicates a factor of 100 larger $\lx/\loiii$ ratio.
    \label{fig:loiii_lx}
    }
\end{figure}

A more quantitative assessment of the AGN activity can be done based on the measured luminosity in the [OIII]$\lambda5007$ line (see Table~\ref{tab:host_line}). This quantity correlates with the X-ray luminosity in Seyfert galaxies \citep{Heckman_2005} and may be used as a proxy of AGN bolometric luminosity (e.g. \citealt{LaMassa_2010}). Typically for Seyfert 1 galaxies (i.e. for AGN whose observed spectra are not significantly affected by intrinsic absorption), the [OIII]$\lambda5007$ luminosity is $\sim 1$--3\% of the luminosity in the standard X-ray band (2--10\,keV) \citep{Heckman_2005}, although there is a substantial scatter around this mean trend. Taking into account that for typical SEDs of Seyferts/quasars \citep{Sazonov_2004} the luminosity in the 0.2--6\,keV band (our working energy range in this study), $\lx$, is a factor of $\approx 2$ higher than in the 2--10\,keV band, we may roughly predict the AGN contribution to the 0.2--6\,keV luminosity of our objects as $\lx\sim 10^2\loiii$. As shown in Fig.~\ref{fig:loiii_lx}, the $\lx/\loiii$ ratios for all of the \srg/\erosita\ transients with detectable [OIII]$\lambda5007$ emission are larger than $10^3$ and concentrate around $10^4$.

We conclude that the few galaxies tentatively classified as LINER/Seyfert according to their optical emission line ratios appear to be relatively low luminosity AGN in which \srg/\erosita\ has registered luminous soft X-ray outbursts associated with stellar tidal disruptions. The low persistent X-ray luminosity expected for these objects based on the [OIII]$\lambda5007$ flux is consistent with their non-detection by \erosita\ during its first scan (eRASS1). However, we cannot completely rule out that AGN do play some role in the properties of the X-ray transients discussed here (see \citealt{Zabludoff2021} for a discussion of the overlap of the observed properties of TDEs and AGN), especially in the case of SRGE\,J144738.4+671821, which showed an atypical X-ray brightening over the 6-month period after its discovery by \erosita. The presence of a few AGN in the current sample of 13 TDEs is not surprising, since roughly every second galaxy in the local Universe appears to have a weakly active nucleus \citep{Ho_2008}.  

\section{Discussion}
\label{s:discussion}

As has been demonstrated in the preceding sections, the totality of existing X-ray and optical data leaves little doubt that the objects under consideration are TDEs. 

\subsection{Properties of the SRG TDE sample}
\label{s:tdeprop}

Table~\ref{tab:properties} summarizes the key properties of the discovered TDEs. Specifically, we provide for each object: (1) coordinates of the host galaxy, (2) the Galactic HI column in this direction, (3) its redshift, (4) TDE intrinsic X-ray luminosity in the 0.2--6\,keV energy band, $\lx$, or the corresponding upper limit in eRASS2 and eRASS3, (5) TDE rest-frame $g$-band luminosity, $\lopt$, or the corresponding upper limit, and (6) the volume of the Universe, $\vmax$, within which the TDE could be detected during eRASS2 (see \S\ref{s:rate} below). 

The quoted X-ray luminosities and upper limits were determined from the best-fitting \textsc{diskbb} models (Tables~\ref{tab:bestfit} and \ref{tab:bestfit_e3}) and corrected for the Galactic absorption in those cases where the available number of \erosita\ photons had allowed us to perform a spectral analysis; otherwise the X-ray luminosities/upper limits for eRASS3 were estimated from the measured count rates adopting the best-fitting spectral model from eRASS2 (i.e. during the ``bright phase'' of the TDE). The estimates and upper limits on $\lopt$ are adopted from the last three columns in Table~\ref{tab:optlim}, i.e. allowing for the characteristic blackbody temperature to range between $1.3\times 10^4$ and $4.0\times 10^4$\,K.

\begin{table*}
  \caption{TDE properties.} 
  \label{tab:properties}
  \begin{tabular}{lrrcccccc}
  \hline
  \multicolumn{1}{c}{Object (SRGE)} &
  \multicolumn{2}{c}{Optical position} &
  \multicolumn{1}{c}{$\nhgal$} &
  \multicolumn{1}{c}{$z$} &
  \multicolumn{2}{c}{$\lx^1$, $10^{43}$\,erg\,s$^{-1}$} & 
  \multicolumn{1}{c}{$\lopt^2$} &
  \multicolumn{1}{c}{$\vmax$} \\
  & 
  \multicolumn{1}{c}{RA} & 
  \multicolumn{1}{c}{Dec} & 
  \multicolumn{1}{c}{$10^{20}$\,cm$^{-2}$} 
  & 
  & 
  \multicolumn{1}{c}{eRASS2} & 
  \multicolumn{1}{c}{eRASS3} & 
  \multicolumn{1}{c}{$10^{43}$\,erg\,s$^{-1}$} &
  \multicolumn{1}{c}{Gpc$^3$} \\
  \hline
J135514.8+311605 & 208.812579 & 31.268121 & 1.21 & $ 0.1989 \pm 0.0004 $ & $ 5.8_{-1.4}^{+1.6} $ & $ <0.3 $ & $<0.4$ & 1.90 \\[0.1cm]
J013204.6+122236 & 23.018675 & 12.376562 & 4.80 & $ 0.132 \pm 0.001 $ & $ 4.2_{-1.4}^{+2.2} $ & $ <0.4 $ & $<0.22$ & 0.439 \\[0.1cm]
J153503.4+455056 & 233.763172 & 45.848598 & 1.22 & $ 0.2314 \pm 0.0004 $ & $ 8.8_{-1.7}^{+2.1} $ & $ 3.1_{-0.9}^{+1.6} $ & 0.9--2.2 & 1.87\\[0.1cm]
J163831.7+534020 & 249.633401 & 53.672931 & 2.93 & $ 0.581 \pm 0.001 $ & $ 25_{-6}^{+7} $ & $ <1.8 $ & $<3.5$ & 10.6 \\[0.1cm]
J163030.2+470125 & 247.626052 & 47.023730 & 1.62 & $ 0.294 \pm 0.001 $ & $ 20_{-4}^{+5} $ & $ <1.0 $ & 0.9--2.4 & 4.16 \\[0.1cm]
J021939.9+361819 & 34.916264 & 36.305054 & 5.26 & $ 0.3879 \pm 0.0002 $ & $ 25_{-8}^{+12} $ & $15_{-3}^{+3}$ & $<2.3$ & 8.43\\[0.1cm]
J161001.2+330121 & 242.505920 & 33.022416 & 1.56 & $ 0.1309 \pm 0.0006 $ & $ 1.2_{-0.3}^{+0.4} $ & $ <0.15 $ & $<0.2$ & 0.235 \\[0.1cm]
J171423.6+085236 & 258.598393 & 8.876918 & 5.39 & $ 0.036 \pm 0.001 $ & $ 0.53_{-0.07}^{+0.09} $ & $0.03_{-0.01}^{+0.01}$ & $<0.07$ & 0.0965 \\[0.1cm]
J071310.6+725627 & 108.293835 & 72.940751 & 3.53 & $ 0.104 \pm 0.001 $ & $ 11_{-2}^{+3} $ & $ 1.9_{-0.8}^{+1.6} $ & $<0.15$ & 1.06 \\[0.1cm]
J095928.6+643023 & 149.868660 & 64.506053 & 3.07 & $ 0.454 \pm 0.001 $ & $ 89_{-32}^{+49} $ & $87_{-35}^{+65}$ & 1.2--2.6 & 13.8 \\[0.1cm]
J091747.6+524821 & 139.447492 & 52.805635 & 1.52 & $ 0.1876 \pm 0.0002 $ & $ 48_{-16}^{+23} $ & $ <5 $ & 0.5--1.2 & 0.747 \\[0.1cm]
J133053.3+734824 & 202.720918 & 73.806739 & 1.59 & $ 0.1505 \pm 0.0002 $ & $ 3.4_{-0.9}^{+1.2} $ & $ <0.2 $ & $<0.24$ & 0.477 \\[0.1cm]
J144738.4+671821 & 221.912771 & 67.305094 & 0.88 & $ 0.1250 \pm 0.0005 $ & $ 2.7_{-0.7}^{+1.0} $ & $3.2_{-0.6}^{+0.6}$ & $<0.6$ & 0.179 \\[0.1cm]
  \hline
  \end{tabular}
  
Notes: 
$^1$: Luminosity in the rest-frame 0.2--6\,keV energy range, corrected for Galactic absorption. $^2$: Rest-frame $g$-band luminosity. 
\end{table*}

\begin{figure}
\centering
\includegraphics[width=\columnwidth]{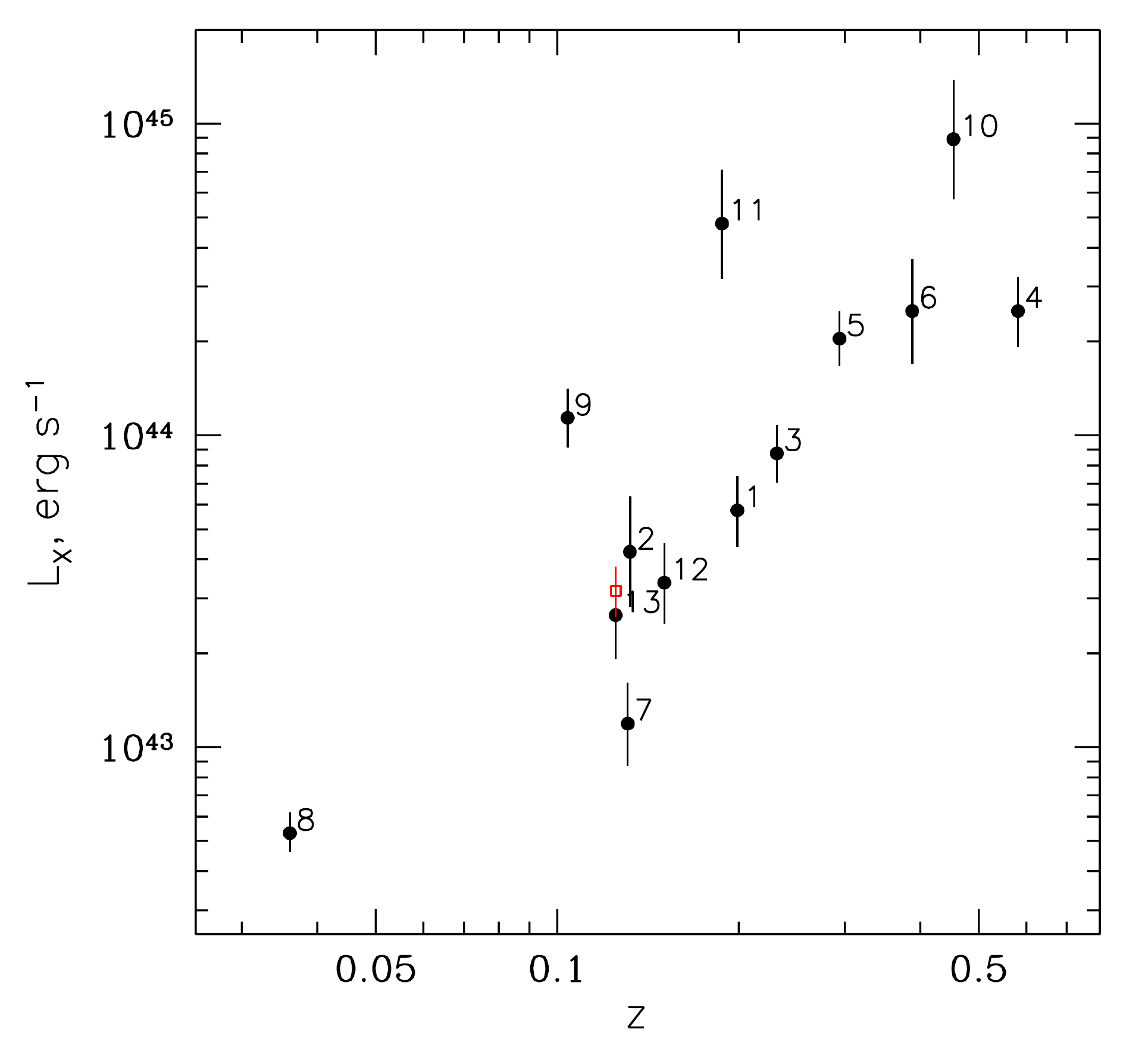}
    \caption{Intrinsic luminosity in the 0.2--6\,keV energy band as a function of redshift for \srg\ TDEs (labeled by their number in Table~\ref{tab:sample}). The luminosities are based on the eRASS2 measurements. For SRGE\,J144738.4+671821, we also show its X-ray luminosity during eRASS3 (red square), when it became brighter than in eRASS2.
    \label{fig:z_lx}
    }
\end{figure}

Figure~\ref{fig:z_lx} shows the distribution of the TDEs over redshift and X-ray luminosity. Thanks to the high sensitivity of the \srg/\erosita\ all-sky survey, the effective ``horizon'' of TDE observability in X-rays has moved out to $z\sim 0.6$ from $z\sim 0.15$, where it was during the \rosat\ all-sky survey \citep{Komossa_2015}. Moreover, our current sample is based on a conservative, high detection threshold (see \S\ref{s:selection}). The latter can be lowered in future work by a factor of $\sim 2$, which should lead to the discovery of TDEs at even higher redshifts. Therefore, thanks to \srg\ we are starting to explore the TDE phenomenon beyond the low-redshift Universe.

\begin{figure}
\centering
\includegraphics[width=\columnwidth]{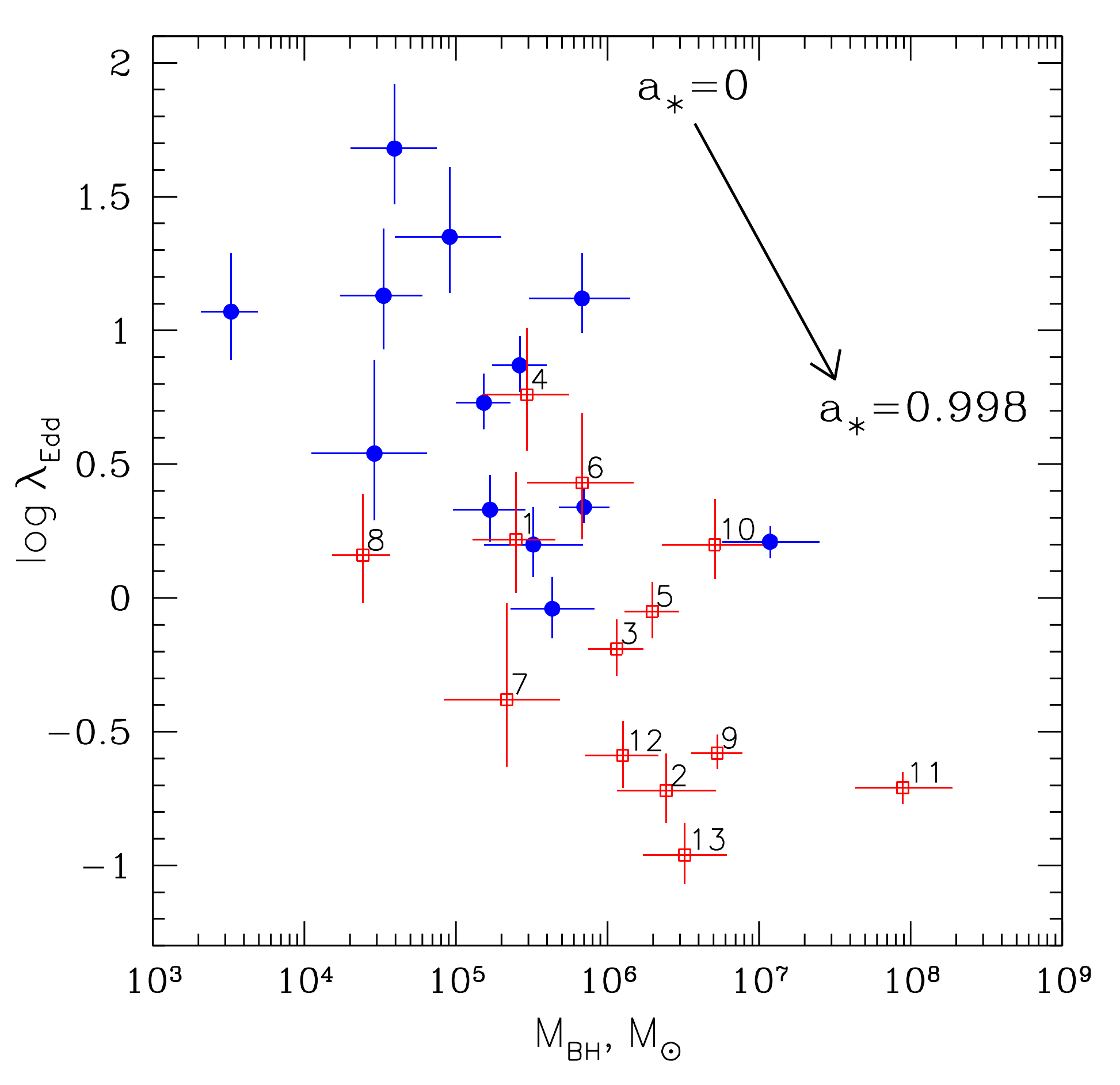}
    \caption{Black hole masses vs. Eddington ratios for the \srg\ TDEs (labeled by the internal numbers in Table~\ref{tab:sample}), estimated from the X-ray spectra assuming a standard accretion disk around a SMBH with $\spin=0$ (blue points) and $\spin=0.998$ (red points). The arrow illustrates the range of ($\mbh$, $\ledd$) values allowed for a given TDE depending on $\spin$.}
    \label{fig:mbh_ledd}
\end{figure}

In \S\ref{s:xrayspec} we made an attempt to estimate the masses ($\mbh$) and Eddington ratios ($\ledd$) of the black holes associated with the TDEs (see Table~\ref{tab:optxagnf}) from their X-ray spectra measured by \erosita, based on the assumption of a standard accretion disk. As noted before, this assumption may fail for at least some of these events if \erosita\ caught them in a super-Eddington accretion phase, when a slim rather than thin accretion disk would be expected. Furthermore, such estimates strongly depend on the adopted spin of the black hole. Bearing in mind these uncertainties, we plot in Fig.~\ref{fig:mbh_ledd} the inferred values of $\mbh$ and $\ledd$ for two extreme cases: $\spin=0$ (a Schwarzschild black hole) and $\spin=0.998$ (a maximally rotating Kerr black hole). In the case of a slowly rotating black hole, most of the TDEs would have been in a super-Eddington accretion phase at the epoch of their first detection by \erosita, while in the $\spin=0.998$ case, the deduced accretion rates are nearly critical. Regardless of the actual black hole spins, all the inferred black hole masses are consistent with being below the theoretical upper limits for nonspinning and rapidly spinning black holes in TDEs of $\sim 10^8\,\msun$ and $\sim 7\times 10^8\,\msun$, respectively \citep{rees1988,Kesden_2012}. 

\subsection{Host galaxy properties} \label{subsec:host_properties}

Following the procedures described by \citet{Mendel2014} and \citet{Velzen_2021}, we can estimate the stellar mass of the host galaxies of the \srg\ TDEs using pre-transient photometry from GALEX \citep{Martin2005, Million2016}, PS2, SDSS, the Two Micron All-Sky Survey (2MASS; \citealt{Skrutskie2006}), and the Near-Earth Object WISE Reactivation Mission (NEOWISE; \citealt{Mainzer2014}). 

Briefly speaking, we employ a simple flexible stellar population systhesis (FSPS; \citealt{Conroy2009}) model to fit the UV--MIR broad-band SEDs. The five free parameters are the total stellar mass ($M_{\rm gal}$), the e-folding time of the star formation history ($\tau_{\rm sfh}$), the age of the stellar population, the metalicity ($Z$), and the \citet{Calzetti2000} dust model optical depth. We measure the Galactic extinction-corrected, rest-frame $u-r$ color ($^{0.0}u-r$) from the synthetic photometry. The broad-band SEDs and fitted parameters are given in Appendix~\ref{sec:hostfit}. 

\begin{figure}
    \centering
    \includegraphics[width=\columnwidth]{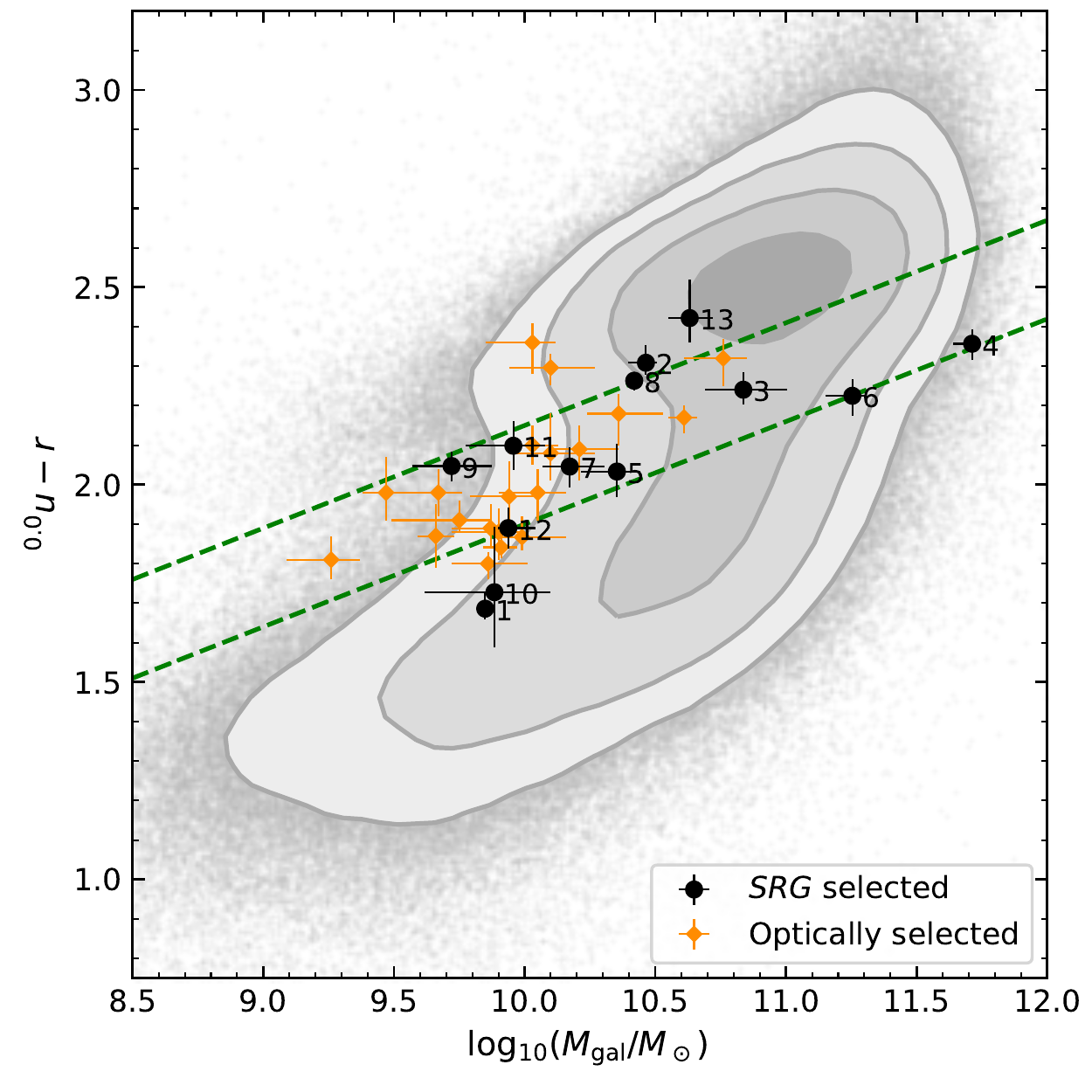}
    \caption{The Galactic extinction-corrected, synthetic rest-frame $u-r$ color of TDE host galaxies. The 13 black data points are from this work (labeled by the internal numbers in Table~\ref{tab:sample}). The 20 orange data points include 17 TDEs selected form the first 1.5\,yr of ZTF \citep{Velzen_2021} and 3 events reported by \citet{Hammerstein2021} afterwards. The green valley is denoted by the dashed green lines, following the boundaries used by \citet{Velzen_2021}. The contours enclose a comparison sample of galaxies from SDSS DR7 \citep{Mendel2014}. \label{fig:color_mass}}
\end{figure}

Figure~\ref{fig:color_mass} shows $^{0.0}u-r$ vs. $M_{\rm gal}$ for the host galaxies of the TDEs discovered by \srg/\erosita\ and discussed in this work, together with a comparison sample of $\approx650,000$ galaxies from SDSS \citep{Mendel2014}. For comparison, we also show a sample of optically selected TDEs \citep{Velzen_2021, Hammerstein2021}. The majority of \srg\ TDE hosts reside in the green valley. This result is expected since previous studies on optically selected TDEs also show that green galaxies are over-represented in TDE hosts \citep{LawSmith2017, Hammerstein2021}. We note, however, that \srg\ TDEs are systematically hosted by galaxies with greater $M_{\rm gal}$. 

\begin{figure}
\centering
\includegraphics[width=\columnwidth]{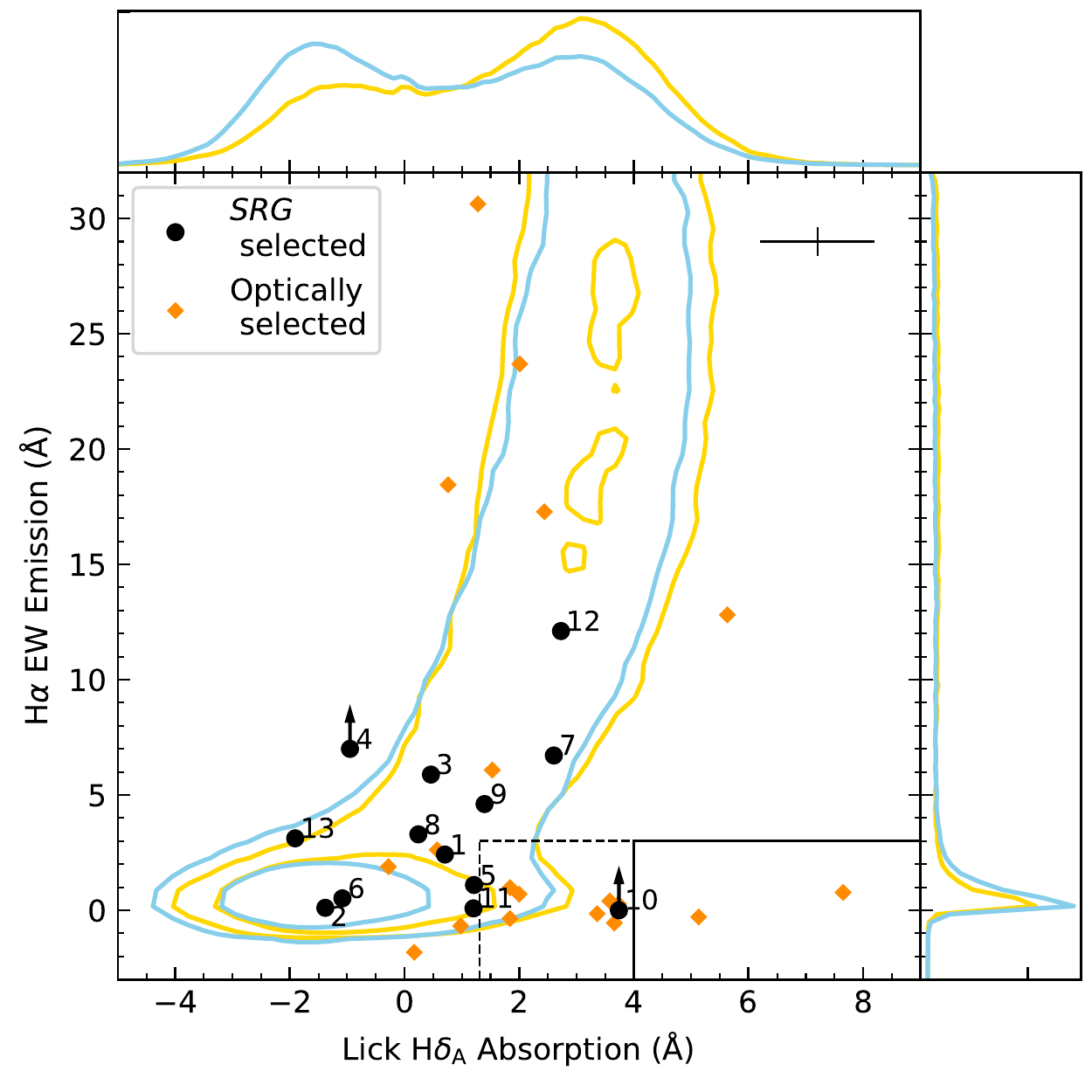}
    \caption{The H$\delta_{\rm A}$ absorption index versus the H$\alpha$ EW emission of TDE host galaxies. The 13 black data points are from this work (labeled by their internal numbers in Table~\ref{tab:sample}). The 19 orange data points are from \citet{Hammerstein2021}. Median uncertainties are shown in the top right. Following the definition in \citet{Hammerstein2021}, the solid black line marks the region of E+A galaxies ($\rm H\delta_{\rm A}-\sigma(H\delta_{\rm A})>4.0$, $\rm EW(H\alpha_{\rm em})<3.0$), and the dashed black line marks the region of QBS galaxies ($\rm H\delta_{\rm A}>1.31$, $\rm EW(H\alpha_{\rm em})<3.0$). We show two comparison samples  with the distribution of $M_{\rm gal}$ similar to \srg\ selected TDE hosts (blue lines) and optically selected TDE hosts (yellow lines). For each comparison sample, the contours enclose 30\% and 85\% of the galaxies. \label{fig:hdelta_halpha}}
\end{figure}

Moreover, it has been found that optically selected TDEs are preferentially hosted by rare quiescent post-starburst (QBS) galaxies or E+A galaxies characterized by strong H$\delta$ absorption and weak H$\alpha$ emission \citep{French2016, LawSmith2017, Hammerstein2021}. In order to see if QBS galaxies are also over-represented in X-ray TDEs, we show the distribution of \srg\ selected TDEs and optically selected TDEs on the H$\delta_{\rm A}$ absorption index versus EW(H$\alpha_{\rm em}$) diagram in Figure~\ref{fig:hdelta_halpha}. In addition, we collect a sample of $\approx 516,000$ SDSS galaxies from the MPA+JHU catalogs \citep{Brinchmann2004} by requiring the H$\alpha$ EW error $0<$\texttt{H\_ALPHA\_EQW\_ERR}$<4$, the H$\alpha$ continuum $>0$, the redshift $z>0.01$, the median signal-to-noise ratio per pixel of the whole spectrum $>10$, and entries in the \citet{Mendel2014} catalog. We construct two comparison samples with the distribution of $M_{\rm gal}$ similar to the \srg\ selected TDE hosts and the ZTF TDE hosts (see the blue and yellow contours in Fig.~\ref{fig:hdelta_halpha}).

Fig.~\ref{fig:hdelta_halpha} shows that 0--8\% of \srg\ TDEs are hosted by QBS galaxies. In contrast, 42\% of ZTF TDEs are hosted by QBS galaxies. Compared with ZTF TDE hosts, \srg\ TDE hosts typically exhibit lower values of H$\delta_{\rm A}$, and there appears to be a dearth of host galaxies with strong H$\alpha$ emission. These differences might be partially accounted for by the greater $M_{\rm gal}$ of the \srg\ TDEs, as evidenced by the distributions shown in Fig.~\ref{fig:hdelta_halpha}. 

\subsection{TDE occurrence rate}
\label{s:rate}

Although the first \srg\ sample of TDEs presented in this work is fairly small, comprising 13 objects, it is already well suited for drawing some inferences about the statistical properties of TDEs in the $z\lesssim 0.6$ Universe. 

The detectability of a TDE in the \srg/\erosita\ all-sky survey largely depends on the following physical parameters: $z$ -- TDE redshift, $\lxmax$ -- X-ray luminosity (in the rest-frame 0.2--6\,keV energy band) at the peak of TDE brightness, $\tdet-t_0$ -- time passed between TDE onset and its detection by \erosita, $\tin$ -- characteristic temperature of the TDE X-ray spectrum assuming multi-blackbody accretion disk emission, $\becl$ -- ecliptic latitude (since the \srg\ all-survey sensitivity increases toward the ecliptic poles), and $\nhgal$ -- Galactic absorption in the TDE direction. 

Since \srg\ passes each location in the sky every 6 months, only a fragmentary X-ray light curve can be obtained for a given TDE based on \erosita\ data. Nevertheless, as we have seen in \S\ref{s:xraylc}, the X-ray light curves obtained in this work suggest that most of the TDEs have been caught by \erosita\ within 2 months after their onset. In addition, our X-ray spectral modeling indicates that during the discovery of these TDEs their Eddington ratios were all near or greater than 100\% (see Table~\ref{tab:optxagnf}). These facts together  suggest that the X-ray luminosities measured by \erosita\ during eRASS2 should be close to the peak X-ray luminosities of these TDEs (i.e. $\lx\approx\lxmax$). We note in passing that it is hardly possible to predict the $\tdet-t_0$ delay based on existing theoretical models, since they have too many parameters (see \citealt{Rossi_2021} for a recent review). 

\begin{figure}
\centering
\includegraphics[width=\columnwidth]{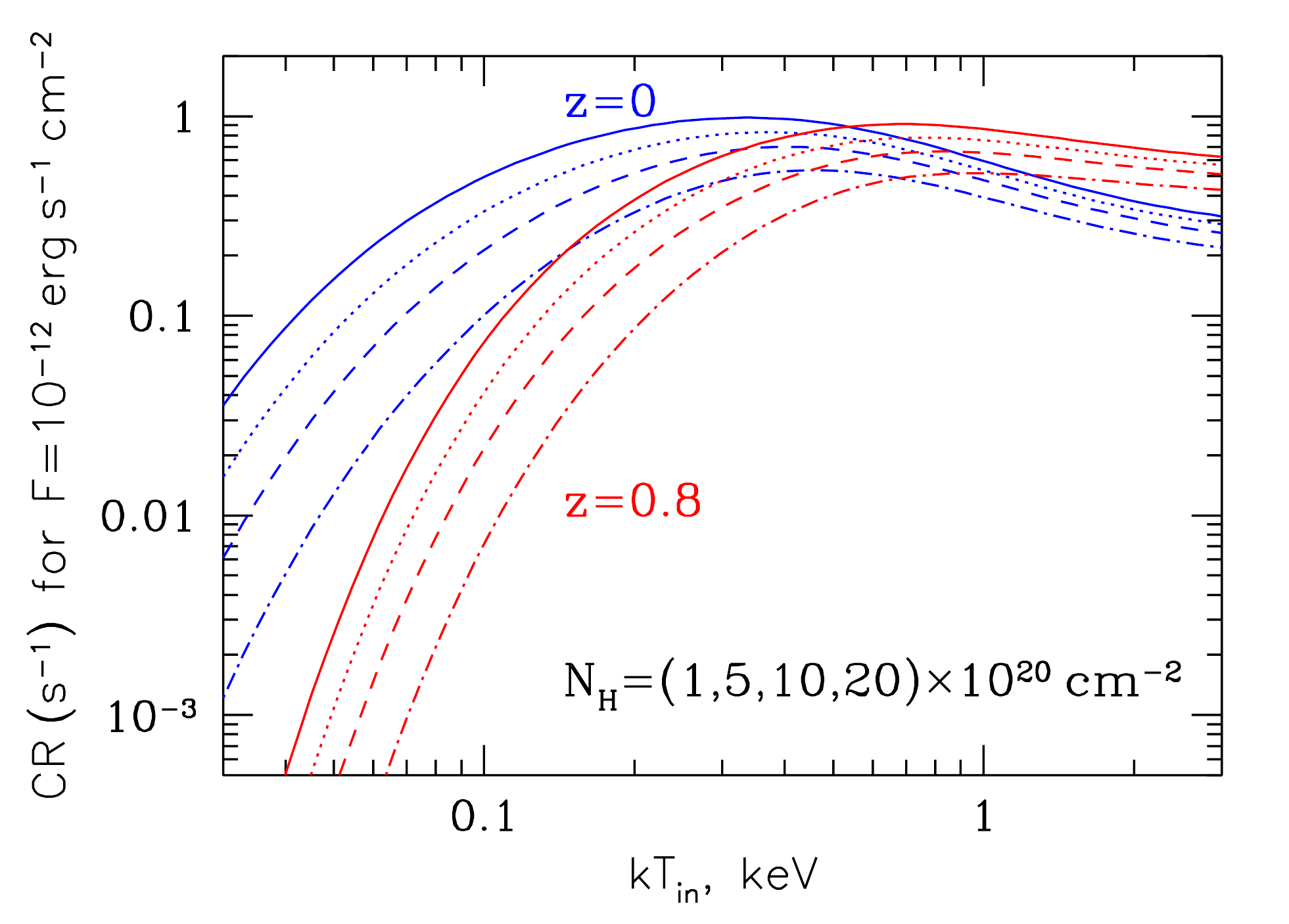}
\includegraphics[width=\columnwidth]{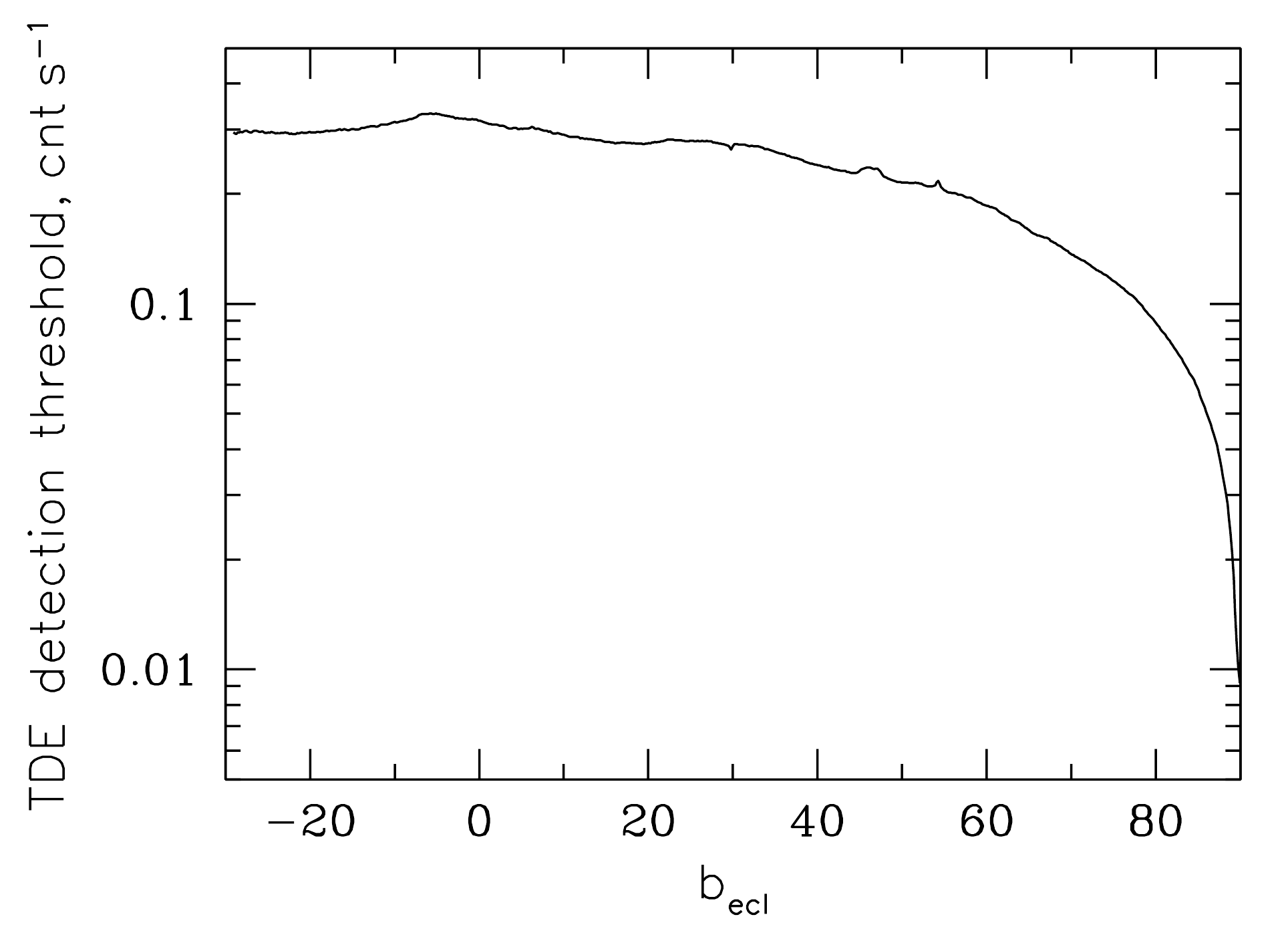}
    \caption{{\it Top:} \erosita\ count rate in the 0.3--2.2\,keV energy range as a function of the temperature of the multicolor accretion disk, $\tin$, for an intrinsic flux of $10^{-12}$\,erg\,s$^{-1}$\,cm$^{-2}$ in the 0.2--6\,keV energy range, for different absorption columns (curves from top to bottom, $\nh=10^{20}$, $5\times 10^{20}$, $10^{21}$, and $2\times 10^{21}$\,cm$^{-2}$. {\it Bottom:} Minimum count rate in the 0.3--2.2\,keV energy band required for TDE detection as a function of ecliptic latitude, averaged over the $0<l<180^\circ$ hemisphere.
    }
    \label{fig:sens}
\end{figure}

The detectability of TDEs in the \srg\ survey is strongly affected by their X-ray spectral hardness/softness. As we have seen in \S\ref{s:xrayspec}, the analyzed \erosita\ spectra can be described fairly well in terms of multi-blackbody accretion disk emission, with their shape characterized by the single parameter $\tin$. Figure~\ref{fig:sens} (upper panel) shows the expected \erosita\ count rate in the 0.3--2.2\,keV energy range (actually used for TDE detection) as a function of $\tin$ for a fixed intrinsic flux of $10^{-12}$\,erg\,s$^{-1}$\,cm$^{-2}$ in the rest-frame 0.2--6\,keV energy range. For nearby TDEs ($z\approx 0$), the sensitivity in the directions with low Galactic absorption does not vary by more than a factor of 2 for $0.1<k\tin<1$\,keV, while it decreases dramatically below 0.1\,keV. This effect is further strengthened if there is a significant column of cold gas ($\nh\gtrsim 5\times 10^{20}$\,cm$^{-2}$) in the direction of the source, and with redshift.

All but one (SRGE\,J091747.6+524821, with $k\tin\approx 0.05$\,keV) TDEs in our sample have $0.1\lesssim k\tin\lesssim 0.5$\,keV, i.e. fall within the temperature range favorable for detection at $z<0.6$ (we recall that the most distant TDE in our sample is located at $z=0.58$). While it is unlikely that we are missing TDEs with very hard spectra ($k\tin\gtrsim 1$\,keV), there is certainly a strong selection effect against TDEs with $k\tin\lesssim 0.1$\,keV. Within the standard TDE paradigm of nearly critical accretion onto a SMBH, such low temperatures correspond to black holes with $\mbh\gtrsim 10^7\,\msun$. Therefore, the current \srg/\erosita\ TDE sample is well suited for estimating the rate of TDEs near SMBHs with $\lesssim 10^7\,\msun$ in the $z\lesssim 0.6$ Universe, while our working energy range (0.3--2.2\,keV) is too hard for systematic exploration of softer TDEs, presumably associated with more massive black holes. 

We next need to know the minimum count rate, $\crmin$, required for detection of TDEs in the 0.3--2.2\,keV band during eRASS2 according to the criteria outlined at the beginning of \S\ref{s:selection}, namely that a candidate transient must be at least 10 times brighter during eRASS2 compared to the upper limit on its flux during eRASS1. This threshold depends primarily on the ecliptic latitude, since the exposure per point accumulated during the \srg\ all-sky survey is inversely proportional to $\cos\becl$. Figure~\ref{fig:sens} (lower panel) shows the exact behavior of the minimum count rate on $\becl$; the details of this computation will be presented elsewhere. Count rates of at least $\crmin\sim 0.3$ and $\sim 0.2$\,counts\,s$^{-1}$ are needed to satisfy our TDE detection criterion at $\becl=0$ and $\becl=60^\circ$, respectively. Note that a typical (vignetting corrected) exposure during one \erosita\ all-sky survey is $\sim 120$\,s at $\becl=0$, and $\sim 240$\,s at $\becl=60^\circ$, so that the quoted thresholds correspond to $\sim 40$ and $\sim 50$ counts for $\becl=0$ and $\becl=60^\circ$, respectively.

\subsubsection{X-ray luminosity function}
\label{s:xlf}

Having discussed the key properties of the \srg/\erosita\ TDE sample and possible selection biases associated with it, we now proceed to evaluation of the TDE X-ray luminosity function (XLF). To this end, we may use the classical $1/\vmax$ method \citep{Schmidt_1968}.

The maximum observable volume $\vmax$ for a given TDE in the studied sample depends on its intrinsic luminosity in the 0.2--6\,keV energy band ($\lx$) and its intrinsic spectral shape, characterized by $\tin$. Given these quantities, the computation of $\vmax$ is straightforward using the tabulated dependencies $CR/F$ ($\tin$, $\nh$, $z$) and $\crmin$ ($\becl$) discussed above. Specifically, we should integrate the comoving volume over the $0<l<180^\circ$ hemisphere (probed by this study) out to a luminosity distance $\dmax$ (or equivalently a redshift $\zmax$) defined for each position in the sky so that $\lx/(4\pi\dmax^2)\times CR/F (\tin,\,\nhgal,\zmax)=\crmin(\becl)$. The Galactic absorption column $\nhgal$ entering this equation varies across the sky, and we adopt it from \citep{hi4pi2016}. 

The computed $\vmax$ values for the studied TDEs are given in the last column of Table~\ref{tab:properties}. We have not tried to estimate the corresponding uncertainties associated with the X-ray spectral analysis (i.e., the uncertainties in $\lx$ and $\tin$) since these are less important compared to the scatter in the $1/\vmax$ values of individual objects used in the calculation of the XLF. To calculate the TDE XLF, we just need to sum the derived $1/\vmax$ values of individual TDEs in specified luminosity bins. To this end, we use 5 equal intervals in $\log\lx$ between 42.5 and 45. The uncertainty within a given bin is found as $\sqrt{\sum(1/\vmax)^2_i}$, where the summation is done over the objects within that bin. 

We next have to take into account that TDEs are transients. As discussed above, \erosita\ appears to typically discover TDEs within 2 months after their onset. Adopting as a fiducial value $\tdet-t_0=2$\,months (ignoring subtleties associated with cosmological time dilation) and taking into account that the studied TDE sample was collected over a period of 6\,months, we should multiply our $1/\vmax$ based estimates by $6/2\times 12/6=6$ to evaluate the volume rate of TDEs per year. 

We finally note that the current sample of 13 TDEs is not statistically complete. Indeed, as already mentioned, there are 3 additional TDEs discovered during eRASS2 based on the same criteria, which are discussed elsewhere (Gilfanov et al., in prep.) because of their pronounced activity in the optical band. To a first approximation, we can take this into account by multiplying the XLF by a factor of 16/13. 

\begin{figure}
\centering
\includegraphics[width=\columnwidth]{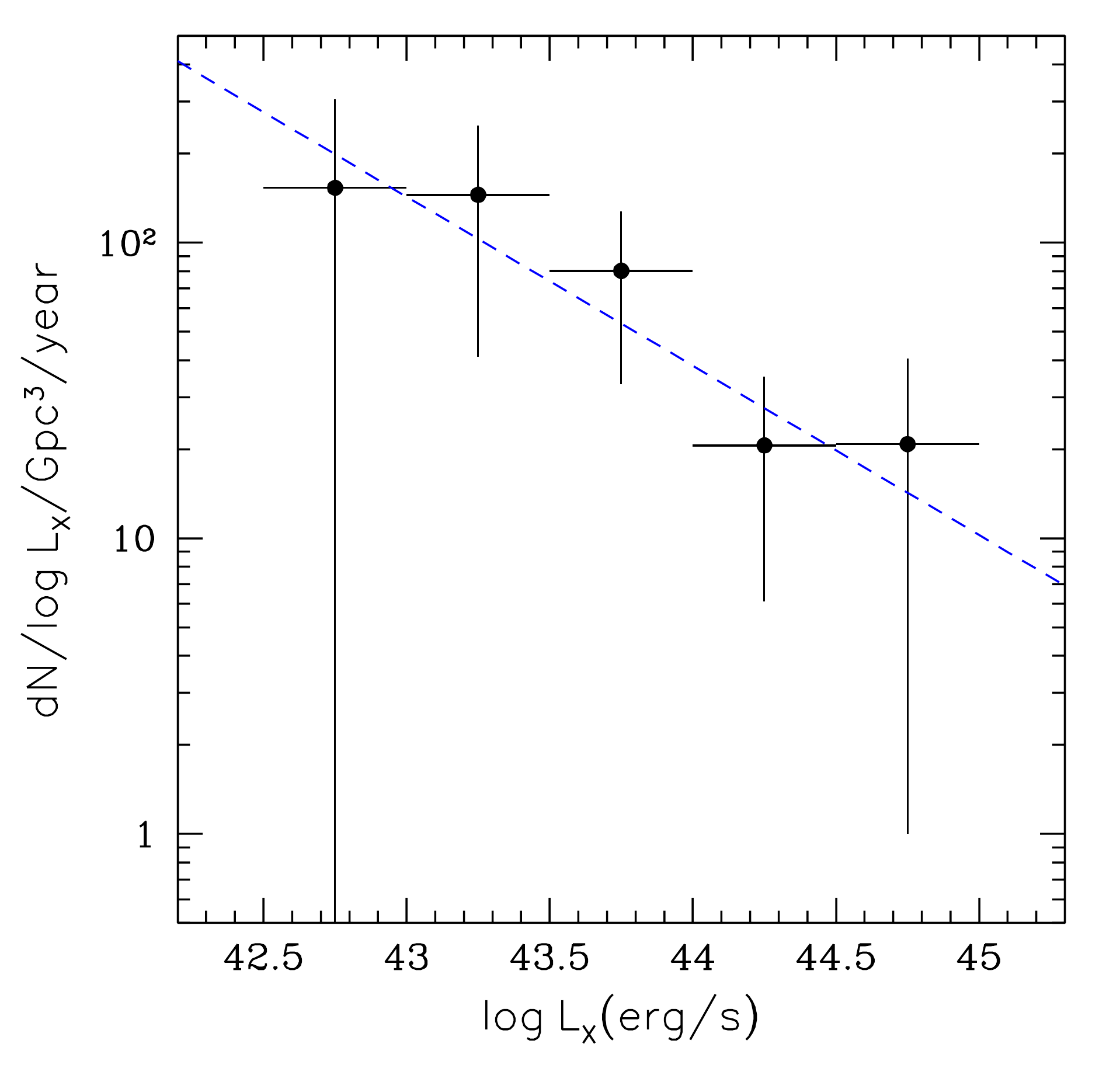}
    \caption{TDE X-ray (0.2--6\,keV) luminosity function. The blue dashed line shows its best bit by a power law.}
    \label{fig:xlf}
\end{figure}

The computed TDE XLF is shown in Fig.~\ref{fig:xlf}. The largest uncertainty is associated with the lowest luminosity bin, $10^{42.5}<\log\lx<10^{43}$\,erg\,s$^{-1}$, where there is just one, nearby TDE (SRGE\,J171423.6+085236 at $z=0.036$) in the current \srg/\erosita\ sample. There is a clear indication of the TDE volume rate declining with increasing X-ray luminosity. Fitting the XLF (using $\chi^2$ minimization) by a power law, we obtain (see Fig.~\ref{fig:xlf}):
\begin{equation}
\frac{dN_{\rm TDE}}{d\log\lx\,dV\,dt}=(1.4\pm 0.8)\times 10^{-7}\left(\frac{\lx}{10^{43}\,{\rm erg}\,{\rm s}^{-1}}\right)^{-0.6\pm0.2}\,{\rm Mpc}^{-3}\,{\rm yr}^{-1}.
\label{eq:pl}
\end{equation}
If we exclude the $10^{42.5}<\log\lx<10^{43}$\,erg\,s$^{-1}$ bin from this calculation, the slope of the power law becomes somewhat steeper, $\alpha=-0.8\pm 0.3$, but consistent with the slope in equation~(\ref{eq:pl}). 

According to theoretical predictions, TDEs should occur more frequently in lower mass galaxies with correspondingly lighter central black holes   (\citealt{Magorrian_1999,Wang_2004,Kesden_2012}; see \citealt{Stone_2020} for a recent review). This, together with the higher abundance of low-mass galaxies compared to high-mass ones, the approximately Eddington luminosities of TDEs, and the expectation that TDEs associated with smaller black holes should emit a larger fraction of their bolometric luminosity in X-rays (due to their hotter accretion disks), suggests that the X-ray TDE rate should increase within decreasing $\lx$. Apparently, we are starting to see this trend in the data of the \srg\ all-sky survey. 

We note in this connection that the nearest and least X-ray luminous TDE in our sample, SRGE\,J171423.6+085236, has the hardest X-ray spectrum ($\tin\sim 0.5$\,keV). This suggests that it is also associated with the smallest black hole in the sample, with $\mbh\sim 10^{4}\,\msun$, albeit with a large uncertainty in this estimate based on our simplistic X-ray spectral modeling (see Table~\ref{tab:optxagnf}). On the other hand, its host galaxy is fairly massive, with stellar mass $M_{\rm gal}\approx 1.7\times 10^{10}\,\msun$ (see Table~\ref{tab:host}).

The declining trend of the TDE rate with increasing X-ray luminosity inferred here is similar to the trend noticed before for optically-UV selected TDEs. In that case, the measured slope of the luminosity function is $\alpha=-1.3\pm 0.3$ \citep{Velzen_2018}, somewhat steeper than inferred here for X-ray selected TDEs. However, this difference is only marginally significant.

\subsubsection{Total rate}
\label{s:totalrate}

By integrating the XLF over the entire luminosity range probed here ($10^{42.5}<\log\lx<10^{45}$\,erg\,s$^{-1}$) we can estimate the average TDE volumetric rate in the $z=0$--0.6 Universe: $(2.1\pm 1.0)\times 10^{-7}$\,Mpc$^{-3}$\,year$^{-1}$. Given the total galaxy volume density of $\sim 2\times 10^{-2}$\,Mpc$^{-3}$ \citep{Bell_2003}, this translates to a rate of $R=(1.1\pm 0.5)\times 10^{-5}$ TDEs per galaxy. 

This estimated specific TDE rate is consistent with the earliest estimate of $R\sim 9\times 10^{-6}$\,yr$^{-1}$ per galaxy, based on 3 TDEs detected during the \rosat\ all-sky survey and followed-up during \rosat\ pointed observations \citep{Donley_2002}. Another published estimate is much higher, $R\sim 2\times 10^{-4}$\,yr$^{-1}$ per galaxy \citep{Esquej_2008}, but it is based on just 2 TDEs discovered during the \xmm\ Slew Survey \citep{Saxton_2008} and undetected previously during the \rosat\ all-sky survey. Yet another estimate was presented by \cite{Khabibullin_2014a}, who used a sample of 4 candidate TDEs selected based on their brightness during the \rosat\ all-sky survey and non-detection in subsequent \xmm\ pointed observations: $R\sim 3\times 10^{-5}$\,yr$^{-1}$ per galaxy. In summary, all these previous estimates were based on extremely small samples of X-ray selected TDEs, and within their large (and poorly defined) uncertainties appear to be in line with the new result obtained here based on the first \srg\ TDE sample. 

For comparison, the existing estimates of the total rate of optically-UV selected TDEs are $\sim 10^{-4}$\,yr$^{-1}$ per galaxy \citep{Velzen_2020}, which is significantly larger than our X-ray-based estimate, and is in better agreement with theoretical ``loss cone'' predictions \citep{Stone2016}. This possibly indicates that the XLF shown in Fig.~\ref{fig:xlf} continues to rise below $\lx\sim 10^{42.5}$\,erg\,s$^{-1}$. However, the optical TDE rate, currently determined for $\lopt>10^{42.5}$\,erg\,s$^{-1}$, may also increase if less luminous TDEs are included.  

The lower volumetric rate of X-ray TDEs compared to that of optical TDEs may also imply that X-ray bright TDEs constitute a minority of all TDEs. The latter possibility would provide support to TDE models (e.g. \citealt{dai2018,Curd_2019}) that predict a strong dependence of the optical/X-ray brightness ratio on the viewing angle: namely, that we can only observe TDEs (during their luminous early phase) in X-rays from directions close to the axis of a thick accretion disk (formed from the debris of the disrupted star) with a powerful wind, while at larger inclination angles one can only observe the reprocessed optical-UV emission. In this connection, we also note that we have found no evidence of intrinsic absorption in the X-ray spectra of the 13 studied TDEs. However, since not only the viewing direction but also the black hole mass is expected to play a key role in diversity of TDE types \citep{Mummery_2021}, a more elaborate comparison between the TDE population properties inferred from this work and from optical studies should be done. In addition, slow or inefficient circularization of the stellar debris \citep{Piran_2015,Shiokawa2015} could also lead to a paucity of X-ray TDEs.

In future work, we plan to lower our X-ray detection threshold by a factor of $\sim 2$, which should increase the TDE discovery rate by \erosita\ by a factor of $\sim 3$ (assuming that the TDE XLF continues to grow to lower luminosities, so that the resulting \erosita\ sample is dominated by low-redshift events). Given that our current sample of bright TDEs detected on one half of the sky over a period of half a year comprises 16 confirmed events, this implies that $\sim 200$\,TDEs per year can be detected by \erosita\ over the whole sky and a total of $\sim 700$\,TDEs can be discovered by the end of the 4-year \srg\ survey (excluding from this calculation the first of the 8 planned sky scans). These numbers are consistent within an order of magnitude with the predictions done for the \srg/\erosita\ all-sky survey prior to its beginning \citep{khabibullin2014b,Jonker_2020}, which, however, strongly depend on the distribution of spins of the black holes associated with TDEs. Continued search during the \srg\ survey will allow us to further narrow down the rate of stellar disruptions in galactic nuclei and place tighter constraints on the underlying population of black holes. 

\subsection{Optical faintness}
\label{s:xrayopt}

\begin{figure}
\centering
\includegraphics[width=\columnwidth]{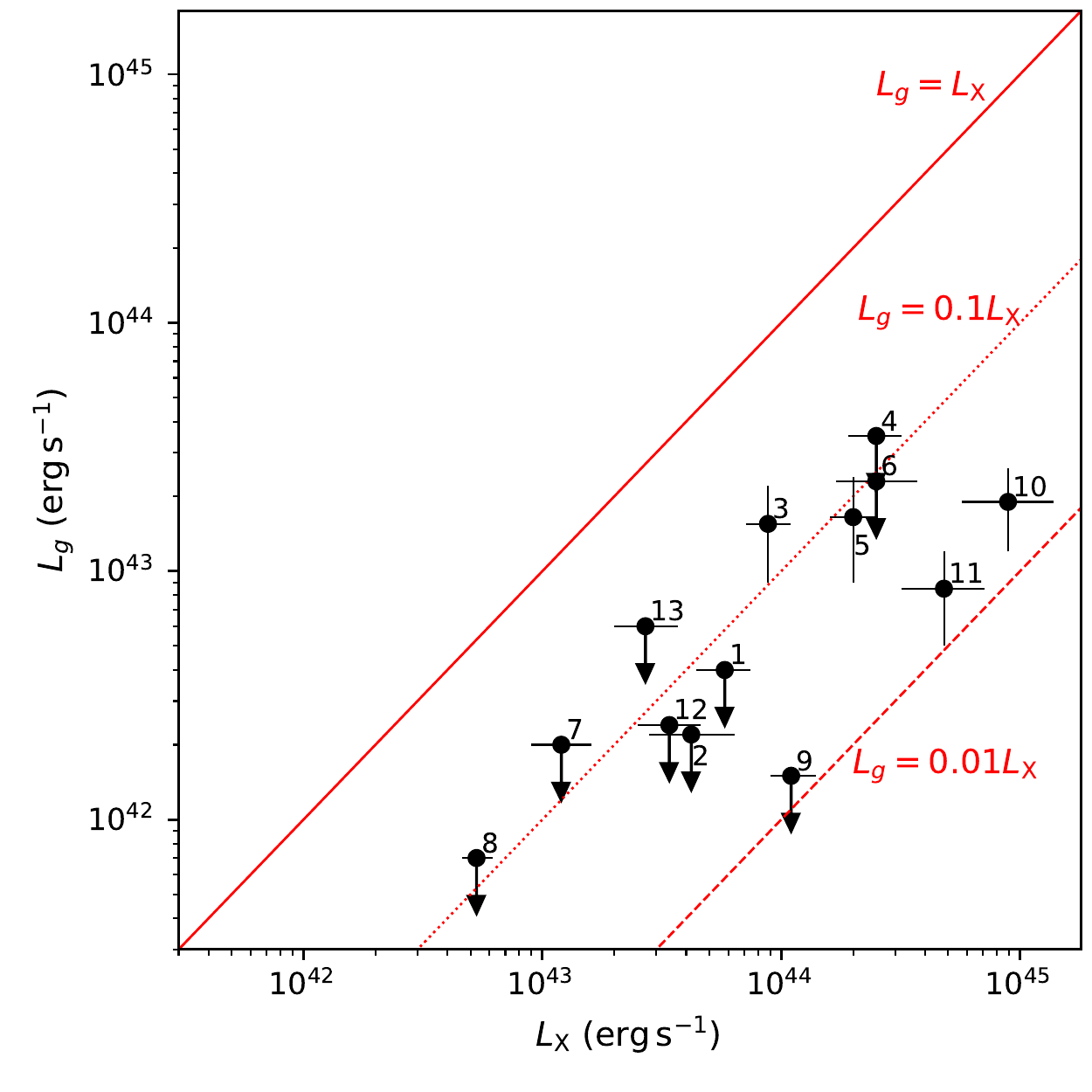}
    \caption{Rest-frame $g$-band luminosity vs. rest-frame 0.2--6\,keV luminosity for the \srg\ TDEs (labeled by the internal numbers in Table~\ref{tab:sample}).}
    \label{fig:lx_lopt}
\end{figure}

A salient feature of the TDEs discovered by \srg/\erosita\ and studied here is their optical faintness. Figure~\ref{fig:lx_lopt} shows the inferred optical vs. X-ray luminosities (from Table~\ref{tab:properties}) for our sample. For the four TDEs that have shown noticeable optical flares in addition to X-ray activity, the $\lopt/\lx$ ratio is constrained between 0.01 and 0.3. For the remaining TDEs, $\lopt/\lx<0.3$ and for most of them $\lopt/\lx<0.1$. 

These constraints are conservative. Indeed, the possibility that the peak of TDE optical emission has been missed is already taken into account in our $\lopt$ estimates (see \S\ref{s:optlum}). On the other hand, as discussed before, our $\lx$ estimates should be considered lower limits on the maximum X-ray luminosity. Therefore, the $\lopt/\lx$ ratios can actually be  smaller than shown in Fig.~\ref{fig:lx_lopt}.

The \srg\ TDEs studied here are similar to the very first TDEs that came to light thanks to the \rosat\ X-ray observatory three decades ago, but drastically different from those discovered recently in the optical/UV, for which typically $\lopt>\lx$ (e.g. \citealt{Gezari2021}).
Although it is certainly too early to draw firm conclusions on the underlying physical picture, it is possible that we are dealing here with the same orientation effect that was discussed above in relation to the observed TDE volumetric rate. Namely, that most of the TDEs detected by \srg\ are observed from small angles with respect to the accretion disk axis, whereas optical-UV surveys catch TDEs from quasi-random inclination angles, so that most of them prove to be X-ray faint. 

We note, however, that apart from optically faint TDEs \srg/\erosita\ has also discovered TDEs with prominent optical activity \citep{Gilfanov_2020,Gilfanov_2021}. Several first events of this kind will be discussed by Gilfanov et al. (in prep.). Postponing further discussion of the relationship between optically bright and optically faint X-ray selected TDEs to that paper, we note that the former appear to constitute $\sim 20$\% of TDEs found during the the \srg\ all-sky survey.  

\section{Conclusions}
\label{s:conclusion}

We have presented the first sample of TDEs discovered during the on-going \srg\ all-sky survey. These 13 events were selected among the multitude of X-ray transients detected by \erosita\ during its second scan of the sky (June 10 -- December 14, 2020) at a level exceeding at least tenfold the upper limit on the flux in the first scan and confirmed as TDEs by our optical follow-up observations. The most distant of these events (SRG\,J163831.7+534020) occurred at $z=0.581$. Therefore, the \srg\ survey has already expanded the horizon of TDE X-ray detectability by a factor of $\sim 4$ compared to the \rosat\ all-sky survey, conducted 30 years ago. 

The properties of these events are similar to those of TDEs detected (in small numbers) by previous X-ray missions. Namely, the X-ray light curves, currently consisting of a flux measured by \erosita\ during the second sky survey and a second measurement/upper limit obtained 6 months later, are in most cases consistent with a $t^{-5/3}$ decline that started shortly before the \erosita\ discovery. Particularly interesting is the TDE SRGE\,J144738.4+671821, which continued to brighten after its discovery for at least another 6 months. 

The early (eRASS2) X-ray spectra of these TDEs can be described by a multi-blackbody accretion disk model with $k\tin$ between $\approx 0.05$ and $0.5$\,keV, which is consistent with nearly critical accretion onto black holes with masses between a few $\times 10^3$ and $\sim 10^8\,\msun$ depending on their (unknown) spins. In reality, we may be dealing with supercritical accretion in most of these events. No evidence of intrinsic absorption is found in the X-ray spectra. Four TDEs remained sufficiently bright in eRASS3 to allow us to analyze their spectra taken by \erosita\ at this later epoch. In two cases, we observe a clear spectral hardening, possibly indicating the formation of an accretion disk corona. 

Four TDEs have shown a brightening in their optical light curves, concurring with or preceding the X-ray outburst. One of these events (SRGE\,J163030.2+470125) also exhibited a blue excess in an optical spectrum taken two months after the \erosita\ discovery, which disappeared in a spectrum obtained eight months later. The other nine TDEs show no signs of optical activity associated with stellar tidal disruption in existing photometric and spectroscopic data. 

All of the TDEs (including those with optical flares) are optically faint in the sense that the estimated optical/X-ray luminosity ratio is less than 0.3 and in most cases $\lopt/\lx<0.1$. In this respect, this sample is drastically different from TDEs selected at optical-UV wavelengths, which typically have $\lopt/\lx>1$. However, apart from the events presented in this work, \srg/\erosita\ has also discovered a few TDEs that demonstrate prominent activity in the optical-UV band (Gilfanov et al., in prep.).

The \srg\ TDEs are mostly hosted by galaxies located in the green valley on the color-stellar mass diagram, similarly to optically selected TDEs. However, no excess of quiescent post-starburst galaxies or E+A galaxies is observed among the hosts of the \srg\ TDEs, unlike in some previous studies of optically selected TDEs. This may be partially accounted for by the fact that the host galaxies of the \srg\ TDEs are more massive on average. 

We have constructed the X-ray luminosity function in the range from $10^{42.5}$ to $10^{45}$\,erg\,s$^{-1}$, the first of its kind to our knowledge. It clearly shows that the TDE volumetric rate decreases with increasing X-ray luminosity. This trend can be described by a power law with a slope of $\alpha=-0.6\pm 0.2$ ($dN/d\log\lx\propto \lx^\alpha$), or $\alpha=-0.8\pm 0.3$ if we exclude from consideration the lowest luminosity bin of $10^{42.5}<\log\lx<10^{43}$\,erg\,s$^{-1}$, which contains just one TDE. This behavior is similar to the trend previously observed for optically selected TDEs, but in that case the decline is marginally steeper, with $\alpha=-1.3\pm0.3$.

The total rate of X-ray TDEs is estimated at $(1.1\pm 0.5)\times 10^{-5}$ events per galaxy per year. This is an order of magnitude lower than recent estimates of optical TDEs, which possibly indicates that the TDE XLF continues to rise below $\lx\sim 10^{42.5}$\,erg\,s$^{-1}$ and/or that X-ray bright events constitute a minority of TDEs. The latter possibility would provide support to the TDE models that predict a strong dependence on the viewing angle, namely, that TDEs can only be observed in X-rays from directions close to the axis of a thick accretion disk formed from the debris of the disrupted star. 

The \srg\ all-sky survey is to continue until the end of 2023, opening up exciting prospects for TDE studies. In particular, we plan to lower our detection threshold for such events by a factor of $\sim 2$, which should increase the \erosita\ TDE discovery rate by a factor of $\sim 3$. This implies that $\sim 700$ TDEs can be found by the end of the 4-year \srg\ survey over the entire sky. Furthermore, regular (every 6 months) visits by \erosita\ of previously discovered TDEs will provide valuable information on their long-term X-ray evolution. 

To conclude, a continued search for TDEs during the \srg\ all-sky survey should allow us to narrow down the rate of stellar disruptions in galactic nuclei, tighten constraints on the properties of the underlying population of supermassive and, perhaps, intermediate-mass black holes, and shed light on the physics of near- and super-critical accretion onto such objects. 

\section*{Acknowledgements}
This work is based on observations with the \erosita\ telescope on board the \srg\ observatory. The \srg\ observatory was built by Roskosmos in the interests of the Russian Academy of Sciences represented by its Space Research Institute (IKI) in the framework of the Russian Federal Space Program, with the participation of the Deutsches Zentrum f\"{u}r Luft- und Raumfahrt (DLR). The \srg/\erosita\ X-ray telescope was built by a consortium of German Institutes led by MPE, and supported by DLR. The \srg\ spacecraft was designed, built, launched, and is operated by the Lavochkin Association and its subcontractors. The science data are downlinked via the Deep Space Network Antennae in Bear Lakes, Ussurijsk, and Baykonur, funded by Roskosmos. The \erosita\ data used in this work were processed using the eSASS software system developed by the German \erosita\ consortium and proprietary data reduction and analysis software developed by the Russian \erosita\ Consortium. 

The observations at the 6-m telescope of the Special Astrophysical Observatory, Russian Academy of Sciences, were carried out with the financial support of the Ministry of Science and Higher Education of the Russian Federation (including agreement No. 05.619.21.0016, project ID RFMEFI61919X0016). The observations at the AZT-33IK telescope were performed within the basic financing of the FNI II.16 program, using the equipment of the Angara sharing center\footnote{http://ckp-rf.ru/ckp/3056/}. CMO SAI MSU observations are supported by the M.V. Lomonosov Moscow State University Program of Development. The authors are grateful to T\"{U}BITAK, IKI, KFU, and AST for their partial support in using RTT150 (the Russian--Turkish 1.5-m telescope in Antalya).

SS, MG, PM, RB, AM, and GH acknowledge the support of this research of by grant 21-12-00343 from the Russian Science Foundation.
Y. Yao thanks the Heising–Simons Foundation for financial support. The work of KAP and AVD is supported by the Ministry of science and higher education of Russia under contract 075-15-2020-778 (observations of objects with extreme energy release) in the framework of the Large scientific projects program within the national project ``Science''. The work of AAB and AMCh was supported by the Scientific and Educational School of M.V. Lomonosov Moscow State University ``Fundamental and applied space research''. The work of IFB and RIG was supported by subsidy 0671-2020 0052 allocated to the Kazan Federal University for state assignment in the sphere of scientific activities.

\section*{Data availability}
X-ray data analysed in this article were used by permission of the Russian \srg/\erosita\ consortium. The data will become publicly available as part of the corresponding \srg/\erosita\ data release along with the appropriate calibration information. Optical data used in the article will be shared on reasonable request to the corresponding author.

\appendix

\section{Short-term X-ray light curves}
\label{s:shortlc}

\begin{figure*}
\centering
\includegraphics[width=0.95\textwidth]{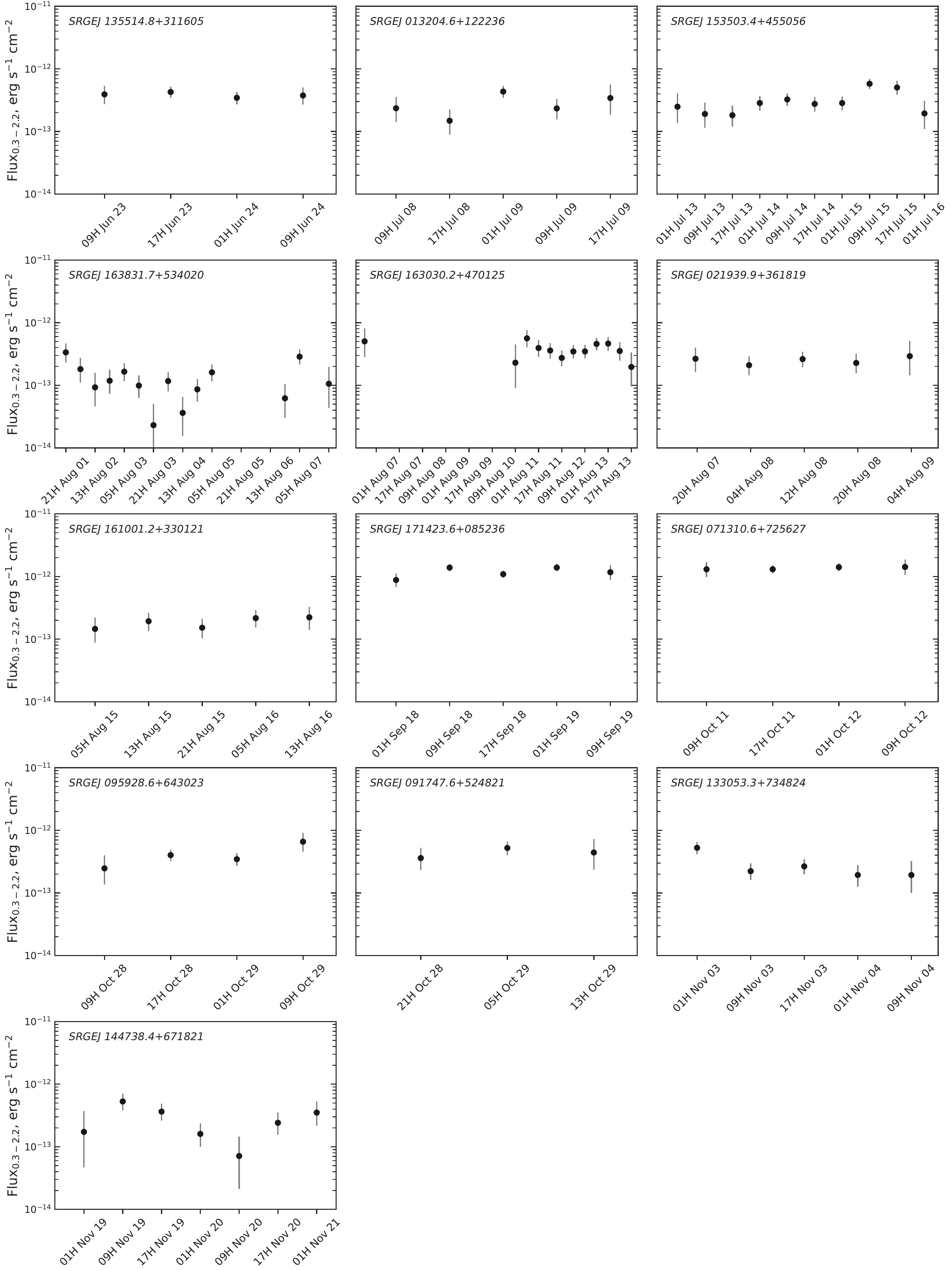}
    \caption{Short-term X-ray light curves of the TDEs obtained by \erosita\ in the 0.3--2.2\,keV energy range in 2020 (during eRASS2).
    }
    \label{fig:lc2}
\end{figure*}

\begin{figure*}
\centering
\includegraphics[width=0.65\textwidth]{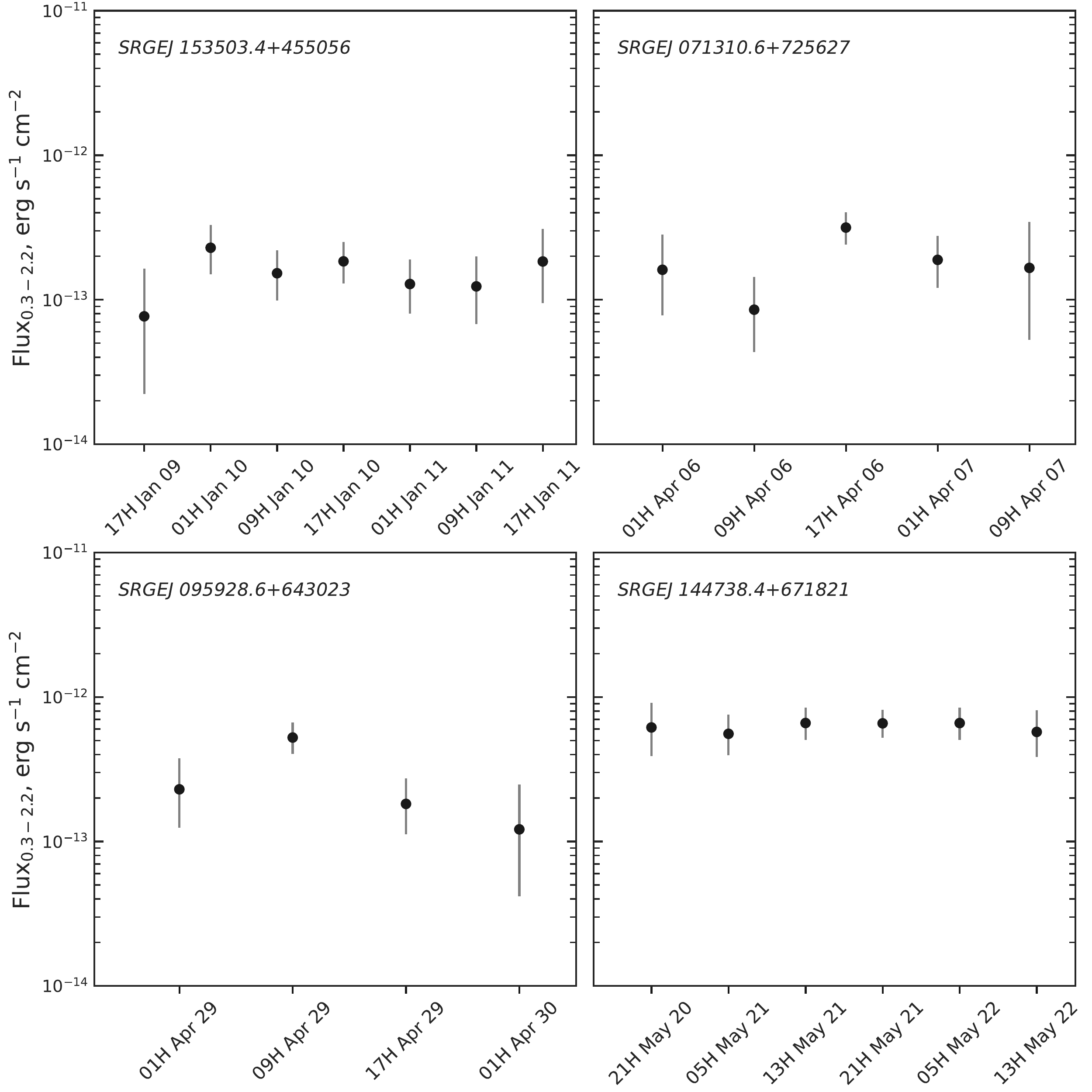}
    \caption{Short-term X-ray light curves of several TDEs obtained by \erosita\ in the 0.3--2.2\,keV energy range in 2021 (during eRASS3).
    }
    \label{fig:lc3}
\end{figure*}

During each \srg\ all-sky survey, \erosita\ monitors any X-ray source in the sky for a period of at least one day. Each such visit consists of a series of $\sim 40$\,s-long measurements taken at 4-hour intervals. Therefore, apart from the long-term X-ray light curves of the TDEs presented in Fig.~\ref{fig:xraylc}, we can also construct their short-term light curves during eRASS2 and eRASS3 (if a given transient remained detectable during eRASS3). These light curves are presented in Figs.~\ref{fig:lc2} and \ref{fig:lc3}. They span between $\sim 1$ and $\sim 7$ days, depending on the ecliptic latitudes, $\becl$, of the TDEs (as a result of the \srg\ survey's strategy, \citealt{Sunyaev_2021}). 

None of the TDEs exhibit substantial X-ray variability on time scales between 4\,hours and a few days. This may place interesting constraints on TDE models. 

\section{Optical observational details and data reduction} 
\label{subsec:details}

All Keck-I/LRIS spectroscopic observations were conducted using the blue side grism 300/3400, the dichroic 560, the red side grating 400/8500, and a slit mask of 1$^{\prime\prime}$. This setup provides a spectral coverage of 3200--10250\,\AA. All LRIS spectra were reduced and extracted using \texttt{Lpipe} \citep{Perley2019lpipe}. 

BTA/SCORPIO-2 spectroscopic observations were conducted with the VPHG940@600 grating, providing a spectral range of 3500--8500\,\AA. The spectrum of SRGE\,J135514.8+311605 was obtained with a slit of 1$^{\prime\prime}$ and that of SRGE\,J163030.2+470125 with a slit of 2$^{\prime\prime}$. During both observations, the seeing was near 1.2\arcsec. SCORPIO-2 photometry was done in a $1024\times1024$~pixel imaging mode, corresponding to a binning of 2$\times$2 ($0\arcsec\!.40$ pixel scale).

Photometry at the CMO’s 2.5-m telescope was conducted using the NBI $4096\times2048\times2$ CCD-photometer with a $0\arcsec\!.155$ pixel scale. Spectroscopy at the 2.5-m telescope was done using the TDS instrument. The spectra were obtained in a range of 3600--7500\,\AA\ with a 1\arcsec\ slit, at a spectral resolution of $\sim 1500$. The spectrograph and data reduction are described in \citep{Potanin_2020}. Photometry at the CMO's RC600 telescope was carried out using the Andor iKon-L camera with $2048\times2048$ pixels and a pixel scale of $0\arcsec\!.67$.

AZT-33IK/ADAM spectra were obtained using a 2\arcsec\ slit and the VPHG600G grating (spectral range 3700--7340\,\AA, resolution 8.8\,\AA). Photometric observations at AZT-33IK were done using the CCD photometer, consisting of a focal reducer and the Andor iKon-M 934 camera with $1024\times1024$ pixels, which provide a $6.3\arcmin \times 6.3\arcmin$ field of view with a $0\arcsec\!.372$ pixel scale. 

Photometric observations at RTT150 were conducted using the TFOSC instrument equipped with an ANDOR iKon-L DZ936N CCD
$2048\times2048$ detector with a $0\arcsec\!.327$ pixel scale. 

Images and spectroscopy from BTA, AZT-33IK, and RTT150 were processed using \textsc{iraf} and our own software. Aperture photometry was done using the \emph{apphot} task from the \textsc{iraf} \emph{digiphot} package. Measurement of source fluxes was done relative to nearby bright stars, with the aperture size for each observation series defined such that the total flux was obtained.

The obtained magnitudes were calibrated using the magnitudes of secondary photometric standards in the source field with a brightness comparable to or greater than the flux of the target object. We used PSF magnitudes for the calibration stars from SDSS \citep{sdssdr16} if available, or from Pan-STARRS1 DR2 \citep{ps1} otherwise.

\section{Host Galaxy Analysis.}
\label{sec:hostfit}

\begin{figure*}
\begin{subfigure}[t]{0.32\textwidth}
\centering
\includegraphics[width=\linewidth]{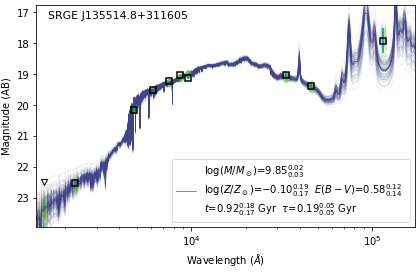}
\end{subfigure}
\begin{subfigure}[t]{0.32\textwidth}
\centering
\includegraphics[width=\linewidth]{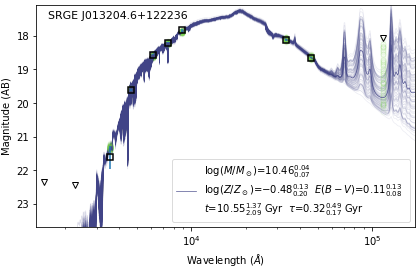}
\end{subfigure}
\begin{subfigure}[t]{0.32\textwidth}
\centering
\includegraphics[width=\linewidth]{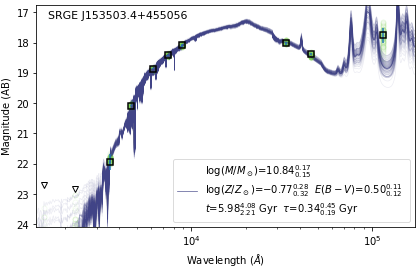}
\end{subfigure}
\begin{subfigure}[t]{0.32\textwidth}
\centering
\includegraphics[width=\linewidth]{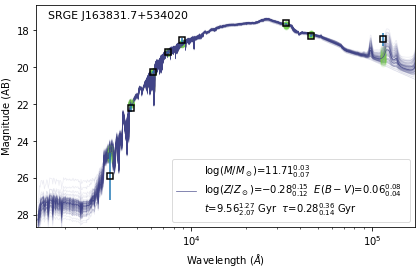}
\end{subfigure}
\begin{subfigure}[t]{0.32\textwidth}
\centering
\includegraphics[width=\linewidth]{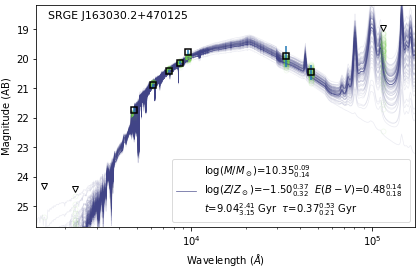}
\end{subfigure}
\begin{subfigure}[t]{0.32\textwidth}
\centering
\includegraphics[width=\linewidth]{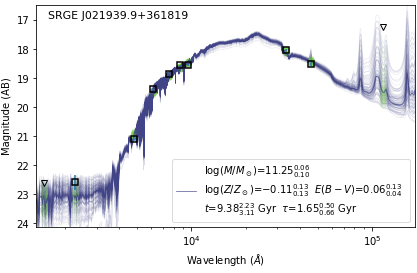}
\end{subfigure}
\begin{subfigure}[t]{0.32\textwidth}
\centering
\includegraphics[width=\linewidth]{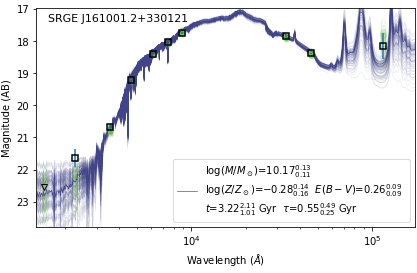}
\end{subfigure}
\begin{subfigure}[t]{0.32\textwidth}
\centering
\includegraphics[width=\linewidth]{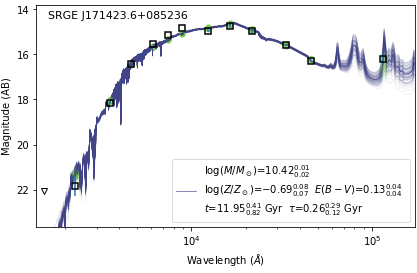}
\end{subfigure}
\begin{subfigure}[t]{0.32\textwidth}
\centering
\includegraphics[width=\linewidth]{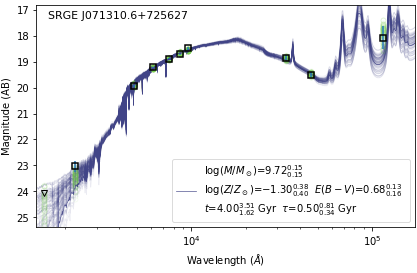}
\end{subfigure}
\begin{subfigure}[t]{0.32\textwidth}
\centering
\includegraphics[width=\linewidth]{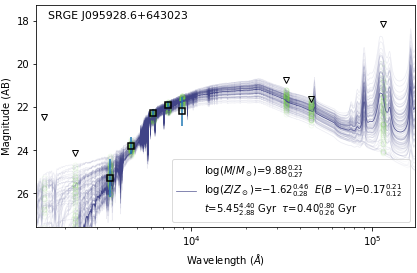}
\end{subfigure}
\begin{subfigure}[t]{0.32\textwidth}
\centering
\includegraphics[width=\linewidth]{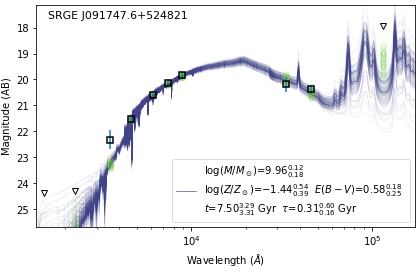}
\end{subfigure}
\begin{subfigure}[t]{0.32\textwidth}
\centering
\includegraphics[width=\linewidth]{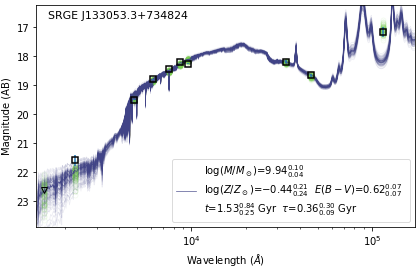}
\end{subfigure}
\begin{subfigure}[t]{0.32\textwidth}
\centering
\includegraphics[width=\linewidth]{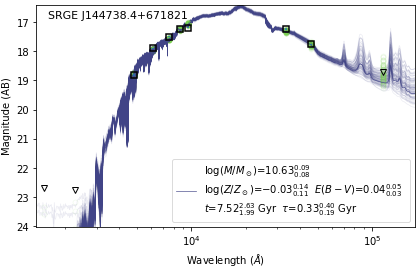}
\end{subfigure}
\caption{SED of TDE host galaxies. Squares (downward triangles) are observed photometric detections (upper limits) corrected by Galactic extinction. For each source, the purple lines show models of the 100 walkers in the Markov Chain Monte Carlo (MCMC) sampler. 
\label{fig:popsyn}}
\end{figure*}

As described in \S\ref{subsec:host_properties}, we fit the historical photometry of the TDE hosts following the approach used by \cite{Velzen_2021}. The FSPS \citep{Conroy2009} and Prospector \citep{Johnson2021} packages are used to find the synthetic galaxy model that is the best description for the host galaxy photometry. Figure~\ref{fig:popsyn} present the host SEDs and the fitted models. The fitted parameters are given in Table~\ref{tab:host}.

\begin{table*}
    \centering
    \caption{Host properties inferred from SED fitting\label{tab:host}}
    \begin{tabular}{ccccccc}
    \hline 
        Object & $M_{\rm gal}$ & $^{0.0}u-r$ & $\tau_{\rm sfh}$ & age & $Z$ & dust\\
         & ${\rm log}_{10}\,M_\odot$ &  & Gyr & Gyr & ${\rm log}_{10}\,Z_\odot$ & $E(B-V)$\\
    \hline
SRGE\,J135514.8+311605 &
$9.85_{-0.03}^{+0.02}$ &
$1.69_{-0.03}^{+0.04}$ &
$0.19_{-0.05}^{+0.05}$ &
$0.92_{-0.17}^{+0.18}$ &
$-0.10_{-0.17}^{+0.19}$ &
$0.58_{-0.14}^{+0.12}$ \\
SRGE\,J013204.6+122236 &
$10.46_{-0.07}^{+0.04}$ &
$2.31_{-0.03}^{+0.05}$ &
$0.32_{-0.17}^{+0.49}$ &
$10.55_{-2.09}^{+1.37}$ &
$-0.48_{-0.20}^{+0.13}$ &
$0.11_{-0.08}^{+0.13}$ \\
SRGE\,J153503.4+455056 &
$10.84_{-0.15}^{+0.17}$ &
$2.24_{-0.04}^{+0.04}$ &
$0.34_{-0.19}^{+0.45}$ &
$5.98_{-2.21}^{+4.08}$ &
$-0.77_{-0.32}^{+0.28}$ &
$0.50_{-0.12}^{+0.11}$ \\
SRGE\,J163831.7+534020 &
$11.71_{-0.07}^{+0.03}$ &
$2.36_{-0.04}^{+0.04}$ &
$0.28_{-0.14}^{+0.36}$ &
$9.56_{-2.07}^{+1.27}$ &
$-0.28_{-0.12}^{+0.15}$ &
$0.06_{-0.04}^{+0.08}$ \\
SRGE\,J163030.2+470125 &
$10.35_{-0.14}^{+0.09}$ &
$2.03_{-0.06}^{+0.07}$ &
$0.37_{-0.21}^{+0.53}$ &
$9.04_{-3.15}^{+2.41}$ &
$-1.50_{-0.32}^{+0.37}$ &
$0.48_{-0.18}^{+0.14}$ \\
SRGE\,J021939.9+361819 &
$11.25_{-0.10}^{+0.06}$ &
$2.23_{-0.05}^{+0.04}$ &
$1.65_{-0.66}^{+0.50}$ &
$9.38_{-3.11}^{+2.23}$ &
$-0.11_{-0.13}^{+0.13}$ &
$0.06_{-0.04}^{+0.13}$ \\
SRGE\,J161001.2+330121 &
$10.17_{-0.11}^{+0.13}$ &
$2.05_{-0.05}^{+0.05}$ &
$0.55_{-0.25}^{+0.49}$ &
$3.22_{-1.01}^{+2.11}$ &
$-0.28_{-0.16}^{+0.14}$ &
$0.26_{-0.09}^{+0.09}$ \\
SRGE\,J171423.6+085236 &
$10.42_{-0.02}^{+0.01}$ &
$2.26_{-0.02}^{+0.02}$ &
$0.26_{-0.12}^{+0.29}$ &
$11.95_{-0.82}^{+0.41}$ &
$-0.69_{-0.07}^{+0.08}$ &
$0.13_{-0.04}^{+0.04}$ \\
SRGE\,J071310.6+725627 &
$9.72_{-0.15}^{+0.15}$ &
$2.05_{-0.04}^{+0.04}$ &
$0.50_{-0.34}^{+0.81}$ &
$4.00_{-1.62}^{+3.51}$ &
$-1.30_{-0.40}^{+0.38}$ &
$0.68_{-0.16}^{+0.13}$ \\
SRGE\,J095928.6+643023 &
$9.88_{-0.27}^{+0.21}$ &
$1.73_{-0.14}^{+0.17}$ &
$0.40_{-0.26}^{+0.80}$ &
$5.45_{-2.88}^{+4.40}$ &
$-1.62_{-0.28}^{+0.46}$ &
$0.17_{-0.12}^{+0.21}$ \\
SRGE\,J091747.6+524821 &
$9.96_{-0.18}^{+0.12}$ &
$2.10_{-0.06}^{+0.06}$ &
$0.31_{-0.16}^{+0.60}$ &
$7.50_{-3.31}^{+3.29}$ &
$-1.44_{-0.39}^{+0.54}$ &
$0.58_{-0.25}^{+0.18}$ \\
SRGE\,J133053.3+734824 &
$9.94_{-0.04}^{+0.10}$ &
$1.89_{-0.05}^{+0.05}$ &
$0.36_{-0.09}^{+0.30}$ &
$1.53_{-0.25}^{+0.84}$ &
$-0.44_{-0.24}^{+0.21}$ &
$0.62_{-0.07}^{+0.07}$ \\
SRGE\,J144738.4+671821 &
$10.63_{-0.08}^{+0.09}$ &
$2.42_{-0.06}^{+0.10}$ &
$0.33_{-0.19}^{+0.40}$ &
$7.52_{-1.99}^{+2.63}$ &
$-0.03_{-0.11}^{+0.14}$ &
$0.04_{-0.03}^{+0.05}$ \\
\hline
    \end{tabular}
\end{table*}

\bibliographystyle{mnras} 
\bibliography{tde}

\end{document}